\documentclass{jfm}
\usepackage{amsmath}
\usepackage{amssymb}
\usepackage{bm}
\usepackage{booktabs}
\usepackage{color}
\usepackage{CJKutf8}
\usepackage{dashrule}
\usepackage{epsfig}
\usepackage{epstopdf}
\usepackage{float}
\usepackage{graphicx}
\usepackage{hyperref}
\hypersetup{
	colorlinks = true,
	urlcolor   = blue,
	citecolor  = blue}
\usepackage{listings}
\usepackage{lipsum}
\usepackage{makecell}
\usepackage{mathtools}
\usepackage{multicol}
\usepackage{multirow}
\usepackage{newtxtext}
\usepackage{newtxmath}
\usepackage{natbib}
\usepackage{overpic}
\usepackage{pgfplots}
\usepackage{pifont}
\usepackage{soul}
\usepackage{subfigure}
\usepackage{tikz,tikz-3dplot,ifthen}
\usetikzlibrary{arrows.meta}
\newcommand\Larrow[1]{\mathrel{\raisebox{0.4mm}{\!\!
			\begin{tikzpicture}[>=stealth]
				\node[inner sep=1ex] (a) {$\scriptstyle #1$};
				\draw[<-, line width = 0.8pt] ($(a.south west)+(0,0.05045)$) --($(a.south east)+(0,0.05045)$);
\end{tikzpicture}}}}

\usetikzlibrary{calc}
\usetikzlibrary{circuits.ee.IEC}
\usetikzlibrary{decorations,decorations.pathreplacing} 
\usepackage{verbatim}
\usepackage[dvipsnames]{xcolor}
\usepackage[svgnames]{xcolor}

\makeatletter
\newsavebox{\@brx}
\newcommand{\llangle}[1][]{\savebox{\@brx}{\(\m@th{#1\langle}\)}%
	\mathopen{\copy\@brx\kern-0.5\wd\@brx\usebox{\@brx}}}
\newcommand{\rrangle}[1][]{\savebox{\@brx}{\(\m@th{#1\rangle}\)}%
	\mathclose{\copy\@brx\kern-0.5\wd\@brx\usebox{\@brx}}}
\makeatother

\newcommand\Mach{\mbox{\textit{Ma}}}  % Reynolds number
\newcommand\Stokes{\mbox{\textit{St}}}  % Stokes number

\newcommand{\RomanNumeralCaps}[1]
\title{Asymmetric particle transport in turbulent flows within concentric annular ducts}
\shorttitle{Journal of Fluid Mechanics}
\shortauthor{T. Wang, C. Zhang and Y. Zhao}

\author{Taiyang Wang\aff{1,2,*},
	Chi Zhang\aff{1,2,*},
	Yaomin Zhao\aff{1,2,\corresp{Yaomin Zhao, \email{yaomin.zhao@pku.edu.cn}}}}

\affiliation{
	\aff{1}HEDPS, Center for Applied Physics and Technology, School of Mechanics and Engineering Science, Peking University, Beijing 100871, China
	\aff{2}State Key Laboratory for Turbulence and Complex Systems, School of Mechanics and Engineering Science, Peking University, Beijing 100871, China
	\aff{*} These authors contributed equally to this work.
}

\begin{document}
\maketitle

\begin{abstract}
We present the first direct numerical simulations of particle-laden turbulent flow in concentric annuli to investigate the effects of transverse curvature over a range of Stokes numbers. 
The results demonstrate that transverse curvature induces asymmetric radial transport, with particles preferentially drifting toward the outer wall.
Unlike canonical planar flows where turbophoresis universally drives near-wall accumulation, the present study identifies a distinct physical regime at the convex inner wall where centrifugal effect competes with turbophoresis.
As a consequence, significant particle depletion is observed near the inner wall under strong curvature, and the transient concentration field exhibits a non-monotonic evolution, with the overshoot generally being more evident at higher Stokes numbers.
By deriving a transport equation and applying Sturm-Liouville modal analysis, we identify the competition between asymmetric transport modes with different decay rates as the physical mechanism driving this non-monotonic evolution, and establish a reduced-order model that captures the dynamics of the particle concentration near the walls.
\end{abstract}


%{\bf MSC Codes }  {\it(Optional)} Please enter your MSC Codes here

\section{Introduction}\label{sec:Introduction}
Particle transport and the resulting concentration in wall-bounded turbulent flows are of central importance to a wide range of industrial and environmental processes \citep{Guha2008Transport,Brandt2022Particle,Marchioli2025Particle}.
The majority of fundamental studies have focused on canonical planar configurations, such as plane channels and flat-plate boundary layers. 
In these geometries, inertial particles are known to drift toward the walls and preferentially concentrate within low-speed streaks \citep{Marchioli2002Mechanisms,Sardina2012Wall}.
The near-wall particle accumulation is primarily dominated by turbophoresis \citep{Caporaloni1975transfer,REEKS1983transport}, which denotes the tendency of particles to migrate from regions of high to low turbulence intensity.

Many practical flows occur in conduits with wall curvature, which introduces inertial forces that can fundamentally alter particle transport.
For example, particles in straight pipe flows show a considerably stronger near-wall concentration compared to plane channels \citep{Picano2009Spatial}.
This is because for the concave wall in turbulent pipe flows, the transverse curvature works in accordance with turbophoresis, as the `centrifugal' force induced by azimuthal velocity fluctuations acts radially toward the outer wall. 
Consequently, both planar and concave geometries exhibit a consistent trend of wallward transport. 

A fundamental gap in this understanding, however, lies in the behaviour of particles near a convex wall, where the effects of turbophoresis and the centrifugal force are expected to compete. 
The canonical configuration to isolate this regime is the concentric annular duct. 
Unlike the pipe, the annulus possesses a convex inner wall where the physics presents a unique contradiction: while turbophoresis drives particles inward, the inertial centrifugal force driven by azimuthal turbulent fluctuations acts away from the inner wall. 
Although the single-phase flow in annuli has been well documented \citep{Chung2002Direct, Bagheri2020Influence, Orlandi2025Effects}, confirming the asymmetry of turbulence statistics, the resulting transport of dispersed phases remains scarcely explored.
Recent work by \citet{Jiang2025Spatial} investigated particles in Taylor-Couette flow, revealing particle depletion at the inner wall. 
However, that system is dominated by imposed mean rotation and Taylor vortices. 
To date, the pure effect of wall curvature on particle transport, isolated from mean rotation, remains unquantified.
It is unknown whether the convex curvature of the inner wall is sufficient to overcome the robust mechanism of turbophoretic deposition. 

Another key topic regarding particle transport is the transient evolution of concentration, which has received far less quantitative scrutiny.  
Particle concentrations are known to require extremely long times to reach a statistical steady state \citep{ROUSON2001preferential, Marchioli2008Statistics, Picano2009Spatial, Sardina2012Wall}, often exceeding the duration of standard DNS. 
While theoretical models for steady-state particle flux exist \citep{Guha2008Transport} and recent DNS-informed models have been proposed \citep{Zahtila2023Particle}, the characteristic time scales governing the non-stationary evolution remain poorly quantified.

In this paper, we present the first DNS of particle-laden turbulent flows within concentric annular ducts. 
We report phenomena absent in channels and pipes, including particle depletion and a non-monotonic evolution of particle concentration at the inner wall, which is driven by the complex interplay of transport mechanisms such as turbophoresis, curvature-induced centrifugal effects, among others. 
We also move beyond phenomenological modelling by deriving a particle concentration transport equation.
Using a Sturm-Liouville modal analysis, we identify the multi-time-scale asymmetric transport modes that govern this process, providing a reduced-order physical model that captures the transient evolution of the concentration field.

\section{Numerical setup}\label{sec:Numerical}

\begin{figure}\small
\centering
	\begin{overpic}[width=\textwidth]{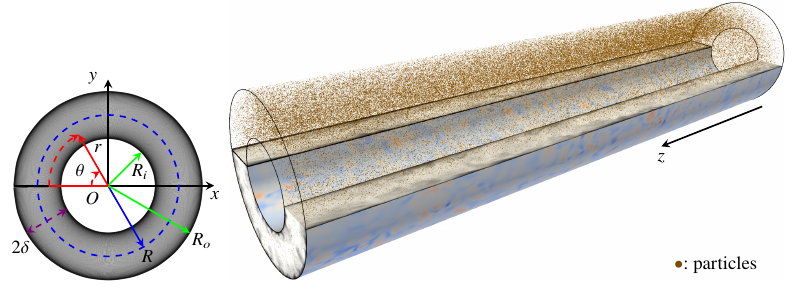}
	\end{overpic}
	\caption{Numerical configuration of the annular duct. The left figure illustrates the cross-section of the duct, while the right figure provides an overview of the computational domain. The contours of axial velocity are shown within the volume domain, and the density contours are displayed on the inner and outer walls.}
	\label{Grids}
\end{figure}

The numerical configuration is illustrated in figure~\ref{Grids}. 
The geometry of the concentric annular duct is described in a cylindrical coordinate system $(r,\theta,z)$, 
where $r$, $\theta$, and $z$ denote the radial, azimuthal, and axial directions, respectively. 
As shown by the cross-section in figure~\ref{Grids}, the inner and outer radii of the annulus are denoted by $R_i$ and $R_o$, respectively. 
Here, $\delta$ represents the half-height of the channel, and $R=(R_i+R_o)/2$ denotes the radial distance from the centre $O$ to the centreline.  
Accordingly, the computational domain extends over $L_r = 2\delta$ in the $r$ direction and $L_\theta = 2\pi R$ in the $\theta$ direction.  
It is noted that the axial domain length $L_z = 12\pi\delta$ is chosen to be sufficiently long to accurately capture near-wall streaks \citep{Bagheri2020Influence}, which have been shown to play an important role in particle distribution \citep{Sardina2012Wall}.
Based on the mapping function $r = R + \delta\tanh(2\zeta)/\tanh(2)$, with $\zeta$ uniformly discretized over the interval $[-1,1]$, the grid in the $r$ direction is refined towards the walls, where no-slip and isothermal boundary conditions are imposed.
Uniform grid spacing is employed in the periodic $\theta$ and $z$ directions.
To isolate curvature effects, two annular configurations with radius ratios $\eta = R_i/R_o = 0.5$ and $0.1$ are considered, representing mild- and strong-curvature cases, respectively \citep{Ishida2016Transitional}.
A plane channel flow is also included as a zero-curvature reference case.
In this situation, $L_\theta$ denotes the spanwise domain size, and the axial length is shortened to $L_z = 4\pi\delta$ 
to reduce the computational cost \citep{Marchioli2008Statistics}.
The parameters of all simulations are summarized in table~\ref{Parameters_continuous}, where case 1 is post-processed using symmetry with respect to the centreline.

\begin{table}
	\centering
	\begin{tabular}{p{30pt}p{45pt}p{10pt}p{55pt}p{50pt}p{48pt}p{48pt}p{48pt}}
		\multicolumn{1}{c}{\multirow{2}{*}{\centering case}} &
		\multicolumn{1}{c}{$\Rey_\tau$} &
		\multicolumn{1}{c}{\multirow{2}{*}{\centering $\eta$}} &
		\multicolumn{1}{c}{\multirow{2}{*}{\centering $L_r,L_\theta,L_z$}} &
		\multicolumn{1}{c}{\multirow{2}{*}{\centering $N_r,N_\theta,N_z$}} &
		\multicolumn{1}{c}{$\Delta r^+$} &
		\multicolumn{1}{c}{$\Delta\theta^+$} &
		\multicolumn{1}{c}{$\Delta z^+$} \\
		\multicolumn{1}{c}{\multirow{2}{*}{\centering }} &
		\centering inner, outer &
		\multicolumn{1}{c}{\multirow{2}{*}{\centering }} &
		\multicolumn{1}{c}{\multirow{2}{*}{\centering }} &
		\multicolumn{1}{c}{\multirow{2}{*}{\centering }} &
		\centering inner, outer &
		\centering inner, outer &
		\multicolumn{1}{c}{\centering inner, outer} \\
		\hline
		\centering $1$ &
		\centering $157$ & 
		\centering $1$ &
		\centering $2\delta,2\pi\delta,4\pi\delta$ &
		\centering $193,192,192$ &
		\centering $0.245$ & 
		\centering $5.147$ &  
		\multicolumn{1}{c}{\centering $10.294$}\\
		\centering $2$ &
		\centering $164,155$ & 
		\centering $0.5$ &
		\centering $2\delta,6\pi\delta,12\pi\delta$ &
		\centering $193,461,470$ &
		\centering $0.255,0.243$ & 
		\centering $4.023,7.642$ &  
		\multicolumn{1}{c}{\centering $13.149,12.488$}\\
		\centering $3,4,5,6$ &
		\centering $194,153$ & 
		\centering $0.1$ &
		\centering $2\delta,\frac{22}{9}\pi\delta,12\pi\delta$ &
		\centering $193,256,566$ &
		\centering $0.305,0.240$ & 
		\centering $1.066,8.404$ &  
		\multicolumn{1}{c}{\centering $13.023,10.263$}
	\end{tabular}
	\caption{Parameters of the different cases. Here, $\Rey_\tau$ is the friction Reynolds number, $N$ and $\Delta^+$ denote the number of grid nodes and the grid resolution in wall units in each direction, respectively. `Inner' and `outer' denote quantities normalized by the viscous scales at the inner and outer walls, respectively.}
	\label{Parameters_continuous}
\end{table}

The turbulent flows are considered to operate at a Reynolds number $\Rey_\mathrm{f} = (\rho_b U_b \delta)/\mu_\infty = 2450$ and a Mach number $\Mach_\mathrm{f} = U_b/a_\infty = 0.25$, where $\rho_b$ and $U_b$ denote the mean bulk density and velocity, respectively. 
Here, $\mu_\infty$ and $a_\infty$ denote the reference viscosity and acoustic velocity. 
The flow field is obtained by solving the Navier-Stokes equations using the finite-difference solver HiPSTAR \citep{Sandberg2015Compressible}, which ensures fourth-order accuracy in both time and space.
Temperature and density fluctuations remain below 0.15\% throughout the domain, as detailed in Appendix A, confirming that compressibility effects are negligible at $\Mach_\mathrm{f}=0.25$.

\begin{figure}\small
	\centering \subfigure{
		\begin{minipage}[c]{0.49\textwidth}{}
			\begin{overpic}[width=\textwidth]{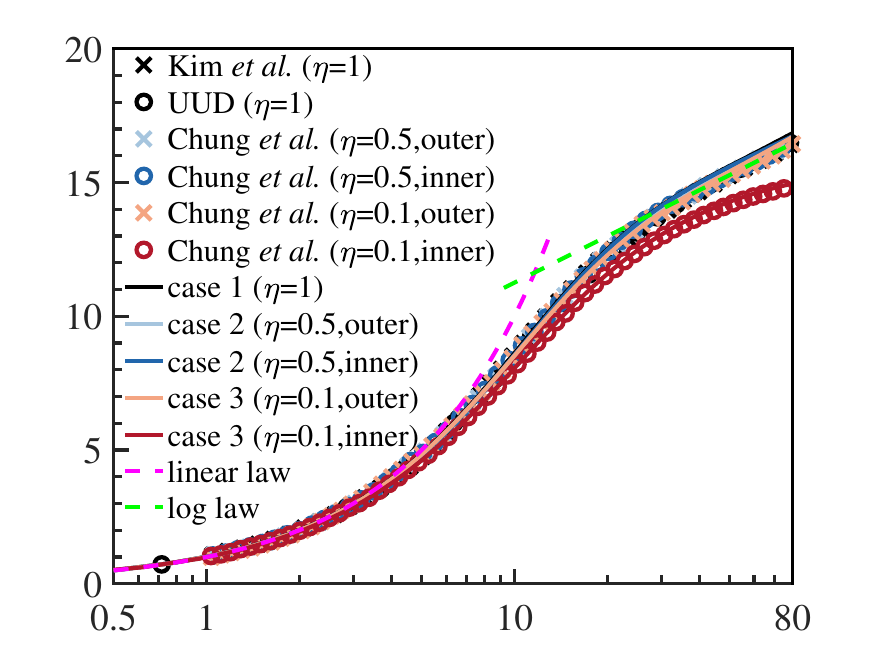}
				\put(-2,68){$(a)$}
				\put(0,37){$u_z^+$}
				\put(50,-2){$r^+$}
			\end{overpic}
		\end{minipage}\label{fluid_uz}}
	\centering \subfigure{
		\begin{minipage}[c]{0.49\textwidth}{}
			\begin{overpic}[width=\textwidth]{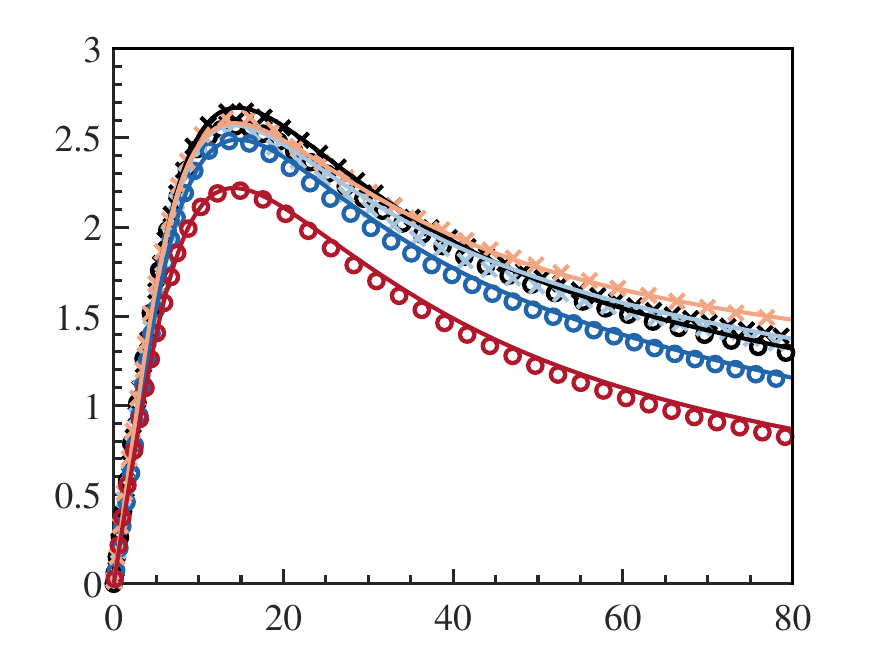}
				\put(-2,68){$(b)$}
				\put(0,33){\rotatebox{90}{$u^{\prime+}_{z,\mathrm{rms}}$}}
				\put(50,-2){$r^+$}
			\end{overpic}
		\end{minipage}\label{fluid_uzuz}}
	\centering \subfigure{
		\begin{minipage}[c]{0.49\textwidth}{}
			\begin{overpic}[width=\textwidth]{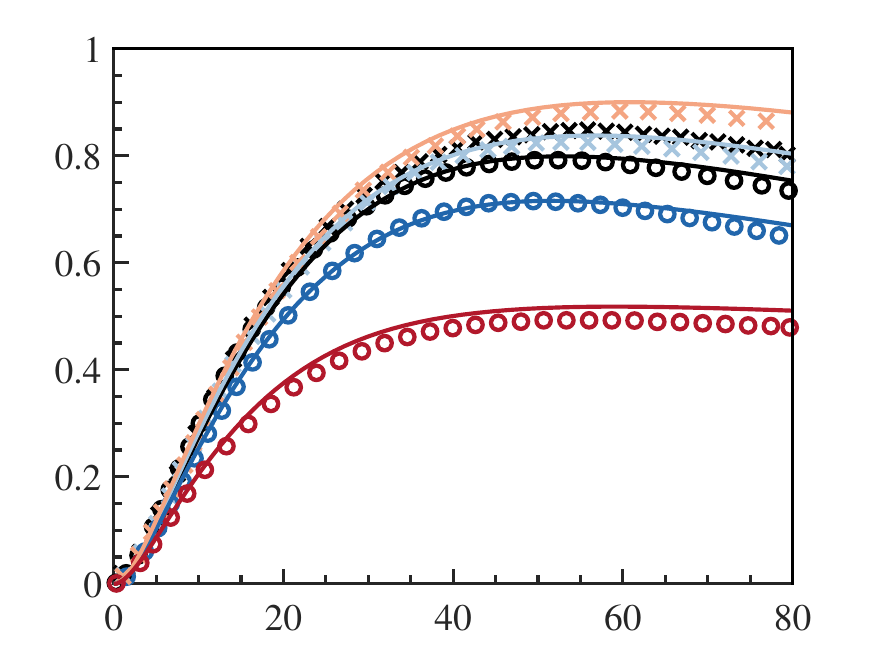}
				\put(-2,68){$(c)$}
				\put(0,33){\rotatebox{90}{$u^{\prime+}_{r,\mathrm{rms}}$}}
				\put(50,-2){$r^+$}
			\end{overpic}
		\end{minipage}\label{fluid_urur}}
	\centering \subfigure{
		\begin{minipage}[c]{0.49\textwidth}{}
			\begin{overpic}[width=\textwidth]{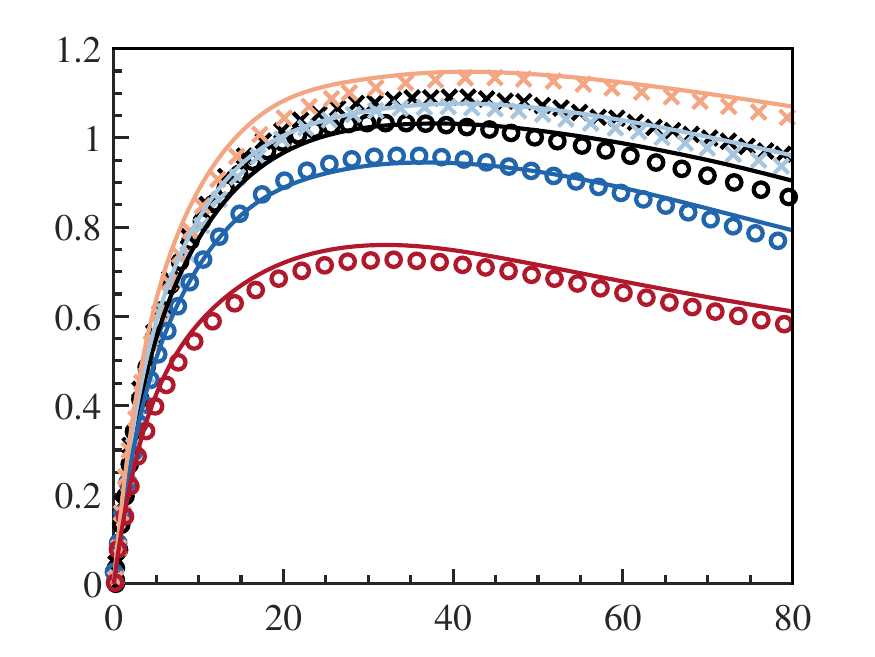}
				\put(-2,68){$(d)$}
				\put(0,33){\rotatebox{90}{$u^{\prime+}_{\theta,\mathrm{rms}}$}}
				\put(50,-2){$r^+$}
			\end{overpic}
		\end{minipage}\label{fluid_utut}}
	\caption{Near-wall fluid velocity statistics: $(a)$ mean axial fluid velocity $u_z^+$, $(b)$ axial rms fluid velocity fluctuation $u^{\prime+}_{z,\mathrm{rms}}$, $(c)$ radial rms fluid velocity fluctuation $u^{\prime+}_{r,\mathrm{rms}}$ and $(d)$ azimuthal rms fluid velocity fluctuation $u^{\prime+}_{\theta,\mathrm{rms}}$. The superscript $+$ denotes viscous scaling, and $r^+$ denotes the wall-normal distance in wall units. The markers denote the data from \citet{Kim1987Turbulence} (plane channel, $\Rey_\mathrm{f}=3300$), the UUD dataset of \citet{Marchioli2008Statistics} (plane channel, $\Rey_\mathrm{f}=2100$) and the data from \citet{Chung2002Direct} (concentric annuli, $\Rey_\mathrm{f}=2225$).}
	\label{validation}
\end{figure}
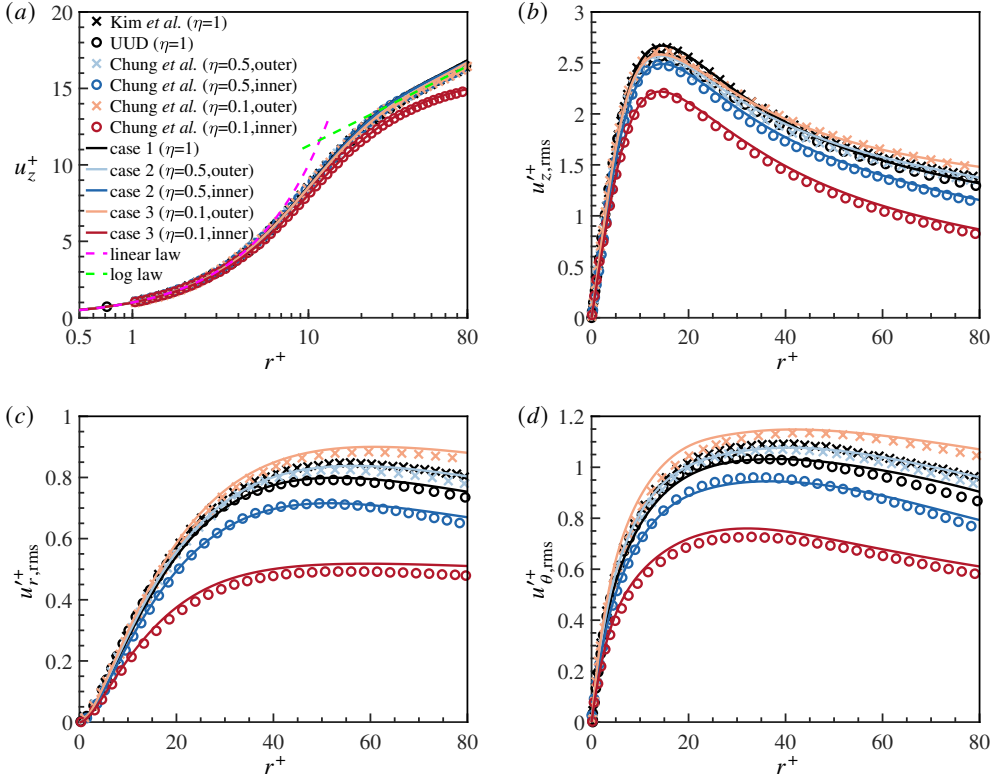

To validate the present carrier flow, figure~\ref{validation} compares the near-wall statistics obtained from the present simulations with those reported in previous incompressible studies conducted under comparable conditions \citep{Kim1987Turbulence,Marchioli2008Statistics,Chung2002Direct}.
Specifically, the comparison includes the mean axial fluid velocity $u^{+}_z$ and the root-mean-square (rms) fluid velocity fluctuations in the axial, radial and azimuthal directions, denoted by $u^{\prime+}_{z,\mathrm{rms}}$, $u^{\prime+}_{r,\mathrm{rms}}$ and $u^{\prime+}_{\theta,\mathrm{rms}}$, respectively.
Here, the superscript $+$ denotes normalization using the viscous velocity scales ($u_{\tau,i}$ and $u_{\tau,o}$) and viscous length scales ($\delta_{\nu,i}$ and $\delta_{\nu,o}$) at the inner and outer walls, respectively.
In particular, $r^+$ denotes the wall-normal distance in wall units, defined as $r^+=(r-R_i)/\delta_{\nu,i}$ in the inner-wall region and $r^+=(R_o-r)/\delta_{\nu,o}$ in the outer-wall region.
The close agreement demonstrates that the present single-phase DNS provides a reliable turbulent carrier flow.

\begin{table}
	\centering
	\begin{tabular}{p{15pt}p{22pt}p{22pt}p{40pt}p{45pt}p{30pt}p{48pt}p{48pt}p{48pt}}
		\multicolumn{1}{c}{\multirow{2}{*}{\centering case}} &
        \multicolumn{1}{c}{\multirow{2}{*}{\centering $d_\mathrm{p}/\delta$}} &
		\multicolumn{1}{c}{\multirow{2}{*}{\centering $\rho_\mathrm{p}/\rho_\mathrm{f}$}} &
		\multicolumn{1}{c}{\multirow{2}{*}{\centering $\tau_\mathrm{p}/(\delta/U_b)$}}&
		\multicolumn{1}{c}{\multirow{2}{*}{\centering $\varPhi_\mathrm{p}$}} &
		\multicolumn{1}{c}{\multirow{2}{*}{\centering $N_\mathrm{p}$}} &
		\multicolumn{1}{c}{$d_\mathrm{p}/\eta_\mathrm{w}$} &
		\multicolumn{1}{c}{$d^+_\mathrm{p}$} &
		\multicolumn{1}{c}{$\Stokes^+$} \\
		\multicolumn{1}{c}{\multirow{2}{*}{\centering }} &
		\multicolumn{1}{c}{\multirow{2}{*}{\centering }} &
		\multicolumn{1}{c}{\multirow{2}{*}{\centering }} &
        \multicolumn{1}{c}{\multirow{2}{*}{\centering }} &
		\multicolumn{1}{c}{\multirow{2}{*}{\centering }} &
		\multicolumn{1}{c}{\multirow{2}{*}{\centering }} &
		\centering inner, outer &
		\centering inner, outer &
		\multicolumn{1}{c}{\centering inner, outer} \\
		\hline
		\centering 1 &
        \centering 0.004 &
        \centering 1200 &
        \centering 2.6133 &
        \centering $2.122\times10^{-5}$ &
        \centering 100000 &
        \centering 0.403 &
		\centering 0.630 &   
		\multicolumn{1}{c}{\centering 26.472}\\
		\centering 2 &  
        \centering 0.004 &
		\centering 1200 &
		\centering 2.6133 &
		\centering $2.122\times10^{-5}$ &
		\centering 900000 &
		\centering $0.399,0.397$ & 
		\centering $0.656,0.623$ &  
		\multicolumn{1}{c}{\centering $25.807,28.611$}\\
		\centering 3 &
        \centering 0.004 &
		\centering 1200 &
		\centering 2.6133 &
		\centering $2.122\times10^{-5}$ &
		\centering 366667 &
		\centering $0.457,0.400$ &  
		\centering $0.782,0.616$ & 
		\multicolumn{1}{c}{\centering $40.535,25.179$}\\
        \centering 4 &
        \centering 0.004 &
		\centering 2400 &
		\centering 5.2267 &
		\centering $2.122\times10^{-5}$ &
		\centering 366667 &
		\centering $0.457,0.400$ & 
		\centering $0.782,0.616$ &  
		\multicolumn{1}{c}{\centering $81.071,50.359$}\\
		\centering 5 &
        \centering 0.004 &
        \centering 600 &
        \centering 1.3067 &
        \centering $2.122\times10^{-5}$ &
        \centering 366667 &
        \centering $0.457,0.400$ &  
		\centering $0.782,0.616$ & 
		\multicolumn{1}{c}{\centering $20.268,12.590$}\\
		\centering 6 &
        \centering 0.004 &
		\centering 300 &
		\centering 0.6533 &
		\centering $2.122\times10^{-5}$ &
		\centering 366667 &
		\centering $0.457,0.400$ & 
		\centering $0.782,0.616$ &  
		\multicolumn{1}{c}{\centering $10.134,6.295$}
	\end{tabular}
	\caption{Parameters of the different cases. Here, $N_\mathrm{p}$ denotes the total particle number, $\varPhi_\mathrm{p}$ denotes the particle volume fraction, and $\eta_\mathrm{w}$ denotes the smallest eddy scale near the wall.}
	\label{Parameters_dispersed}
\end{table}

The dispersed phase is modelled as rigid spheres with uniform diameter $d_{\mathrm{p}}=0.004\delta$ and fixed density $\rho_{\mathrm{p}}$.
For cases 1$\sim$3, the particle density is set to $\rho_{\mathrm{p}}=1200\rho_b$ to quantify the effect of curvature.
For the strong-curvature annular configuration (cases 3$\sim$6), four types of particles with different response times, $\tau_\mathrm{p}=\rho_\mathrm{p}d_\mathrm{p}^2/(18 \mu_\infty)$, are considered to assess the influence of the Stokes number, where $\tau_\mathrm{p}$ is varied by changing $\rho_\mathrm{p}$.
The detailed parameters are summarized in table~\ref{Parameters_dispersed}, and the resulting nominal viscous Stokes number $\Stokes^+=\tau_\mathrm{p}/\tau_\mathrm{f}$ spans a range of 6$\sim$81.
Here, $\tau_\mathrm{f}$ denotes the viscous timescale, defined as $\tau_\mathrm{f}=\delta_{\nu,i}/u_{\tau,i}$ at the inner wall and $\tau_\mathrm{f}=\delta_{\nu,o}/u_{\tau,o}$ at the outer wall.
Note that the particle diameter is smaller than both the Kolmogorov and viscous length scales, thereby justifying the point-particle assumption.
Owing to the large density ratio, particle dynamics are dominated by drag \citep{Marchioli2008Statistics}, with the drag coefficient modelled using the correlation proposed by \citet{Wen1966Mechanics}.
Third-order Lagrange polynomials \citep{Ruan2024shear} are employed for spatial interpolation, and the Velocity-Verlet algorithm \citep{Tartakovsky2016Pairwise} is adopted for time integration.
Particle-wall interactions are assumed to be perfectly elastic \citep{Marchioli2008Statistics}.

\begin{figure}\small
	\centering \subfigure{
		\begin{minipage}[c]{0.49\textwidth}{}
			\begin{overpic}[width=\textwidth]{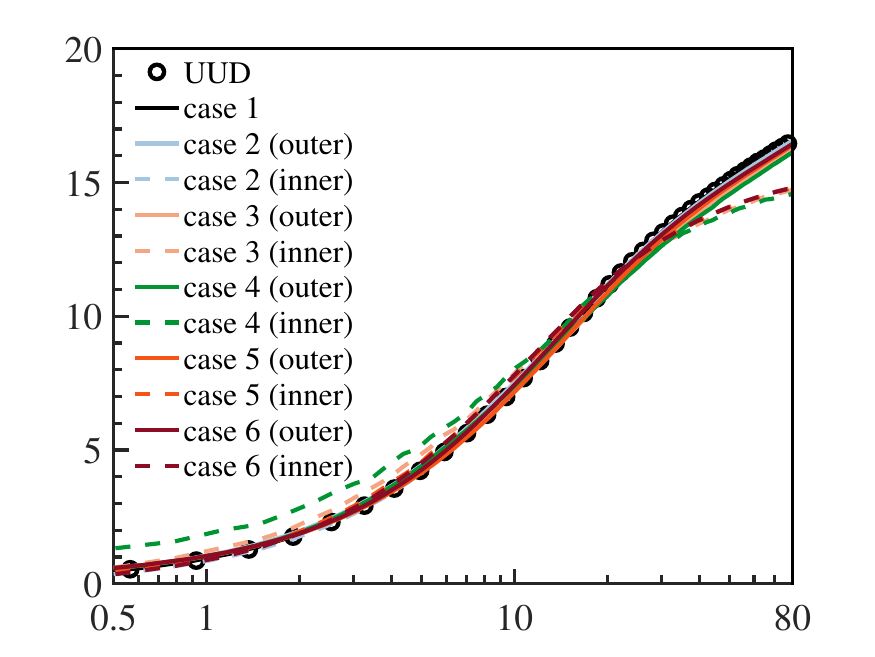}
				\put(-2,68){$(a)$}
				\put(0,37){$v_z^+$}
				\put(50,-2){$r^+$}
			\end{overpic}
		\end{minipage}\label{particle_uz}}
	\centering \subfigure{
		\begin{minipage}[c]{0.49\textwidth}{}
			\begin{overpic}[width=\textwidth]{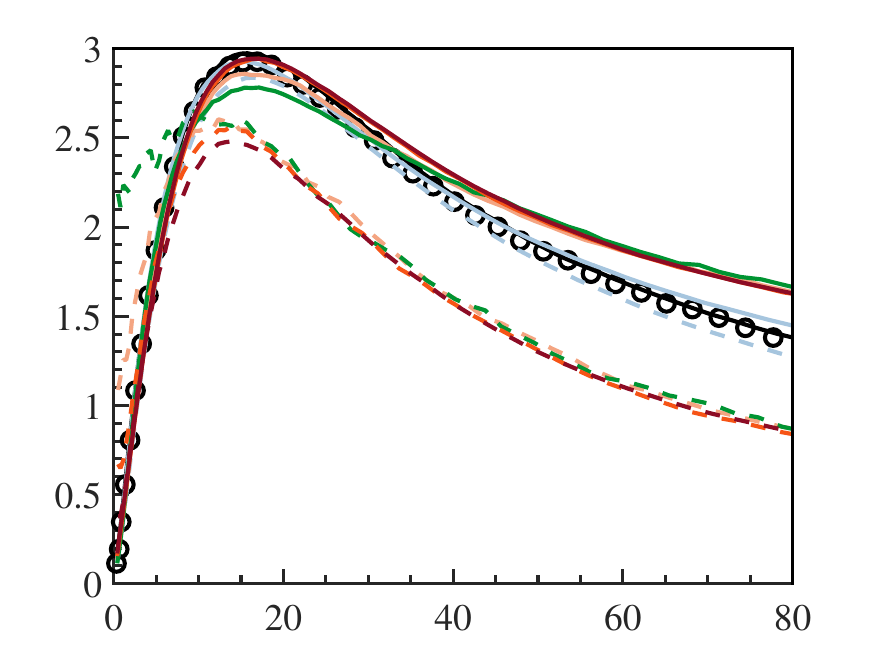}
				\put(-2,68){$(b)$}
				\put(0,33){\rotatebox{90}{$v^{\prime+}_{z,\mathrm{rms}}$}}
				\put(50,-2){$r^+$}
			\end{overpic}
		\end{minipage}\label{particle_uzuz}}
	\centering \subfigure{
		\begin{minipage}[c]{0.49\textwidth}{}
			\begin{overpic}[width=\textwidth]{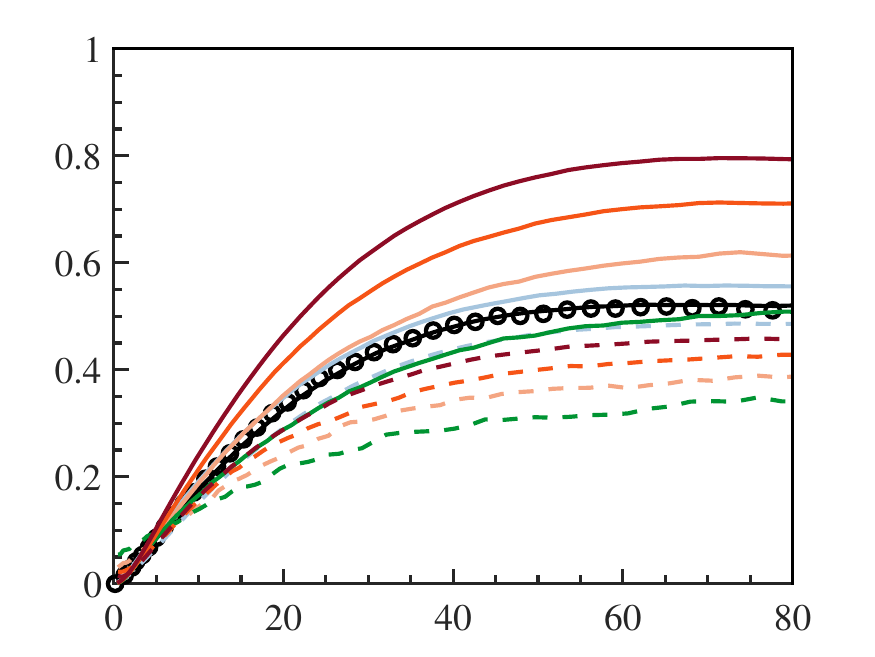}
				\put(-2,68){$(c)$}
				\put(0,33){\rotatebox{90}{$v^{\prime+}_{r,\mathrm{rms}}$}}
				\put(50,-2){$r^+$}
			\end{overpic}
		\end{minipage}\label{particle_urur}}
	\centering \subfigure{
		\begin{minipage}[c]{0.49\textwidth}{}
			\begin{overpic}[width=\textwidth]{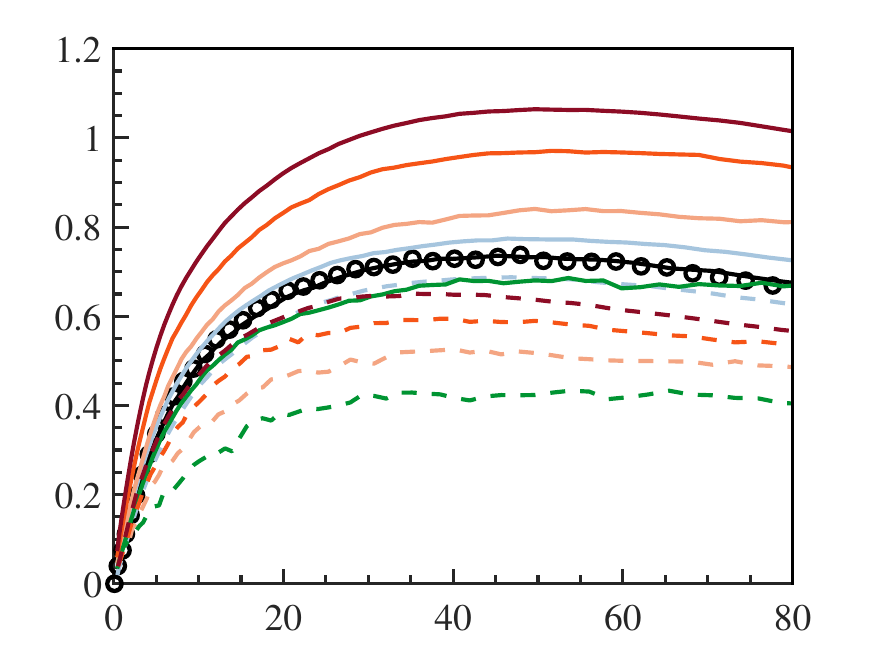}
				\put(-2,68){$(d)$}
				\put(0,33){\rotatebox{90}{$v^{\prime+}_{\theta,\mathrm{rms}}$}}
				\put(50,-2){$r^+$}
			\end{overpic}
		\end{minipage}\label{particle_utut}}
	\caption{Near-wall particle velocity statistics: $(a)$ mean axial particle velocity $v_z^+$, $(b)$ axial rms particle velocity fluctuation $v^{\prime+}_{z,\mathrm{rms}}$, $(c)$ radial rms particle velocity fluctuation $v^{\prime+}_{r,\mathrm{rms}}$ and $(d)$ azimuthal rms particle velocity fluctuation $v^{\prime+}_{\theta,\mathrm{rms}}$. The superscript $+$ denotes viscous scaling, and $r^+$ denotes the wall-normal distance in wall units. The markers denote the UUD dataset of \citet{Marchioli2008Statistics} (plane channel, $\Rey_\mathrm{f}=2100$).}
	\label{particle_statistics}
\end{figure}
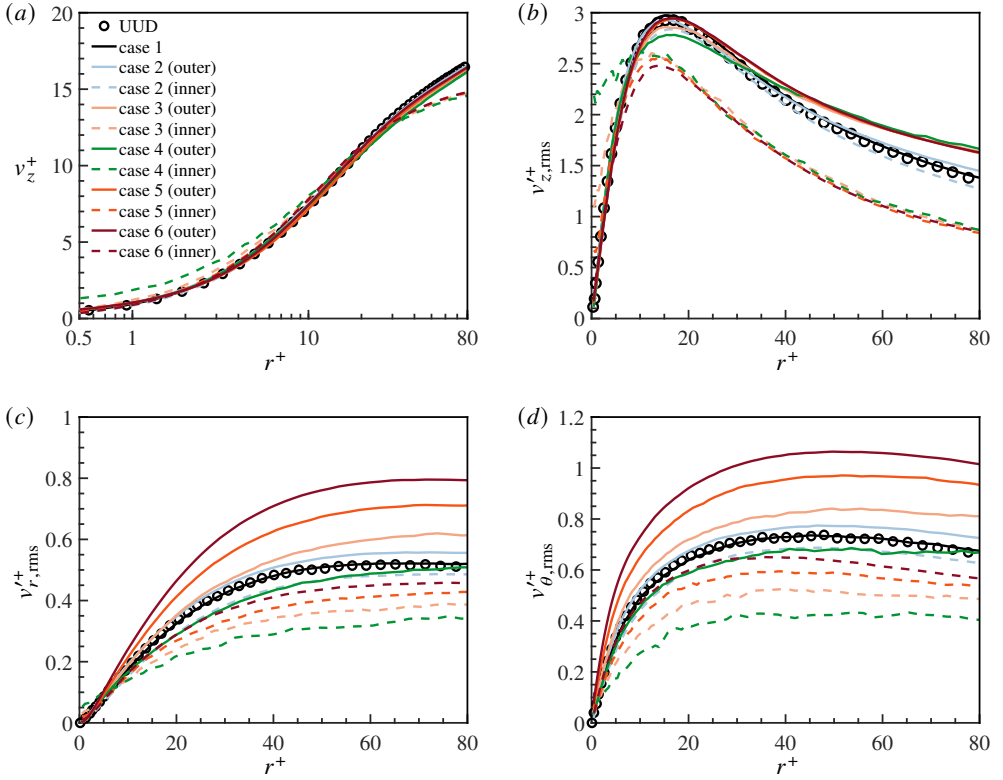

All simulations are performed with the one-way coupling strategy.
Starting from an initially uniform random particle distribution as shown in figure~\ref{Grids}, the particles are initialized to follow the fully turbulent fluid motion at $t=0$, where $t$ denotes the time non-dimensionalized by $\delta/U_b$.
The simulations are carried out for a total non-dimensional duration of $T/(\delta/U_b)=3000$.
Note that this duration is longer than typical channel flow studies \citep{Marchioli2008Statistics} to account for the slow transients of concentration evolution, although, as discussed in \S~\ref{sec:Results}, case 2 exhibits relaxation timescales that challenge even this extended integration time.
To validate the present particle framework \citep{Wang2025high}, the statistics of the `stationary' particle velocity \citep{ROUSON2001preferential} in plane channel flow are compared with the reference data of \citet{Marchioli2008Statistics}, as shown in figure~\ref{particle_statistics}.
The statistics are computed from the data collected over the last 1000 non-dimensional time units, including the mean axial particle velocity $v^{+}_z$ and the rms particle velocity fluctuations in the axial, radial and azimuthal directions, denoted by $v^{\prime+}_{z,\mathrm{rms}}$, $v^{\prime+}_{r,\mathrm{rms}}$ and $v^{\prime+}_{\theta,\mathrm{rms}}$, respectively.
The present plane-channel DNS results agree well with the previous data, which supports the reliability of the current simulations.

Importantly, the near-wall particle statistics in the curved configurations exhibit noticeable differences between the two walls, particularly in the more strongly curved cases, with an additional dependence on the Stokes number. 
This behaviour indicates distinct particle dynamics in the vicinity of the two walls, motivating the subsequent analysis of particle transport.
Note that the particle velocity statistics in cases 3 and 4 appear to show some unexpected trends near the inner wall, especially for $v_z^+$ and $v^{\prime+}_{z,\mathrm{rms}}$ shown in figures~$\ref{particle_uz}$ and $\ref{particle_uzuz}$.
These discrepancies may be due to insufficient sampling, since the inner wall in these high-$\Stokes^+$ and strong-curvature cases shows significant particle depletion, as will be discussed later in figure~$\ref{Cw_inner}$.
In fact, under the present setup, the instantaneous number of particles within the inner-wall viscous sublayer ($r^+<5$) for cases 3 and 4 is only of order $\mathcal{O}(1\sim10)$.

\section{Results and discussion}\label{sec:Results}

\subsection{Curvature-induced asymmetric particle transport and its statistical characterization}\label{subsec:transport}

\begin{figure}\small
	\centering \subfigure{
		\begin{minipage}[c]{0.49\textwidth}{}
			\begin{overpic}[width=\textwidth]{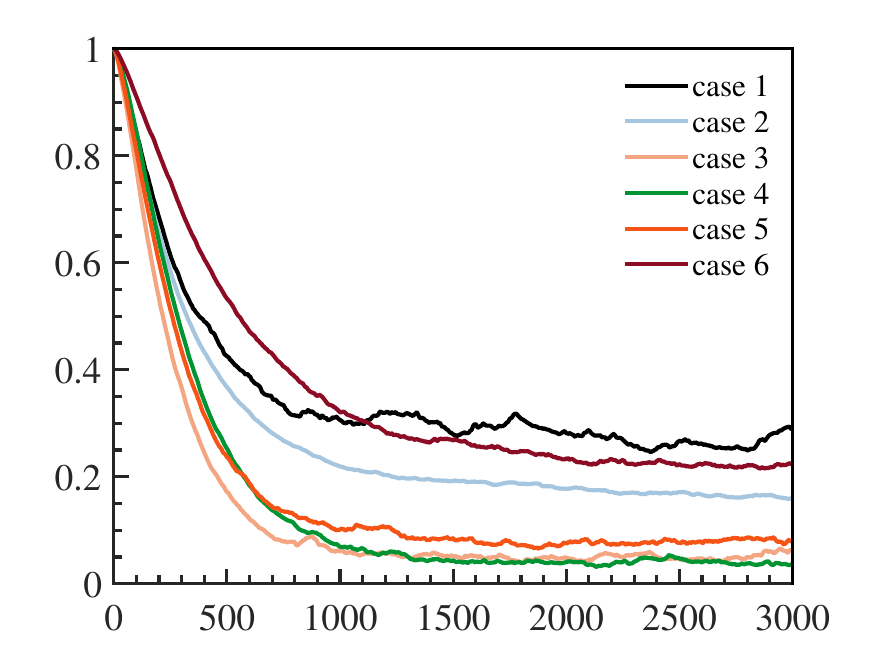}
				\put(-5,37.5){$\dfrac{\mathcal{S}}{\mathcal{S}_{\mathrm{max}}}$}
				\put(51,-2){$t$}
			\end{overpic}
		\end{minipage}}
	\caption{Temporal evolution of the Shannon entropy parameter.}
	\label{entropy}
\end{figure}

To elucidate the radial particle transport within the duct, we employ the Shannon entropy $\mathcal{S}$ \citep{Picano2009Spatial,Bernardini2014Reynolds} as a statistical descriptor.
To obtain the parameter $\mathcal{S}$, the computational domain is first divided radially into $N_r-1$ slabs of equal cross-sectional area \citep{Picano2009Spatial}, where $N_r = 193$ is the number of radial grid nodes.
For each instantaneous snapshot, the probability of finding a particle in the $j$th slab is defined as $p_{\!j}=N_{\!j}/N_\mathrm{p}$, where $N_{\!j}$ is the number of particles contained in that slab and $N_\mathrm{p}$ is the total number of particles.
Accordingly, the Shannon entropy is given by
\begin{equation}
    \mathcal{S}=-\sum_{j=1}^{N_r-1}p_{\!j}\log p_{\!j},
\end{equation}
whose maximum value $\mathcal{S}_{\mathrm{max}}=\log(N_r-1)$ is attained for a uniform particle distribution.
As shown in figure~$\ref{entropy}$, the normalized entropy initially decreases before reaching a plateau, suggesting an increase in the spatial inhomogeneity of the particle concentration.
The comparison among cases 1$\sim$3 indicates that the inhomogeneity is more pronounced in the strong-curvature case, as evidenced by the lower asymptotic entropy value, suggesting enhanced radial particle transport.
As $\Stokes^+$ decreases from cases 3 and 4 to cases 5 and 6, the entropy decays more slowly overall and remains at a relatively higher level, indicating weaker radial particle transport and a more homogeneous radial distribution.

\begin{figure}\small  
	\centering \subfigure{
		\begin{minipage}[c]{0.32\textwidth}{}
			\begin{overpic}[width=\textwidth]{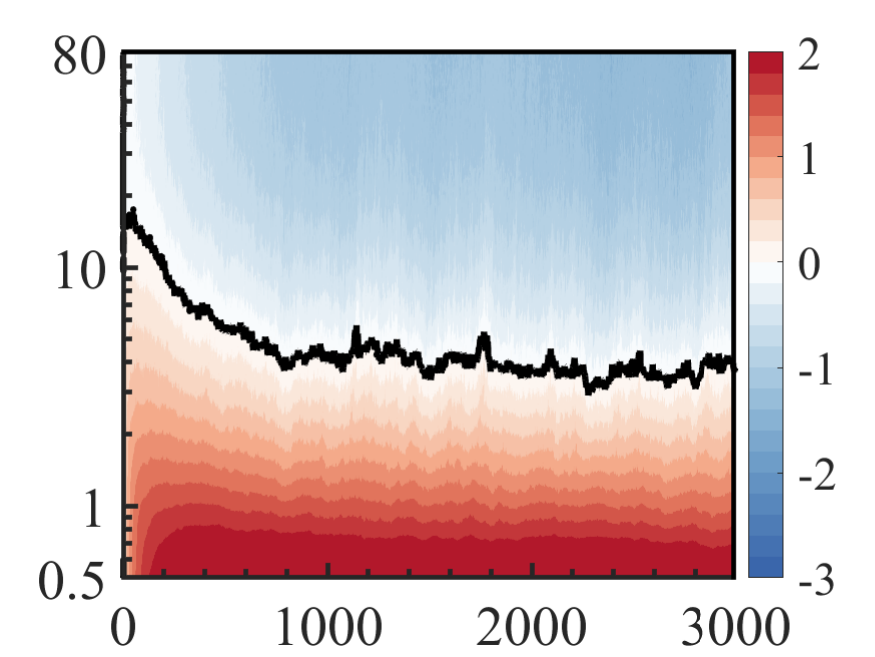}
				\put(-6,67){$(a)$}
				\put(0,36){$r^+$}
				\put(47.5,-4){$t$}
				
				\put(95,29){\rotatebox{90}{\scriptsize$\log_{10}(c)$}}
			\end{overpic}
		\end{minipage}\label{C_eta1_outer}}
	\centering \subfigure{
		\begin{minipage}[c]{0.32\textwidth}{}
			\begin{overpic}[width=\textwidth]{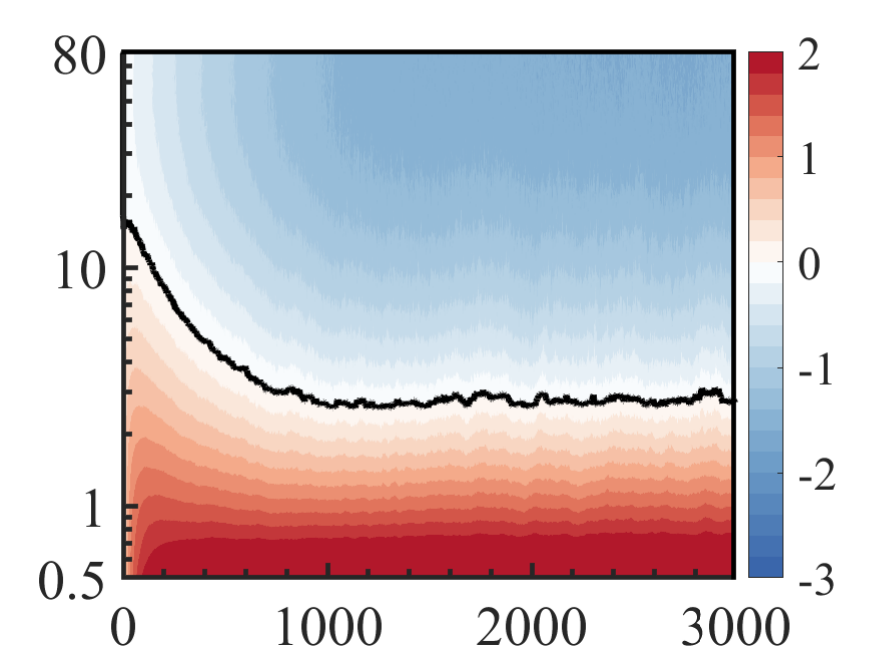}
				\put(-6,67){$(b)$}
				\put(0,36){$r^+$}
				\put(47.5,-4){$t$}
				
				\put(95,29){\rotatebox{90}{\scriptsize$\log_{10}(c)$}}
			\end{overpic}
		\end{minipage}\label{C_eta05_outer}}
	\centering \subfigure{
		\begin{minipage}[c]{0.32\textwidth}{}
			\begin{overpic}[width=\textwidth]{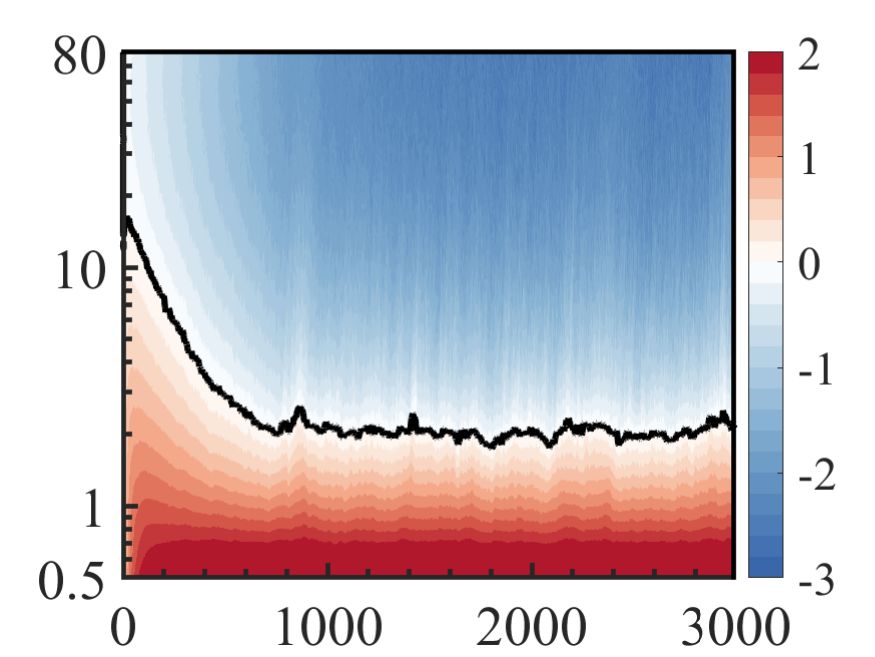}
				\put(-6,67){$(c)$}
				\put(0,36){$r^+$}
				\put(47.5,-4){$t$}
				
				\put(95,29){\rotatebox{90}{\scriptsize$\log_{10}(c)$}}
			\end{overpic}
		\end{minipage}\label{C_eta01_rho1200_outer}}
	\centering \subfigure{
		\begin{minipage}[c]{0.32\textwidth}{}
			\begin{overpic}[width=\textwidth]{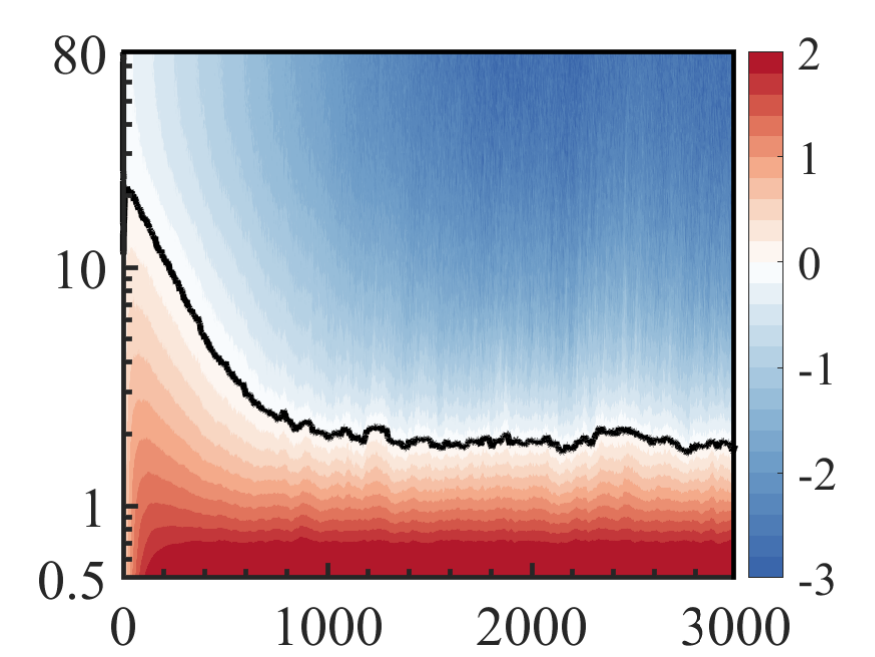}
				\put(-6,67){$(d)$}
				\put(0,36){$r^+$}
				\put(47.5,-4){$t$}
				
				\put(95,29){\rotatebox{90}{\scriptsize$\log_{10}(c)$}}
			\end{overpic}
		\end{minipage}\label{C_eta01_rho2400_outer}}
	\centering \subfigure{
		\begin{minipage}[c]{0.32\textwidth}{}
			\begin{overpic}[width=\textwidth]{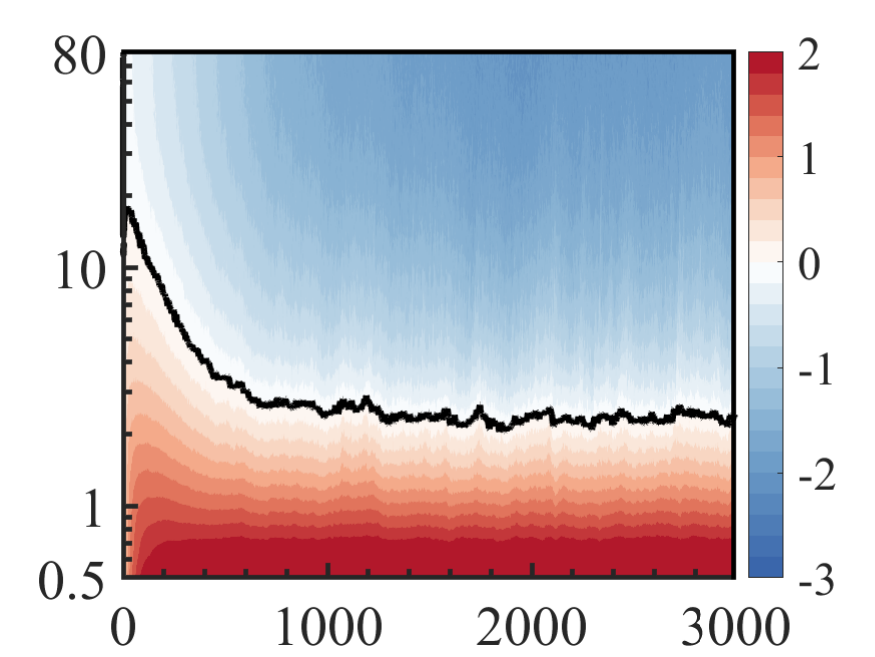}
				\put(-6,67){$(e)$}
				\put(0,36){$r^+$}
				\put(47.5,-4){$t$}
				
				\put(95,29){\rotatebox{90}{\scriptsize$\log_{10}(c)$}}
			\end{overpic}
		\end{minipage}\label{C_eta01_rho0600_outer}}
	\centering \subfigure{
		\begin{minipage}[c]{0.32\textwidth}{}
			\begin{overpic}[width=\textwidth]{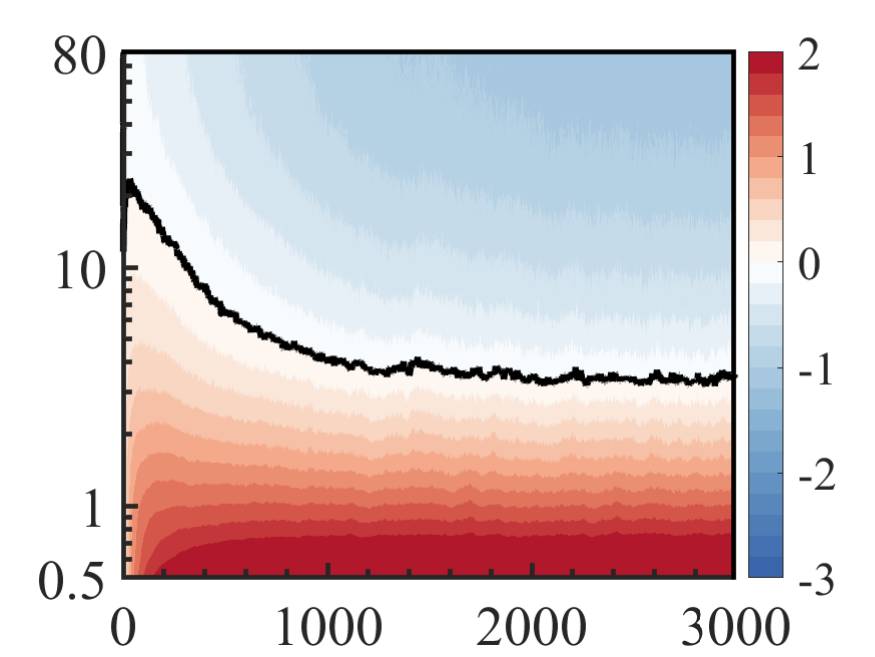}
				\put(-6,67){$(f)$}
				\put(0,36){$r^+$}
				\put(47.5,-4){$t$}
				
				\put(95,29){\rotatebox{90}{\scriptsize$\log_{10}(c)$}}
			\end{overpic}
		\end{minipage}\label{C_eta01_rho0300_outer}}
	\caption{Spatial and temporal evolution of the dimensionless particle concentration $c(r,t)$ near the (outer) wall of $(a)$ case 1, $(b)$ case 2, $(c)$ case 3, $(d)$ case 4, $(e)$ case 5 and $(f)$ case 6. Here, $r^+$ denotes the viscous-scaled radial distance, and contours represent the particle concentration $c(r,t)$, with black iso-lines denoting $c(r,t)=1$.}
	\label{C_evolution_outer}
\end{figure}

\begin{figure}\small   
	\centering \subfigure{
		\begin{minipage}[c]{0.32\textwidth}{}
			\begin{overpic}[width=\textwidth]{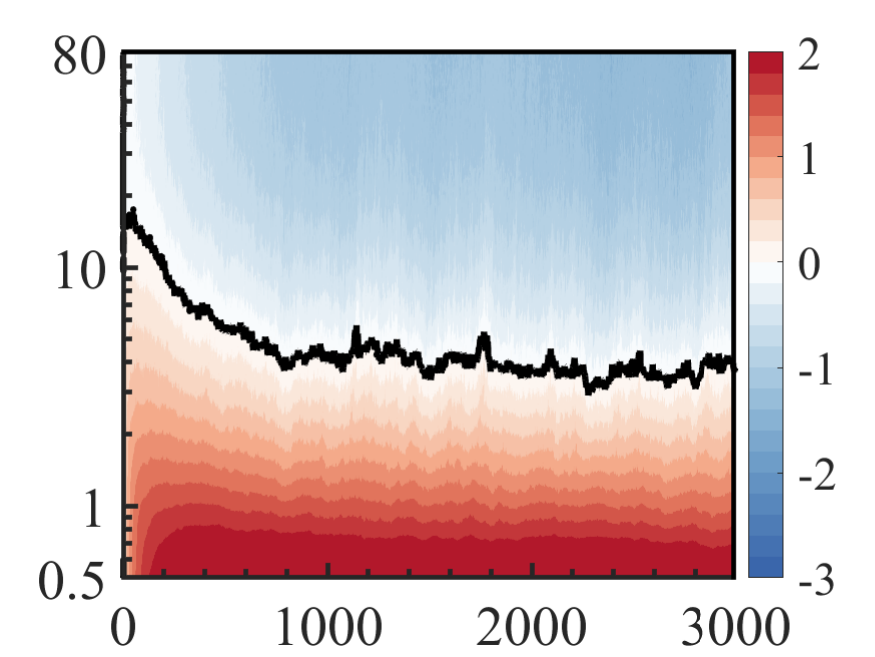}
				\put(-6,67){$(a)$}
				\put(0,36){$r^+$}
				\put(47.5,-4){$t$}
				
				\put(95,29){\rotatebox{90}{\scriptsize$\log_{10}(c)$}}
			\end{overpic}
		\end{minipage}\label{C_eta1_inner}}
	\centering \subfigure{
		\begin{minipage}[c]{0.32\textwidth}{}
			\begin{overpic}[width=\textwidth]{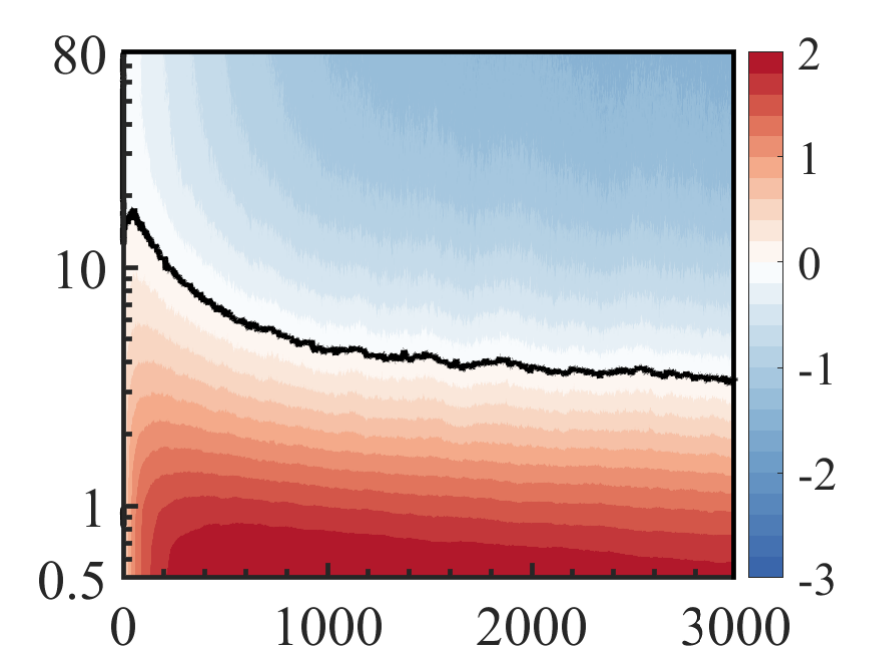}
				\put(-6,67){$(b)$}
				\put(0,36){$r^+$}
				\put(47.5,-4){$t$}
				
				\put(95,29){\rotatebox{90}{\scriptsize$\log_{10}(c)$}}
			\end{overpic}
		\end{minipage}\label{C_eta05_inner}}
	\centering \subfigure{
		\begin{minipage}[c]{0.32\textwidth}{}
			\begin{overpic}[width=\textwidth]{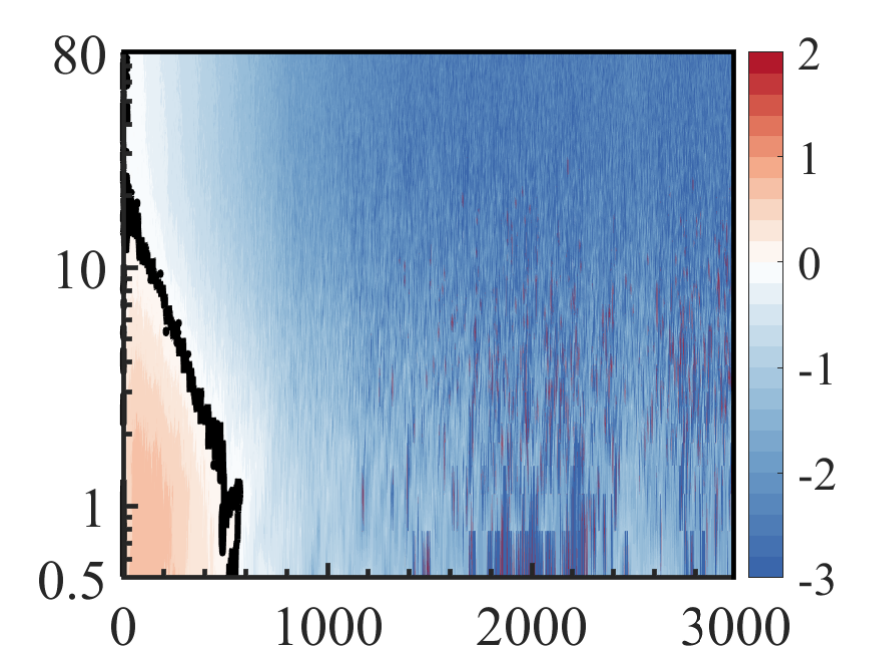}
				\put(-6,67){$(c)$}
				\put(0,36){$r^+$}
				\put(47.5,-4){$t$}
				
				\put(95,29){\rotatebox{90}{\scriptsize$\log_{10}(c)$}}
			\end{overpic}
		\end{minipage}\label{C_eta01_rho1200_inner}}
	\centering \subfigure{
		\begin{minipage}[c]{0.32\textwidth}{}
			\begin{overpic}[width=\textwidth]{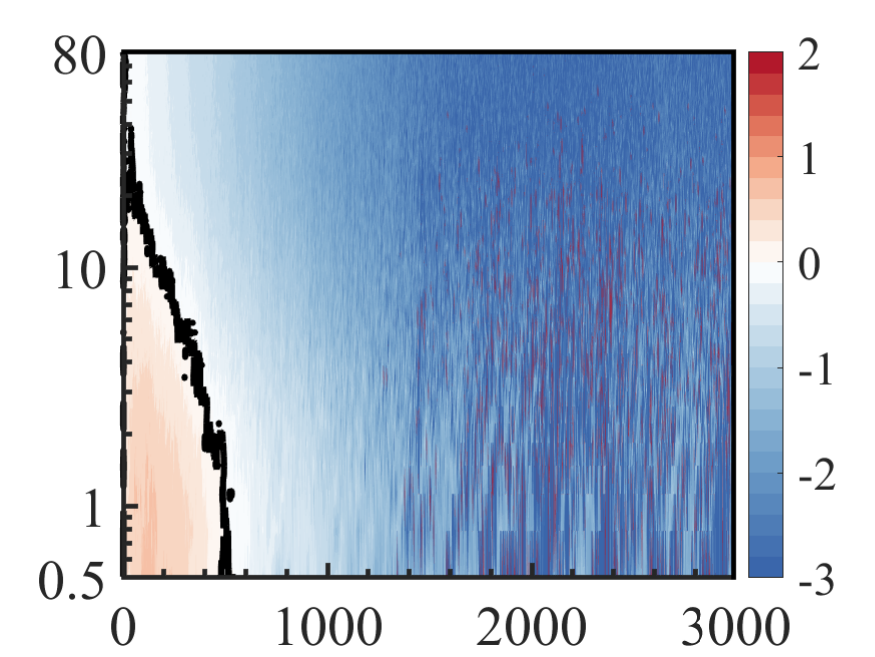}
				\put(-6,67){$(d)$}
				\put(0,36){$r^+$}
				\put(47.5,-4){$t$}
				
				\put(95,29){\rotatebox{90}{\scriptsize$\log_{10}(c)$}}
			\end{overpic}
		\end{minipage}\label{C_eta01_rho2400_inner}}
	\centering \subfigure{
		\begin{minipage}[c]{0.32\textwidth}{}
			\begin{overpic}[width=\textwidth]{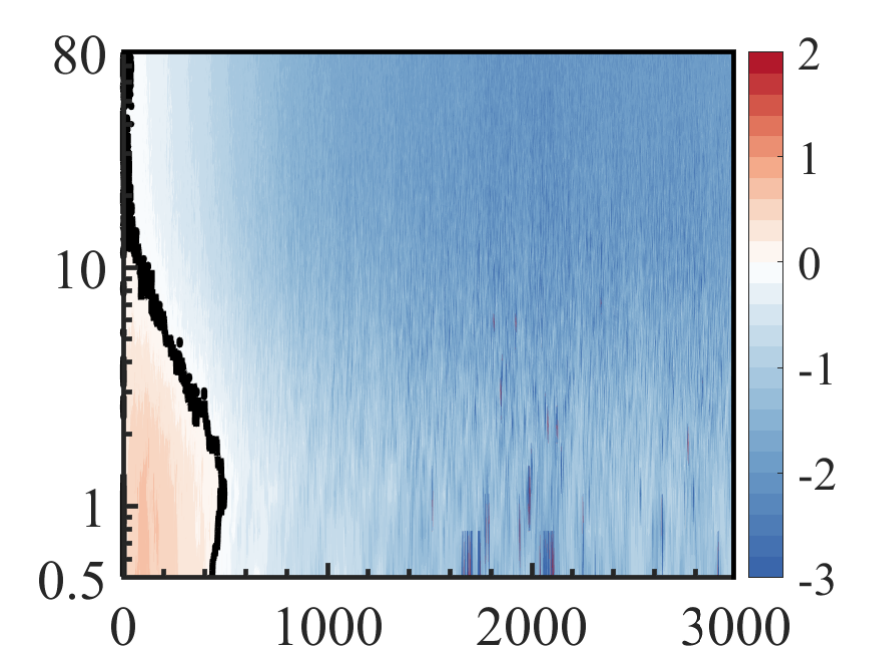}
				\put(-6,67){$(e)$}
				\put(0,36){$r^+$}
				\put(47.5,-4){$t$}
				
				\put(95,29){\rotatebox{90}{\scriptsize$\log_{10}(c)$}}
			\end{overpic}
		\end{minipage}\label{C_eta01_rho0600_inner}}
	\centering \subfigure{
		\begin{minipage}[c]{0.32\textwidth}{}
			\begin{overpic}[width=\textwidth]{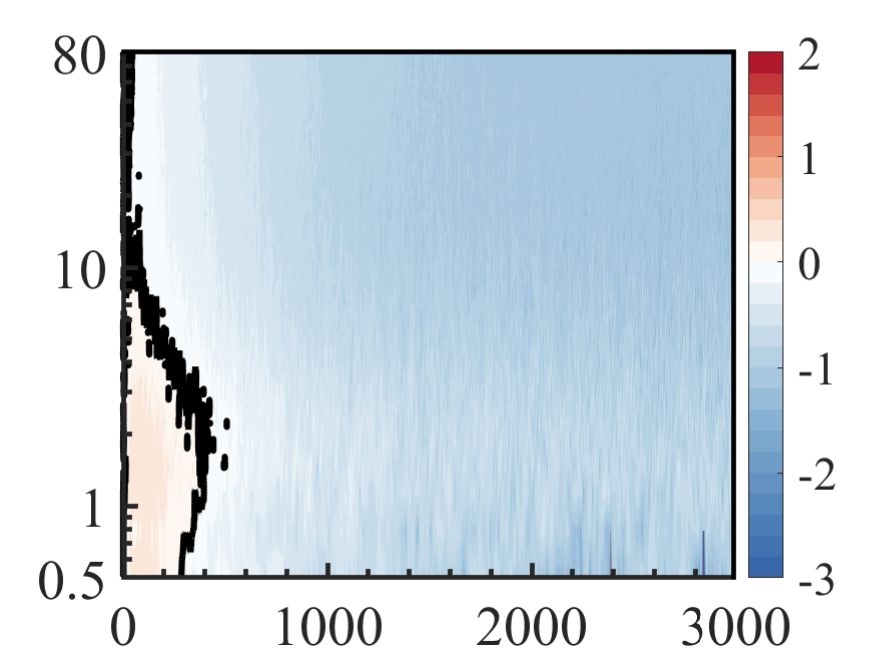}
				\put(-6,67){$(f)$}
				\put(0,36){$r^+$}
				\put(47.5,-4){$t$}
				
				\put(95,29){\rotatebox{90}{\scriptsize$\log_{10}(c)$}}
			\end{overpic}
		\end{minipage}\label{C_eta01_rho0300_inner}}
	\caption{Spatial and temporal evolution of the dimensionless particle concentration $c(r,t)$ near the (inner) wall of $(a)$ case 1, $(b)$ case 2, $(c)$ case 3, $(d)$ case 4, $(e)$ case 5 and $(f)$ case 6. Here, $r^+$ denotes the viscous-scaled radial distance, and contours represent the particle concentration $c(r,t)$, with black iso-lines denoting $c(r,t)=1$.}
	\label{C_evolution_inner}
\end{figure}

\begin{figure}\small
	\centering \subfigure{
		\begin{minipage}[c]{0.49\textwidth}{}
			\begin{overpic}[width=\textwidth]{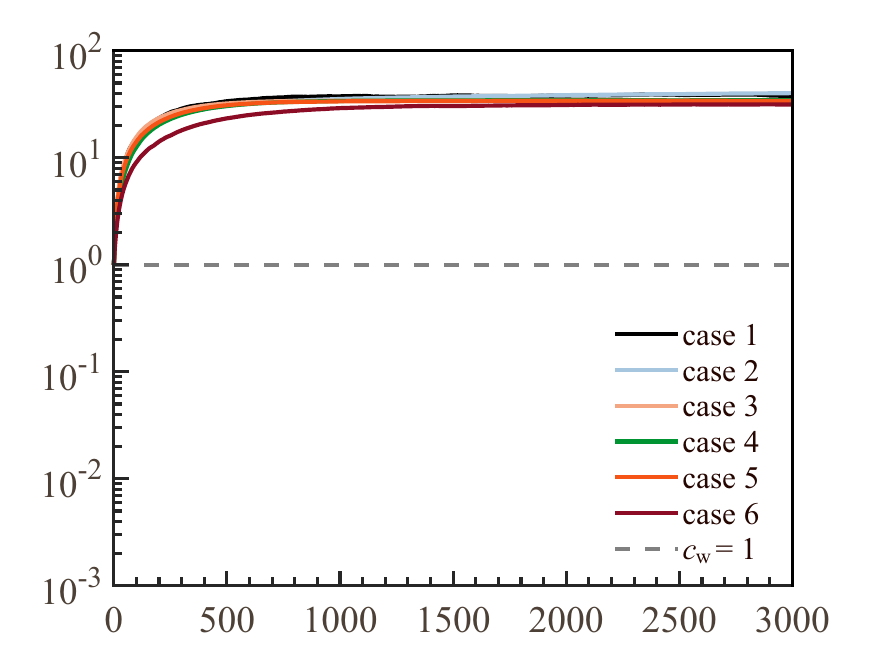}
				\put(-2,68){$(a)$}
				\put(-2,38){$c_\mathrm{w}$}
				\put(51,-2){$t$}
			\end{overpic}
		\end{minipage}\label{Cw_outer}}
    \centering \subfigure{
		\begin{minipage}[c]{0.49\textwidth}{}
			\begin{overpic}[width=\textwidth]{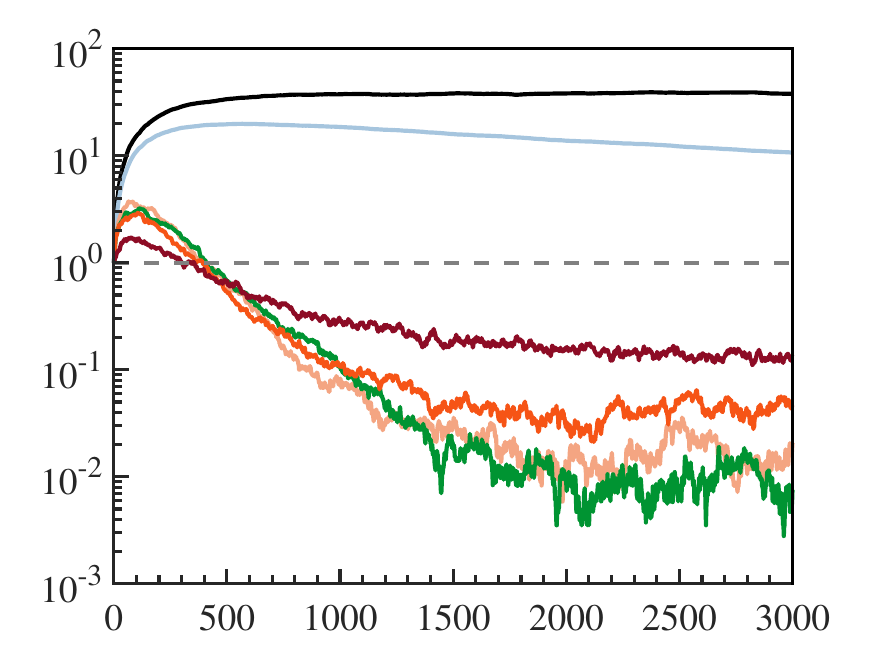}
				\put(-2,68){$(b)$}
                \put(-2,38){$c_\mathrm{w}$}
				\put(51,-2){$t$}

                \put(54.5,12){\rotatebox{90}{$\Larrow{\rule{1.75cm}{0pt}}$}}
                \put(61,11){$\Stokes^+$ increases}
			\end{overpic}
		\end{minipage}\label{Cw_inner}}
	\caption{Dimensionless mean particle concentration within the viscous sublayer ($r^+<5$) near the $(a)$ outer wall and $(b)$ inner wall.}
	\label{Cw}
\end{figure}
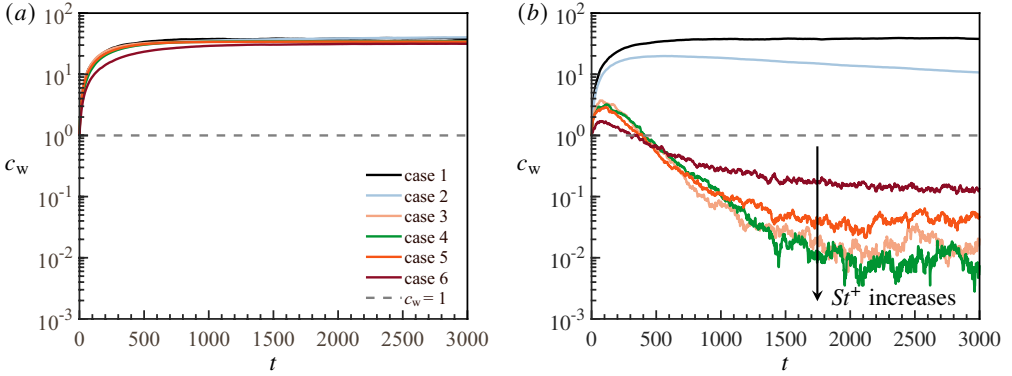

To provide insight into the directionality of the transport process, we further investigate the evolution of the dimensionless concentration
$c(r,t)=c_\mathrm{slice}/c_0$ as shown in figures~$\ref{C_evolution_outer}$ and $\ref{C_evolution_inner}$, where $c_\mathrm{slice}$ and $c_0$ denote the mean particle number densities in each radial slice and in the entire domain, respectively.
Compared to the zero-curvature reference case, the concentration distributions in annular channels exhibit noticeable differences near both walls, indicating asymmetric radial transport of particles.
Similar to the wallward particle accumulation which has been reported in plane channel flows \citep{Sardina2012Wall}, particles show continuous accumulation near the outer wall of concentric annular ducts.
This behaviour is evidenced by the monotonic growth of the dimensionless near-wall concentration $c_\mathrm{w}$ shown in figure~$\ref{Cw_outer}$, where $c_\mathrm{w}$ is defined as the mean concentration within the viscous sublayer ($r^+<5$).


In contrast, $c_\mathrm{w}$ near the inner wall exhibits non-monotonic temporal behaviour, characterized by an initial increase followed by a subsequent decrease before reaching eventual saturation, as presented in figure~$\ref{Cw_inner}$.
Note that $c_\mathrm{w}$ in case 2 shows no obvious sign of saturation, indicating that a statistical steady state has not yet been reached, 
even though the entropy shown in figure~\ref{entropy} eventually becomes nearly flat.
The effects of this slow convergence issue will be further discussed in \S~\ref{subsec:steady-state}.
Importantly, the transient overshoot observed over a range of Stokes number suggests a complex interplay of transport mechanisms intrinsic to curved geometries, thereby challenging standard equilibrium models that predict monotonic accumulation.
The final near-inner-wall concentration shows relative accumulation $(c_\mathrm{w}>1)$ in case 2 with mild curvature, whereas in cases 3$\sim$6, corresponding to strong curvature, a depletion of particles is observed $(c_\mathrm{w}<1)$.
As shown in figure~$\ref{Cw_inner}$, the steady-state $c_\mathrm{w}$ in case 3 is slightly higher than that in case 4, but remains lower than those in cases 5 and 6.
The trend indicates that the inner-wall particle depletion is weakened as $\Stokes^+$ decreases. 
The transient evolution of $c_\mathrm{w}$ is also affected by the Stokes number, as lower-$\Stokes^+$ cases generally exhibit both a less pronounced overshoot and a slower subsequent decrease in $c_\mathrm{w}$.
Nevertheless, the transient overshoot persists in all inertial-particle cases considered here. 
Note that the overshoot is expected to vanish in the tracer-particle limit $\tau_\mathrm{p}\rightarrow 0$, for which the near-wall concentration remains constant at $c_\mathrm{w}=1$ in the incompressible limit.

To quantitatively characterize the asymmetric radial transport of particles, with emphasis on the non-monotonic temporal evolution of the near-inner-wall concentration, we establish a statistical formulation for the concentration field.
Within the single-particle position-velocity PDF formalism \citep{Johnson2020Turbophoresis, Jiang2025Spatial}, the conservation equations for the particle-phase number density and radial momentum are given by
\begin{align}
    &\frac{\partial c(r,t)}{\partial t}+\frac{1}{r}\frac{\partial}{\partial r}[rJ(r,t)]=0,
    \label{eq:C-evo}\\
    &\tau_\mathrm{p}\frac{\partial J(r,t)}{\partial t}+J(r,t)=D_1(r,t)c(r,t)-D_2(r,t)\frac{\partial c(r,t)}{\partial r},
    \label{eq:J-evo}
\end{align}
where $J=\langle v_r|r\rangle c$ denotes the radial particle flux, and the angle brackets $\langle\cdot|r\rangle$ represent ensemble averages conditioned on the particle being located at the radial position $r$. 
Here, the coefficient $D_1$ characterizes the advection velocity, and $D_2$ corresponds to the turbulent diffusivity \citep{Zahtila2023Particle}, both of which are expressed as follows:
\begin{align}
	\frac{D_1(r,t)}{\tau_\mathrm{p}}&=
    \underbrace{\frac{1}{\tau_\mathrm{p}}\langle u_r|r\rangle}_{\text{biased sampling}}
    -\underbrace{\frac{1}{r}\frac{\partial}{\partial r}(r\langle v_r^2|r\rangle)}_{\text{`turbophoresis'}}
	+\underbrace{\left\langle \frac{v_\theta^2}{r}|r\right\rangle}_{\text{centrifugal force}},\\
	D_2(r,t)&=\tau_\mathrm{p}\langle v_r^2|r\rangle .
\end{align}
Here, $u_r$ denotes the radial fluid velocity evaluated at the particle location, while $v_r$ and $v_\theta$ represent the radial and azimuthal components of the particle velocity, respectively.
It should be emphasized that the coefficient $D_1$ reflects the combined contributions of biased sampling, curvature-modulated turbophoresis, and additional centrifugal effects relative to plane channels, extending the framework of \citet{Zahtila2023Particle} which focused primarily on turbophoresis.

By combining \eqref{eq:C-evo} and \eqref{eq:J-evo} and eliminating the radial particle flux $J(r,t)$, a transport equation for the particle concentration $c(r,t)$ is derived
\begin{equation}
		\tau_\mathrm{p}\frac{\partial^2 c(r,t)}{\partial t^2}
		+\frac{\partial c(r,t)}{\partial t}
		=
		-\frac{1}{r}\frac{\partial}{\partial r}
		\left[D_1(r,t)rc(r,t)-D_2(r,t)r\frac{\partial c(r,t)}{\partial r}\right].
		\label{eq:C-conv-diff}
\end{equation}
Under the assumption of perfectly elastic particle-wall interactions, which implies $J=0$ at the particle-accessible boundaries, the corresponding no-flux boundary conditions are given by
\begin{equation}
    \frac{\partial c(r,t)}{\partial r}-\frac{D_1(r,t)}{D_2(r,t)}c(r,t)=0 \text{ at $r=R_1=R_i+d_\mathrm{p}/2$ and $R_2=R_o-d_\mathrm{p}/2$}.
    \label{boundary_condition}
\end{equation}
Non-integrable singularities in the concentration $c$ are avoided, since $D_2$ remains strictly positive over the radial domain $r\in[R_1,R_2]$.
Owing to the initially random particle distribution, the initial conditions are approximated as
\begin{equation}
    \begin{aligned}
	c(r,0)&\approx1,\\
    \frac{\partial c(r,t)}{\partial t}&
    =-\frac{1}{r}\frac{\partial}{\partial r}[r\langle v_r|r\rangle c]
    \approx-\frac{1}{r}\frac{\partial}{\partial r}(r\langle u_r|r\rangle)
    \approx0 \text{ at $t=0$}.
    \end{aligned}
    \label{initial_condition}
\end{equation}
The equation \eqref{eq:C-conv-diff}, along with the boundary conditions \eqref{boundary_condition} and initial conditions \eqref{initial_condition}, are thus used for analysis of the statistical steady state and transient evolution of particles in the following subsections.

\subsection{Particle concentration at statistical steady state}\label{subsec:steady-state}

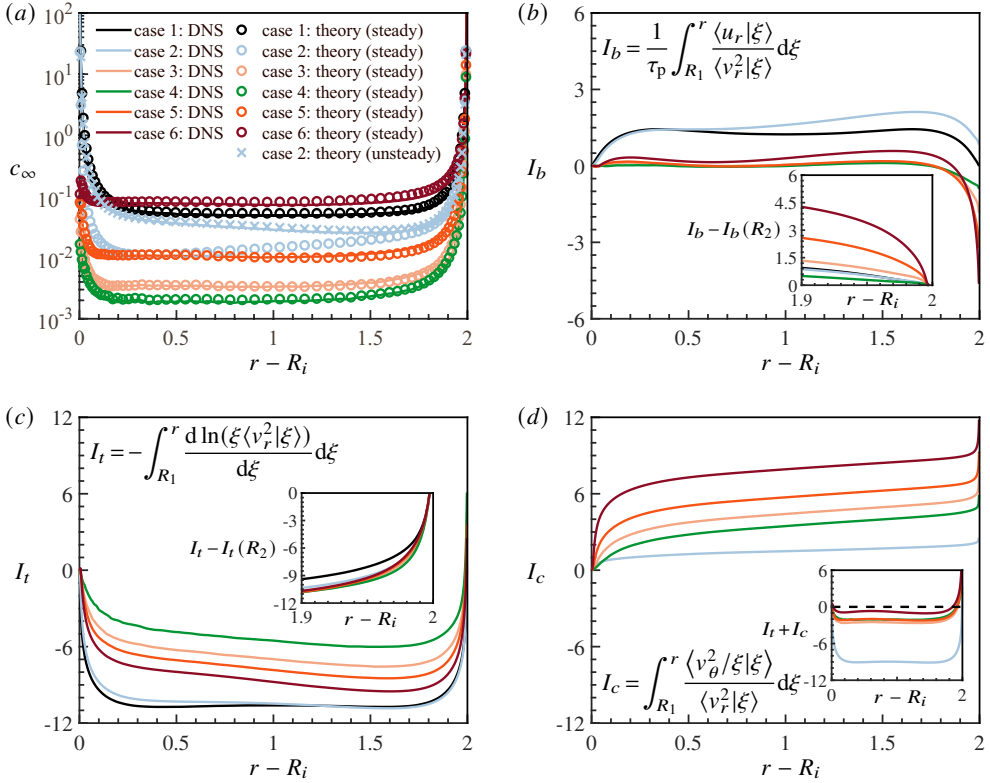
\begin{figure}\small
	\centering \subfigure{
		\begin{minipage}[c]{0.49\textwidth}{}
			\begin{overpic}[width=\textwidth]{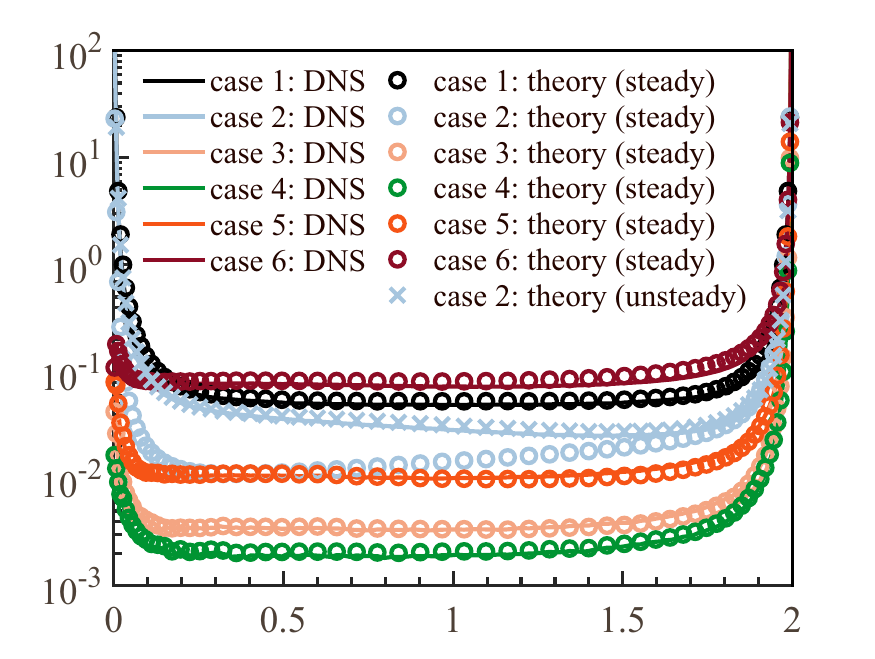}
				\put(-2,68){$(a)$}
				\put(-1,37){$c_\infty$}
				\put(47,-2){$r-R_i$}
			\end{overpic}
		\end{minipage}\label{concentration}}
    \centering \subfigure{
		\begin{minipage}[c]{0.49\textwidth}{}
			\begin{overpic}[width=\textwidth]{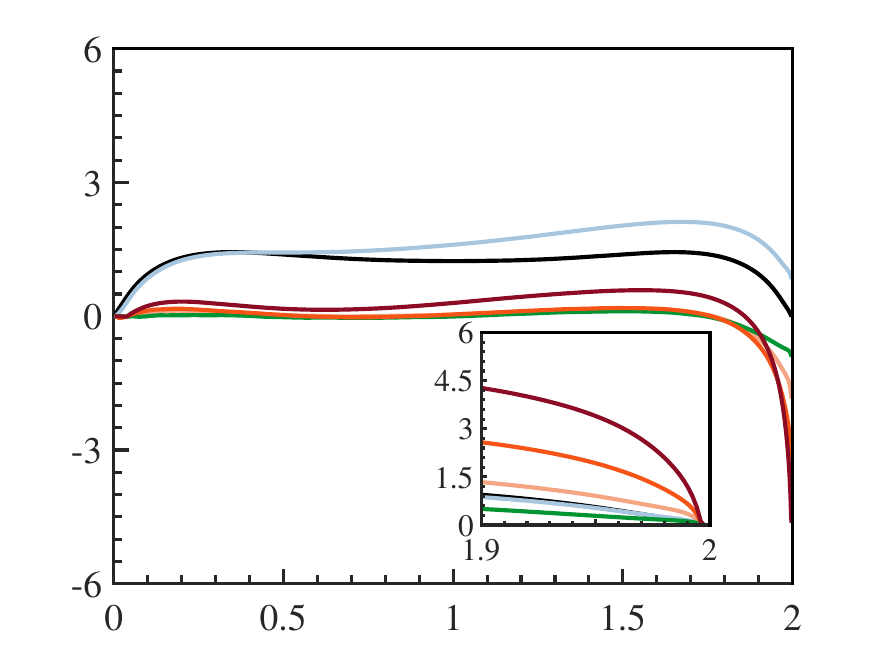}
				\put(-2,68){$(b)$}
                \put(0,37){$I_b$}
				\put(47,-2){$r-R_i$}
				
				\put(32,25){\scriptsize$I_b\!-\!I_b(R_2)$}
				\put(64,11){\scriptsize$r-R_i$}
				
				\put(15,61){$\displaystyle{I_b\!=\!\frac{1}{\tau_\mathrm{p}}\!\int_{R_1}^{r}\!\frac{\langle u_r|\xi\rangle}{\langle v^2_r|\xi\rangle}\mathrm{d}\xi}$}
			\end{overpic}
		\end{minipage}\label{Ib}}
    \centering \subfigure{
	\begin{minipage}[c]{0.49\textwidth}{}
		\begin{overpic}[width=\textwidth]{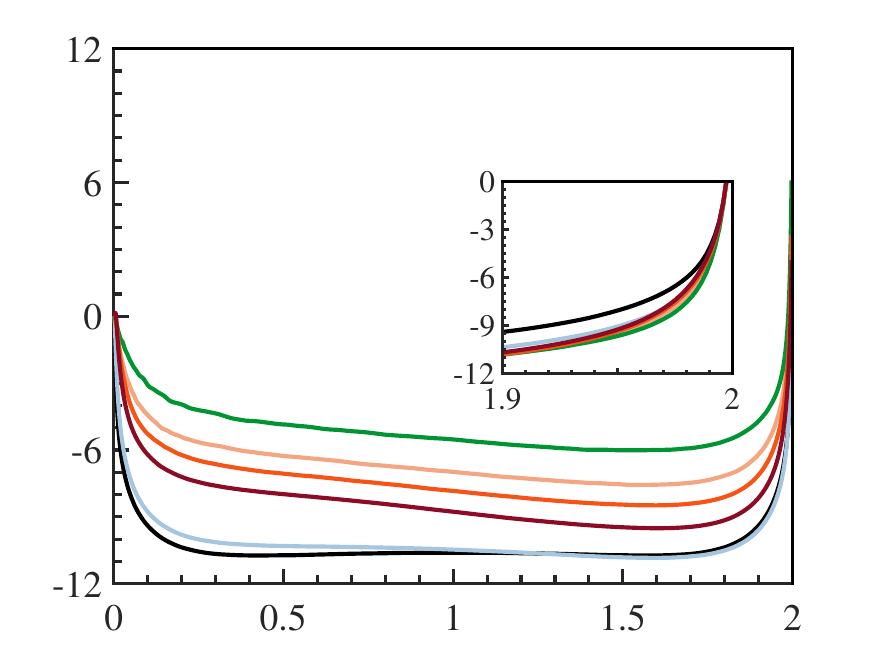}
			\put(-2,68){$(c)$}
			\put(0,37){$I_t$}
			\put(47,-2){$r-R_i$}
			
			\put(35,42){\scriptsize$I_t\!-\!I_t(R_2)$}
			\put(66,28){\scriptsize$r-R_i$}
			
			\put(15,61){$\displaystyle{I_t\!=\!-\!\int_{R_1}^{r}\!\frac{\mathrm{d}\ln(\xi\langle v^2_r|\xi\rangle)}{\mathrm{d}\xi}\mathrm{d}\xi}$}
		\end{overpic}
	\end{minipage}\label{It}}
    \centering \subfigure{
	\begin{minipage}[c]{0.49\textwidth}{}
		\begin{overpic}[width=\textwidth]{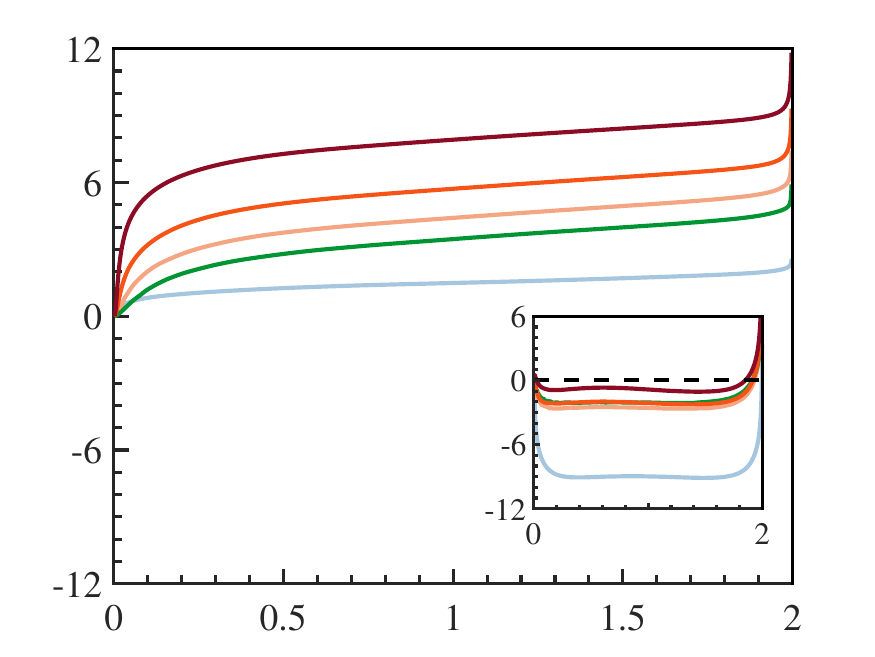}
			\put(-2,68){$(d)$}
			\put(0,37){$I_c$}
			\put(47,-2){$r-R_i$}

            \put(47,26){\scriptsize$I_t\!+\!I_c$}
            \put(69.5,12.25){\scriptsize$r-R_i$}
			
			\put(15,15){$\displaystyle{I_c\!=\!\int_{R_1}^{r}\!\frac{\left\langle v^2_\theta/\xi|\xi\right\rangle}{\langle v^2_r|\xi\rangle}\mathrm{d}\xi}$}
		\end{overpic}
	\end{minipage}\label{Ic}}
	\caption{$(a)$ Radial profiles of the steady-state concentration and contributions of phoresis integrals: $(b)$ biased sampling integral $I_b$ and a zoomed-in view of its variations near the outer wall, $(c)$ turbophoresis integral $I_t$ and a zoomed-in view of its variations near the outer wall, $(d)$ centrifugal integral $I_c$ and a zoomed-in view of $I_t+I_c$, where the black dashed line marks the zero level.}
	\label{mechanism}
\end{figure}

For the statistical steady state, imposing the time-independence condition $\partial/\partial t = 0$, equation~\eqref{eq:J-evo} indicates that $J=0$ can be satisfied through a balance between the advection velocity $D_1$ and turbulent diffusivity $D_2$. 
Equation~\eqref{eq:C-conv-diff} then reduces to an ordinary differential equation governing the steady-state concentration profile $c_{\infty}(r)$, which reads:
\begin{equation}
    c_\infty(r)
    =\mathcal{N}\exp
    \left[
    \frac{1}{\tau_\mathrm{p}}\int_{R_1}^{r}\frac{\langle u_r|\xi\rangle}{\langle v^2_r|\xi\rangle}\mathrm{d}\xi
    -\int_{R_1}^{r}\frac{\mathrm{d}\ln(\xi\langle v^2_r|\xi\rangle)}{\mathrm{d}\xi}\mathrm{d}\xi
    +\int_{R_1}^{r}\frac{\left\langle v^2_\theta/\xi|\xi\right\rangle}{\langle v^2_r|\xi\rangle}\mathrm{d}\xi
    \right],
    \label{eq:concentration}
\end{equation}
where $\mathcal{N}$ is an integration constant, and three phoresis integrals in the bracket are denoted by $I_b$, $I_t$, and $I_c$, corresponding to the contributions of biased sampling, turbophoresis and centrifugal effects, respectively.
As shown in figure~$\ref{concentration}$, the numerical results of $c_{\infty}(r)$ (based on the statistics collected over the last 1000 non-dimensional time units) are well reproduced by equation \eqref{eq:concentration} for a majority of cases, demonstrating the applicability of the statistical approach.
Note, however, that case 2 does not reach a fully steady state till the end of the simulation, as indicated by the continued presence of active radial particle transport shown in figure~$\ref{Cw_inner}$.
Accordingly, the $c_{\infty}(r)$ distribution given by~\eqref{eq:concentration} (blue circles in figure~$\ref{concentration}$) shows non-negligible deviations from the DNS results.
To account for the non-zero radial flux, the steady solution $c_\infty(r)$ in equation~\eqref{eq:concentration} is modified by premultiplying it with the following unsteady correction term
\begin{equation}
    \exp
    \left[
    -\frac{1}{\tau_\mathrm{p}}\int_{R_1}^{r}\frac{\langle v_r|\xi\rangle}{\langle v^2_r|\xi\rangle}\mathrm{d}\xi
    \right],
    \label{eq:concentration_modification}
\end{equation}
which follows from \eqref{eq:J-evo} under the approximation that the term $\tau_\mathrm{p}\partial J(r,t)/\partial t$ is neglected.
It is observed that introducing the modification in equation~\eqref{eq:concentration_modification} significantly refines the results (see the blue crosses in figure~$\ref{concentration}$).
As will be demonstrated rigorously in \S~\ref{subsec:modes}, the improvement introduced by \eqref{eq:concentration_modification} corresponds to the dominant slowly decaying eigenmode of the transport equation.

The relative contributions of individual mechanisms are quantified by the phoresis integrals plotted in figures~$\ref{Ib}$, $\ref{It}$ and $\ref{Ic}$.
Overall, the biased sampling integral $I_b$ remains small compared with the turbophoresis integral $I_t$ and the centrifugal integral $I_c$.
As the curvature increases from case 1 to case 3, the turbophoretic contribution near the inner wall is significantly reduced, as evidenced by the diminished near-wall variation of $I_t$ at $r-R_i<0.1$ in figure~$\ref{It}$.
This is consistent with the decreasing peak radial velocity fluctuation shown in figure~$\ref{particle_urur}$, presumably related to the suppression of turbulent fluctuations by the convex curvature near the inner wall in figure~$\ref{fluid_urur}$.
In contrast, the variations of $I_t$ near the outer wall, measured by $I_t-I_t(R_2)$, remain largely unchanged across different cases, as shown in the inset of figure~$\ref{It}$.
This behaviour is consistent with the limited influence of curvature on the turbulence statistics in the present concave boundary layer, as evidenced by figure~\ref{validation}.
As shown in figure~$\ref{Ic}$, comparison between cases 2 and 3 further reveals that the centrifugal effect is more pronounced in the case with stronger curvature, driving particles toward the outer wall.
The turbophoresis and centrifugal effects contribute synergistically near the outer wall, while competing near the inner wall.
Accordingly, particles are expected to accumulate near the outer wall, while the inner wall boundary layer shows particle depletion in the strong curvature case. 

We further examine the effects of varying Stokes number by considering cases 3$\sim$6. 
As shown in figure~$\ref{concentration}$, cases with smaller Stokes number feature a  more uniform distribution accompanied by a relatively higher inner-wall concentration, consistent with figure~$\ref{Cw_inner}$.
Further quantifying the phoresis integrals, the biased sampling contribution becomes slightly stronger as the Stokes number decreases, as reflected by the increased amplitude of $I_b$ shown in figure~$\ref{Ib}$.
The inner-wall turbophoretic contribution is also enhanced, whereas the outer-wall turbophoretic contribution remains largely unchanged, as indicated by figure~$\ref{It}$.
The centrifugal integral $I_c$ also becomes stronger with decreasing Stokes number.  
Despite the increase in the individual turbophoretic and centrifugal contributions, the combined effect, measured by $I_t+I_c$, becomes weaker in magnitude with decreasing Stokes number, as shown in the inset of figure~$\ref{Ic}$.
This suggests that the net redistribution driven by turbophoresis and centrifugal effects is weakened at lower Stokes number.
As a result, the relative role of biased sampling becomes more important, which is consistent with the more uniform spatial distribution observed for case 6 in figure~$\ref{concentration}$.

\subsection{Slowly decaying asymmetric particle transport modes}\label{subsec:modes}

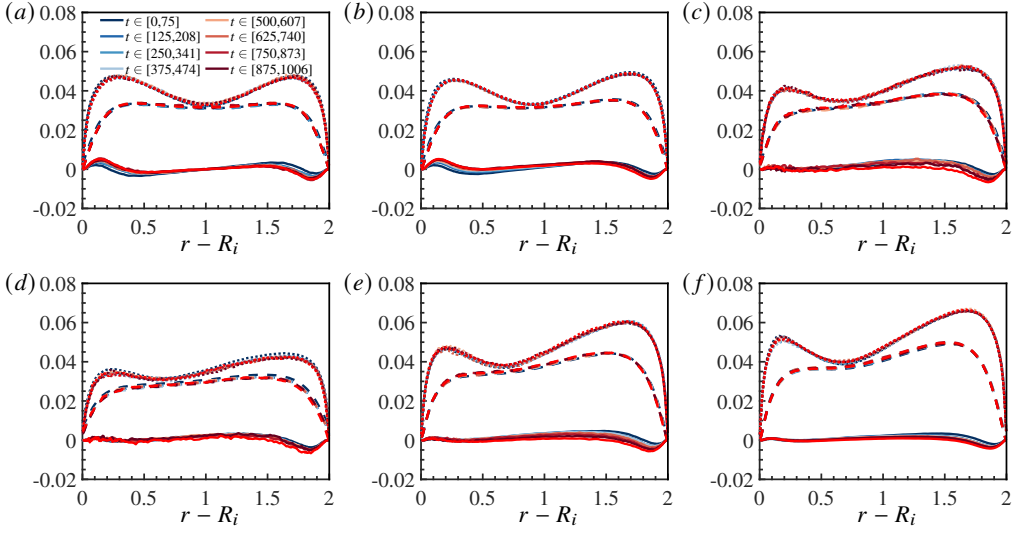
\begin{figure}\small
	\centering \subfigure{
		\begin{minipage}[c]{0.32\textwidth}{}
			\begin{overpic}[width=\textwidth]{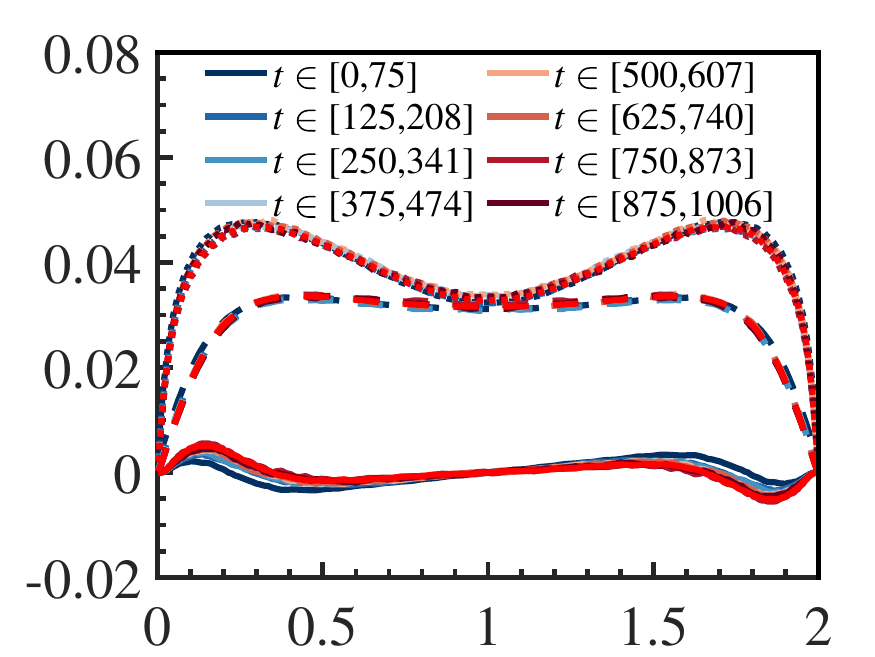}
				\put(-6,67){$(a)$}
				\put(48,-4){$r-R_i$}
			\end{overpic}
		\end{minipage}\label{evolution_eta1}}
	\centering \subfigure{
		\begin{minipage}[c]{0.32\textwidth}{}
			\begin{overpic}[width=\textwidth]{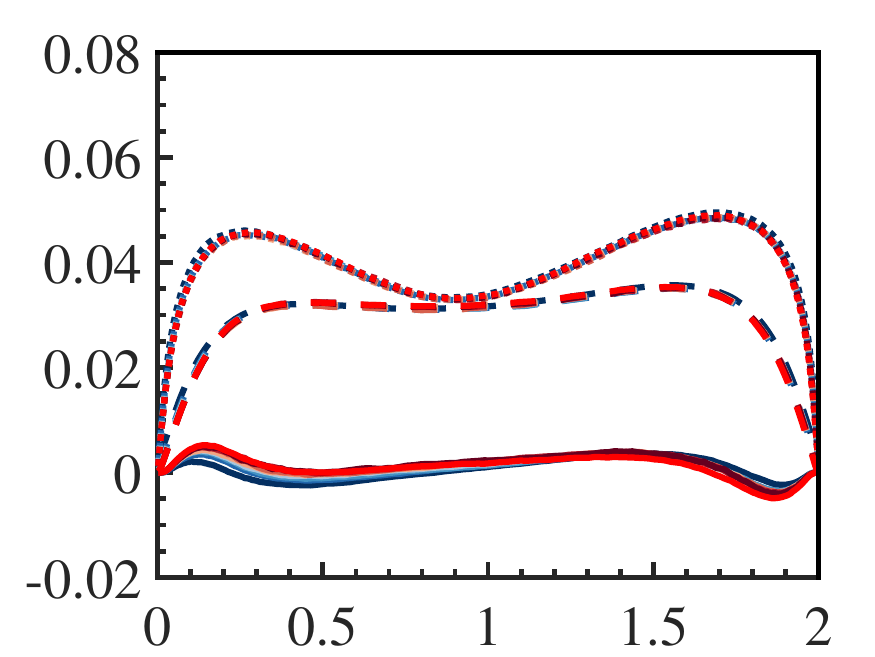}
				\put(-6,67){$(b)$}
				\put(48,-4){$r-R_i$}
			\end{overpic}
		\end{minipage}\label{evolution_eta05}}
	\centering \subfigure{
		\begin{minipage}[c]{0.32\textwidth}{}
			\begin{overpic}[width=\textwidth]{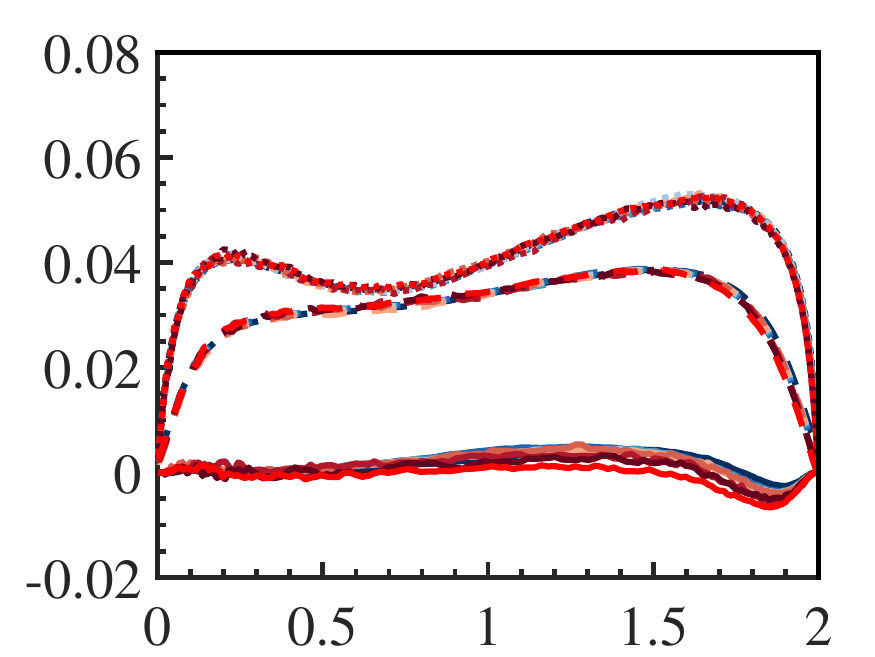}
				\put(-6,67){$(c)$}
				\put(48,-4){$r-R_i$}
			\end{overpic}
		\end{minipage}\label{evolution_eta01_rho1200}}  
	\centering \subfigure{
		\begin{minipage}[c]{0.32\textwidth}{}
			\begin{overpic}[width=\textwidth]{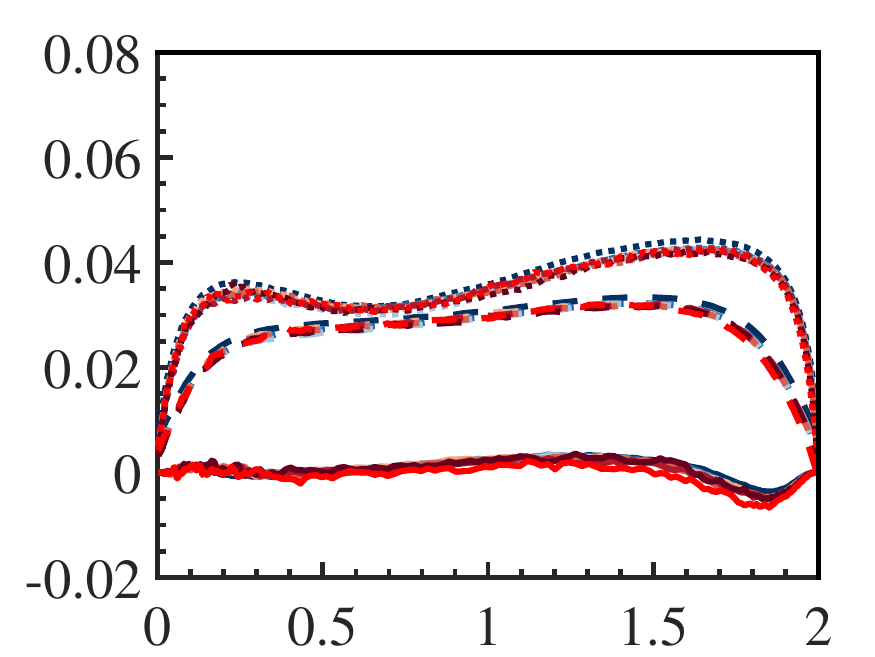}
				\put(-6,67){$(d)$}
				\put(48,-4){$r-R_i$}
			\end{overpic}
		\end{minipage}\label{evolution_eta01_rho2400}}  
	\centering \subfigure{
		\begin{minipage}[c]{0.32\textwidth}{}
			\begin{overpic}[width=\textwidth]{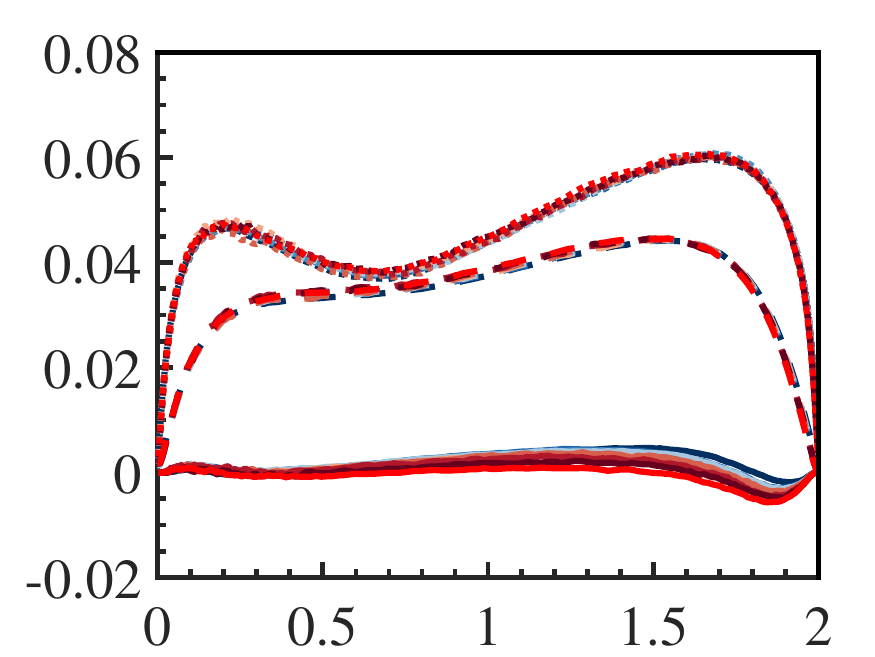}
				\put(-6,67){$(e)$}
				\put(48,-4){$r-R_i$}
			\end{overpic}
		\end{minipage}\label{evolution_eta01_rho0600}}  
	\centering \subfigure{
		\begin{minipage}[c]{0.32\textwidth}{}
			\begin{overpic}[width=\textwidth]{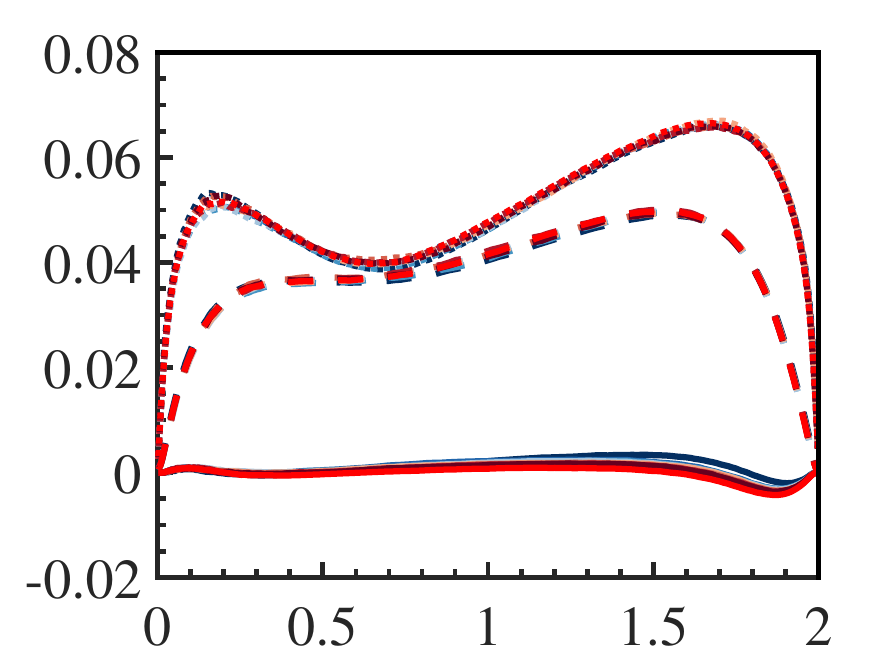}
				\put(-6,67){$(f)$}
				\put(48,-4){$r-R_i$}
			\end{overpic}
		\end{minipage}\label{evolution_eta01_rho0300}}    
	\caption{Temporal evolution of the short-time-averaged particle-sampled velocity quantities associated with $D_1$ and $D_2$: radial fluid velocity $\langle u_r|r\rangle$ (solid lines), radial rms particle velocity $\langle v_r^2|r\rangle^{1/2}$ (dashed lines) and azimuthal rms particle velocity $\langle v_{\theta}^2|r\rangle^{1/2}$ (dotted lines) for $(a)$ case 1, $(b)$ case 2, $(c)$ case 3, $(d)$ case 4, $(e)$ case 5 and $(f)$ case 6. The red lines denote the results at statistical steady state. In the plane channel case, $\langle u_r|r\rangle$, $\langle v_r^2|r\rangle^{1/2}$ and $\langle v_{\theta}^2|r\rangle^{1/2}$ reduce to particle-sampled normal fluid velocity, radial rms particle velocity, and spanwise rms particle velocity, respectively.}
	\label{steady_closure}
\end{figure}

Asymmetric particle transport is analysed quantitatively based on equation \eqref{eq:C-conv-diff}.
Given the much longer convergence time of particle concentration compared to particle velocity statistics \citep{ROUSON2001preferential,Marchioli2008Statistics, Picano2009Spatial, Sardina2012Wall}, a stationary closure, $D_1(r,t)=D_{1,\infty}(r)$ and $D_2(r,t)=D_{2,\infty}(r)$ is adopted for simplicity.
The validity of this time-independent assumption is assessed through the temporal evolution of the short-time-averaged particle-sampled velocity statistics, which are shown in figure~\ref{steady_closure}.
As discussed in \S~\ref{subsec:steady-state}, the near-wall particle concentration approaches its statistical steady state only for $t>2000$, with pronounced temporal evolution during the transient stage, as also reflected by its much slower convergence in figures~\ref{C_evolution_outer} and \ref{C_evolution_inner}.
By contrast, the averaged radial and azimuthal particle velocity fluctuations in the first window, $t\in[0,75]$, are already in close agreement with their statistical steady-state values.
The particle-sampled radial fluid velocity $\langle u_r|r\rangle$ requires a relatively longer time to converge, while its amplitude varies only weakly over time.
This separation of timescales suggests that $D_1$ and $D_2$ can be treated as approximately constant during the slow evolution of particle concentration, thereby supporting the adopted stationary closure.

The concentration is then expressed as $c(r,t)=c_\infty(r)\zeta(r,t)$, where $\zeta(r,t)$ represents the transient modulation describing the relaxation toward the steady-state profile $c_\infty(r)$.
Accordingly, the evolution of $\zeta(r,t)$ is governed by the equation derived from \eqref{eq:C-conv-diff} and \eqref{boundary_condition},
\begin{equation}
		\tau_\mathrm{p}\frac{\partial^2 \zeta(r,t)}{\partial t^2}
		+\frac{\partial \zeta(r,t)}{\partial t}
		=
		\frac{1}{rc_\infty(r)}\frac{\partial}{\partial r}
        \left[
		rD_{2,\infty}(r)c_\infty(r)\frac{\partial \zeta(r,t)}{\partial r}
		\right],~
        \left.\frac{\partial \zeta(r,t)}{\partial r}\right|_{R_1,R_2}=0.
	\label{eq:zeta-evolution}
\end{equation}
Applying separation of variables, $\zeta(r,t)=\mathcal{R}(r)\mathcal{T}(t)$, the radial function $\mathcal{R}(r)$ satisfies the following self-adjoint Sturm-Liouville eigenvalue problem:
\begin{equation}
\frac{\mathrm{d}}{\mathrm{d}r}
\left\{
[rD_{2,\infty}(r)c_\infty(r)]\frac{\mathrm{d}\mathcal{R}_n(r)}{\mathrm dr}
\right\}
=-\lambda_n[rc_\infty(r)]\mathcal{R}_n(r),~
\left.\frac{\mathrm{d}\mathcal{R}_n(r)}{\mathrm{d}r}\right|_{R_1,R_2}=0,
\label{eq:radial_function}
\end{equation}
while the temporal function $\mathcal{T}(t)$ is governed by
\begin{equation}
	\tau_\mathrm{p}\frac{\mathrm{d}^2\mathcal{T}_n(t)}{\mathrm{d}t^2}+\frac{\mathrm{d}\mathcal{T}_n(t)}{\mathrm{d}t}+\lambda_n\mathcal{T}_n(t)=0.
\end{equation}

Based on the steady-state concentration profile $c_{\infty}(r)$ evaluated from equation \eqref{eq:concentration}, with the coefficients $D_{1,\infty}(r)$ and $D_{2,\infty}(r)$ extracted a posteriori from the DNS statistics, a complete set of orthonormal eigenfunctions $\{\mathcal{R}_n(r)\}$ and the corresponding eigenvalues $\{\lambda_n\}\geq0$ can be obtained numerically from \eqref{eq:radial_function}.
For simplicity, each eigenfunction $\mathcal{R}_n(r)$ is normalized such that
\begin{equation}
    \int_{R_1}^{R_2}\mathcal{R}_m(r)[rc_\infty(r)]\mathcal{R}_n(r)\mathrm{d}r=\delta_{mn},
\end{equation}
where $\delta_{mn}$ denotes the Kronecker symbol.
The temporal function admits two exponential branches of the form $\mathcal{T}_n(t)\sim\exp(s_n^{\pm}t)$, with decay rates $s_n^{\pm}=(-1\pm\sqrt{\Delta_n})/(2\tau_\mathrm{p})$, where $\Delta_n=1-4\tau_\mathrm{p}\lambda_n$.
Therefore, the particle concentration field can be expressed as a modal expansion of the form $\mathcal{R}_n(r)\mathcal{T}_n(t)$, given by
\begin{equation}
\frac{c(r,t)}{c_\infty(r)}
=\sum_{n=0}^{n_c}\mathcal{R}_n(r)
(\underbrace{a_n\mathrm{e}^{s_n^{+}t}}_{\substack{\text{slowly}\\ \text{decaying}}}
+
\underbrace{b_n\mathrm{e}^{s_n^{-}t}}_{\substack{\text{rapidly}\\ \text{decaying}}})
+\sum_{n=n_c+1}^{\infty}\!\mathcal{R}_n(r)
\underbrace{c_n
\mathrm{e}^{-\frac{t}{2\tau_\mathrm{p}}}
\cos\left(
\frac{\sqrt{|\Delta_n|}}{2\tau_\mathrm{p}}t+\theta_n
\right)}_{\text{oscillatory decaying}},
\label{eq:concentration_evolution}
\end{equation}
where the coefficients $a_n$, $b_n$, $c_n$ and phase $\theta_n$ are determined by the initial conditions \eqref{initial_condition}, and $n_c$ denotes the cutoff index separating slowly and rapidly decaying modes ($\Delta_n \ge 0$) from oscillatory decaying modes ($\Delta_n < 0$).

The decoupled modal expansion quantitatively describes the multi-time-scale relaxation of the concentration field toward its steady state.
The slowly decaying modes correspond to particle redistribution with larger time scales $(>2\tau_\mathrm{p})$, while the rapidly decaying modes correspond to redistribution with smaller time scales $(<2\tau_\mathrm{p})$.
Both types of modes reflect the relaxation of the particle concentration toward the steady state along a fixed direction.
In contrast, the oscillatory decaying modes correspond to oscillatory relaxation components whose direction varies with time, whereas the associated decay time scale remains constant $(=2\tau_\mathrm{p})$.
In particular, for the zero eigenvalue $\lambda_0=0$, the slowly decaying solution degenerates into the zeroth-order mode ($n=0$).
This mode is referred to as the constant mode, because its amplitude remains unchanged over time, and the corresponding eigenfunction $\mathcal{R}_0(r)$ is spatially uniform obtained from equation~\eqref{eq:radial_function}.
This mode represents the equilibrium particle partition, considering all other modes decay to zero for $t\to \infty$.

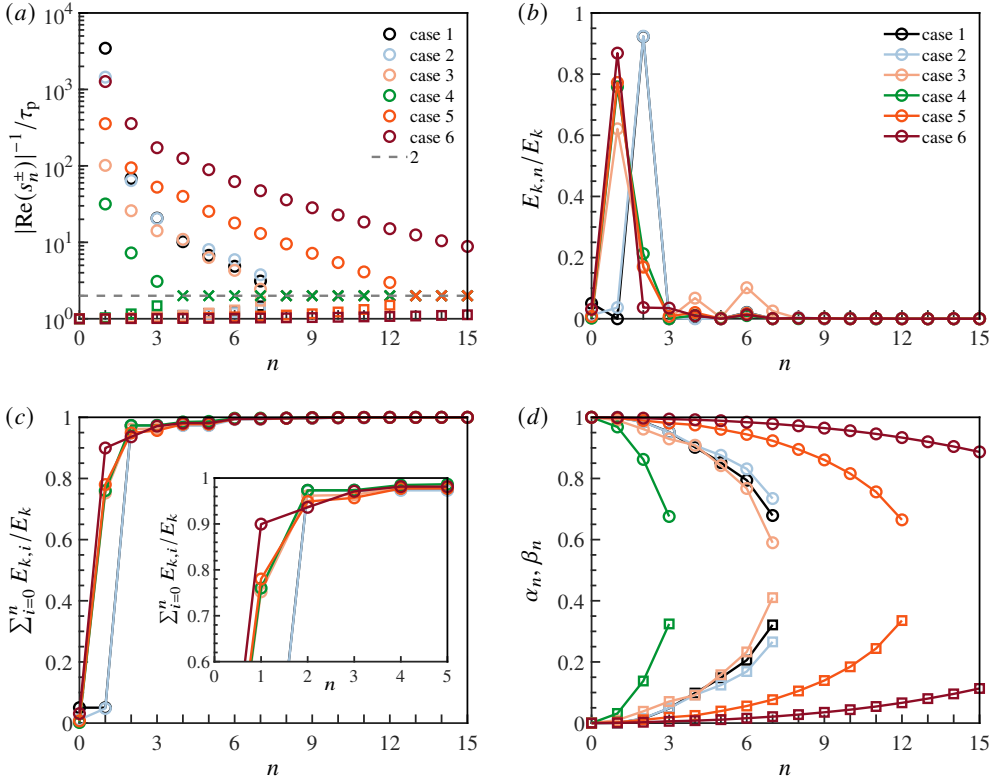
\begin{figure}\small
	\centering \subfigure{
		\begin{minipage}[c]{0.49\textwidth}{}
			\begin{overpic}[width=\textwidth]{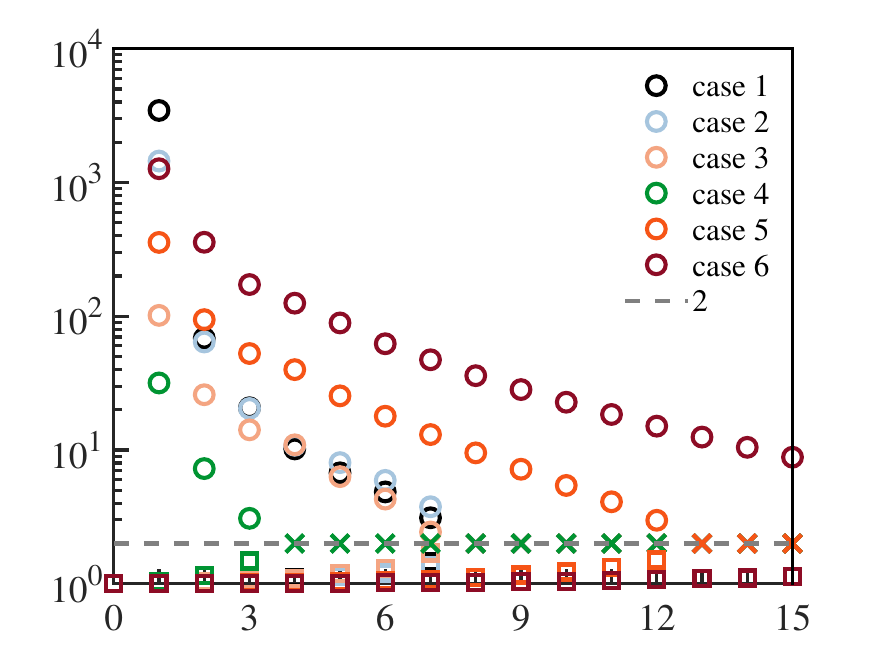}
				\put(-2,68){$(a)$}
				\put(-1,25){\rotatebox{90}{${|\mathrm{Re}(s_n^\pm)|}^{-1}/\tau_\mathrm{p}$}}
				\put(50.5,-2){$n$}
			\end{overpic}
		\end{minipage}\label{times_cales}}
    \centering \subfigure{
		\begin{minipage}[c]{0.49\textwidth}{}
			\begin{overpic}[width=\textwidth]{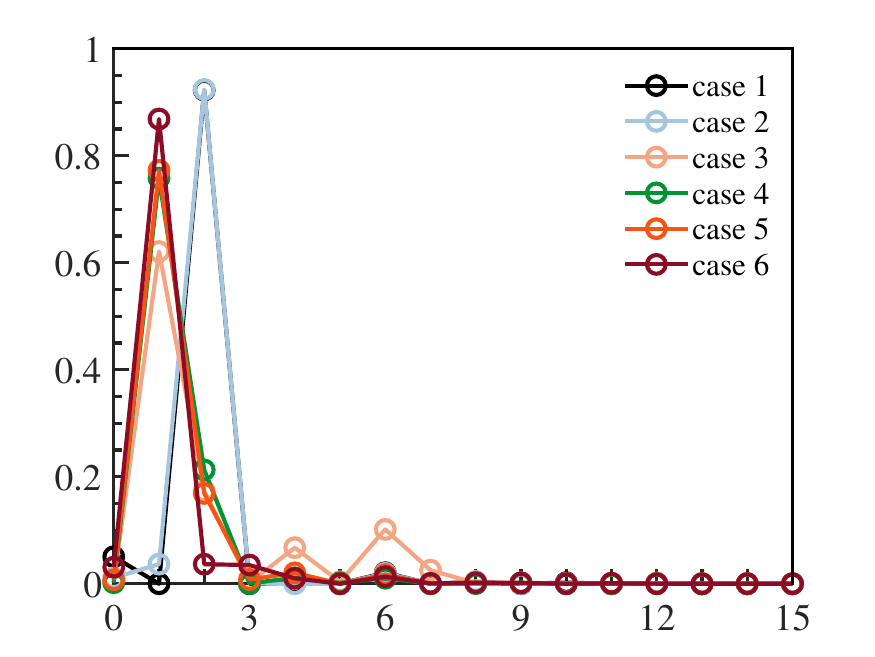}
				\put(-2,68){$(b)$}
                \put(-1,30){\rotatebox{90}{$E_{k,n}/E_k$}}
				\put(50.5,-2){$n$}
			\end{overpic}
		\end{minipage}\label{energy}}    
    \centering \subfigure{
		\begin{minipage}[c]{0.49\textwidth}{}
			\begin{overpic}[width=\textwidth]{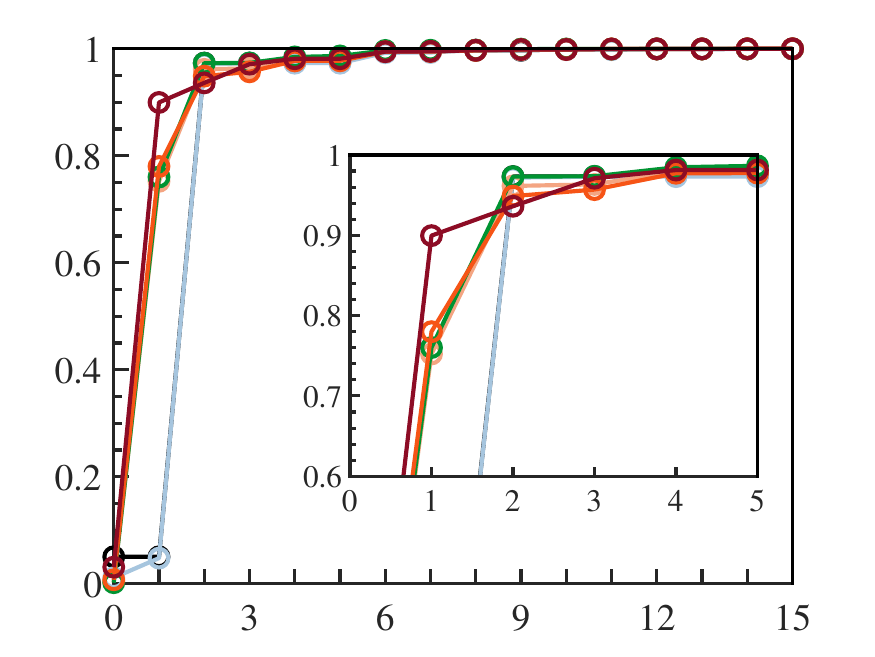}
				\put(-2,68){$(c)$}
                \put(-1,26){\rotatebox{90}{$\sum_{i=0}^{n}E_{k,i}/E_k$}}
				\put(50.5,-2){$n$}

                \put(62,15){\scriptsize$n$}
			    \put(29,28){\scriptsize\rotatebox{90}{$\sum_{i=0}^{n}E_{k,i}/E_k$}}
			\end{overpic}
		\end{minipage}\label{truncation}}    
    \centering \subfigure{
		\begin{minipage}[c]{0.49\textwidth}{}
			\begin{overpic}[width=\textwidth]{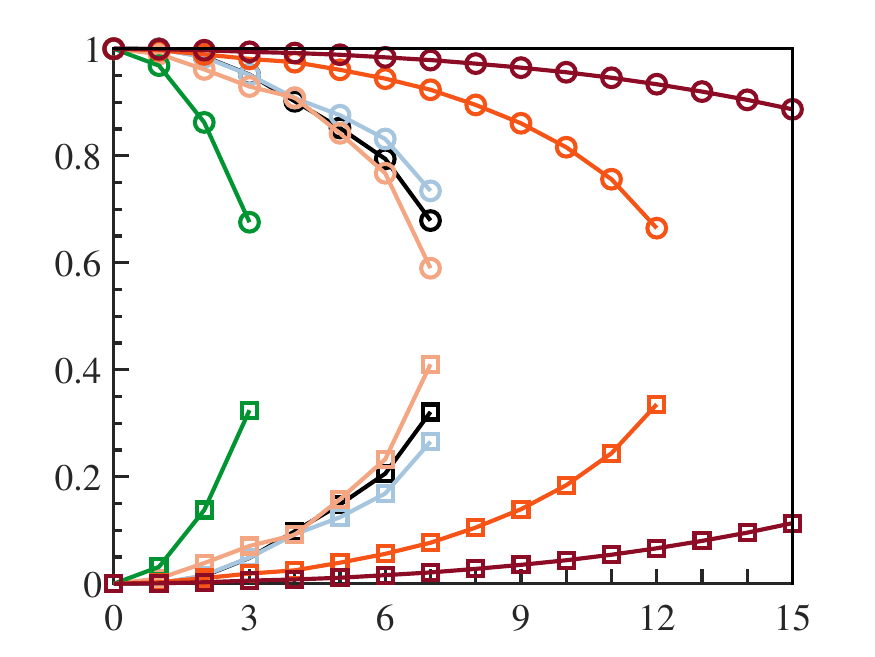}
				\put(-2,68){$(d)$}
                \put(-1,32.5){\rotatebox{90}{$\alpha_n$, $\beta_n$}}
				\put(50.5,-2){$n$}
			\end{overpic}
		\end{minipage}\label{amplititudes_relative}}
	\caption{Characteristics of the modes: $(a)$ the time scales normalized by $\tau_\mathrm{p}$ of the slowly decaying modes (circle), rapidly decaying modes (square), and oscillatory decaying modes (cross); $(b)$ the initial modal energy contribution $E_{k,n}/E_{k}$; $(c)$ the cumulative modal energy contribution $\sum_{i=0}^{n}E_{k,i}/E_k$; $(d)$ the relative amplitudes of slowly decaying modes $\alpha_n=|a_n|/(|a_n|+|b_n|)$ (circle) and rapidly decaying modes $\beta_n=|b_n|/(|a_n|+|b_n|)$ (square).}
	\label{modes}
\end{figure}

As evidenced by the normalized decay time scales shown in figure~$\ref{times_cales}$, the slowest decaying mode evolves on a time scale much larger than $\tau_\mathrm{p}$, indicating a very slow transport process.
The comparison among cases 1$\sim$3 clearly demonstrates that the corresponding time scale decreases with increasing curvature, consistent with the enhanced asymmetric particle transport revealed by comparison of figures~\ref{C_evolution_outer} and \ref{C_evolution_inner}.
The increase in time scale with decreasing Stokes number quantitatively explains the slower concentration evolution in case 6 compared with cases 3$\sim$5, as shown in figure~$\ref{Cw_inner}$.
The initial modal energy contribution $E_{k,n}/E_{k}$ is presented in figure~$\ref{energy}$.
Here, the initial energy associated with the $n$th-order mode, $E_{k,n}$, is defined within the Sturm-Liouville framework as
\begin{equation}
\begin{aligned}
    E_{k,n}&=\left\{\int_{R_1}^{R_2}\zeta(r,0)[rc_\infty(r)]\mathcal{R}_n(r)\mathrm{d}r\right\}^2=
    \left\{
    \begin{aligned}
        &(a_n+b_n)^2  &n&\leq n_c\\
        &c_n^2\cos^2(\theta_n) &n&\geq n_c+1
    \end{aligned}
    \right.\\
    % E_{k}&=\int_{R_1}^{R_2}\zeta^2(r,0)[rc_\infty(r)]\mathrm{d}r
\end{aligned}
\end{equation}
while $E_{k}=\sum_{n=0}^{\infty}E_{k,n}$ denotes the total modal energy.
It is observed that the decaying modes ($n\leq n_c$) account for most of the initial energy, with the first- and second-order modes $(n=1\text{ and }n=2)$ making the largest contributions.
By contrast, the higher-order oscillatory decaying modes ($n\geq n_c+1$) contribute only negligibly.
The dominant role of the leading modes is further illustrated by the cumulative initial modal energy contribution shown in figure~$\ref{truncation}$.
The results indicate that the modes of the first three orders account for more than 90\% of the total initial energy.
Since the decaying mode at each order comprises both a slowly decaying and a rapidly decaying component, as given by equation~\eqref{eq:concentration_evolution}, it is useful to further distinguish their respective contributions.
To this end, the relative amplitudes of the slowly decaying and rapidly decaying modes, defined as $\alpha_n=|a_n|/(|a_n|+|b_n|)$ and $\beta_n=|b_n|/(|a_n|+|b_n|)$, are shown in figure~$\ref{amplititudes_relative}$.
It is observed that the slowly decaying modes possess larger initial amplitudes than the rapidly decaying ones, suggesting that the transient concentration evolution is governed primarily by the slowly decaying modes.

\begin{figure}\small
	\centering \subfigure{
		\begin{minipage}[c]{0.49\textwidth}{}
			\begin{overpic}[width=\textwidth]{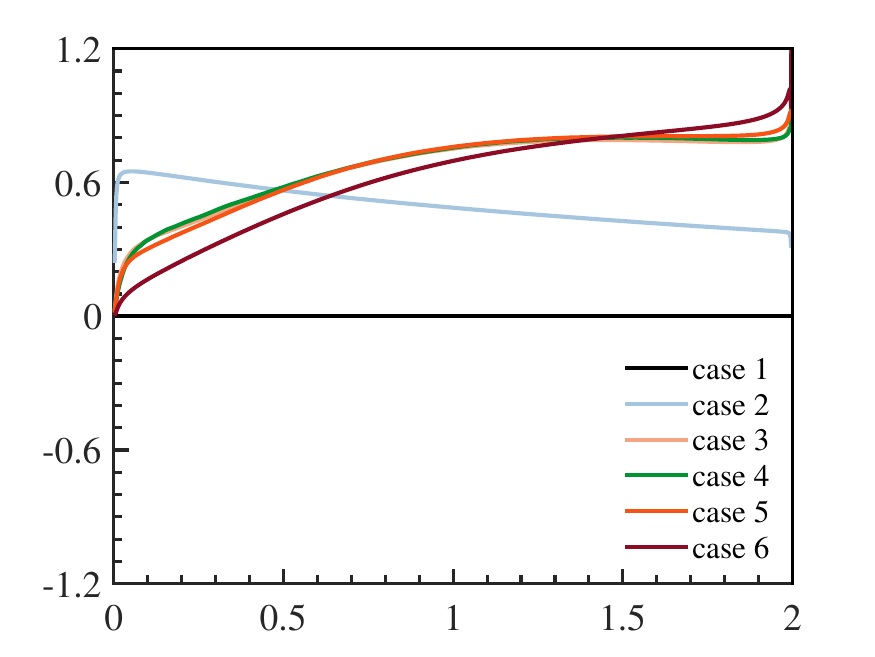}
				\put(-2,68){$(a)$}
				\put(0,31){\rotatebox{90}{$J^+_1(r,0)$}}
				\put(47,-2){$r-R_i$}
			\end{overpic}
		\end{minipage}\label{mode_flux_1}}
	\centering \subfigure{
		\begin{minipage}[c]{0.49\textwidth}{}
			\begin{overpic}[width=\textwidth]{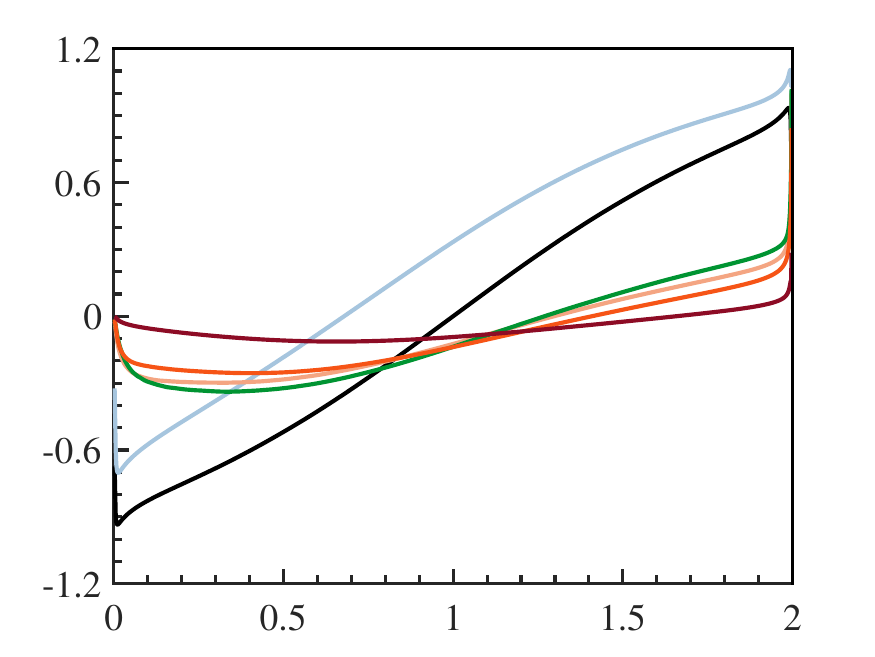}
				\put(-2,68){$(b)$}
				\put(0,31){\rotatebox{90}{$J^+_2(r,0)$}}
				\put(47,-2){$r-R_i$}
			\end{overpic}
		\end{minipage}\label{mode_flux_2}}
	\caption{Radial profile of initial flux $J^+_n(r,0)$ for $(a)$ the outer-wall transport mode and $(b)$ the wallward transport mode.}
	\label{mode_flux}
\end{figure}
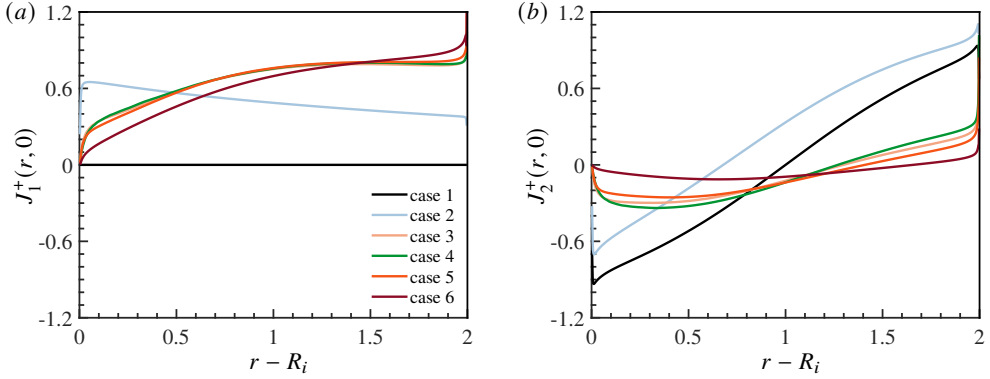

To elucidate the physical role of the dominant first- and second-order modes, the corresponding modal flux $J_n(r,t)$ is examined, where $J_n(r,t)$ is defined according to \eqref{eq:C-evo} as
\begin{equation}
	J_n(r,t)
    =-\frac{1}{r}\!\int_{R_1}^{r}\xi\frac{\partial [c_\infty(\xi)\mathcal{R}_n(\xi)\mathcal{T}_n(t)]}{\partial t}\mathrm{d}\xi
	=\frac{1}{\lambda_n}D_{2,\infty}(r)c_\infty(r)\frac{\mathrm{d}\mathcal{R}_n(r)}{\mathrm{d}r}\frac{\mathrm{d}\mathcal{T}_n(t)}{\mathrm{d}t}.
    \label{eq:particle_flux}
\end{equation}
The flux associated with the slowly decaying modes, $J^+_n(r,t)$, satisfies $J^+_n(r,t) = J^+_n(r,0)\exp(s_n^{+}t)$.
This form corresponds to selecting the branch $\mathcal{T}_n(t)\sim\exp(s_n^+t)$ in equation~\eqref{eq:particle_flux} and implies exponential decay from the initial flux $J^+_n(r,0)$.
We therefore examine its radial profile, as shown in figure~$\ref{mode_flux}$.
The first slowly decaying mode represents an outer-wall-directed transport mode arising from the wall curvature, preferentially transporting particles from the inner-wall region toward the outer wall.
This mode is specific to curved geometries, since it is activated only in annular ducts, as evidenced by its vanishing amplitude in the plane channel case shown in figure~$\ref{mode_flux_1}$.
This observation is also consistent with the zero energy contribution of the first-order mode, $E_{k,1}=0$, in case 1, as shown in figure~$\ref{energy}$.
In contrast, the second slowly decaying mode corresponds to a wallward transport mode that drives particles toward both walls simultaneously, analogous to the classical turbophoretic mechanism, as reflected by the sign of its associated initial flux shown in figure~$\ref{mode_flux_2}$.
Except for case 1, the profiles of $J^+_n(r,0)$ in the other cases exhibit pronounced asymmetry, highlighting the intrinsic effect of geometric curvature.

As also indicated by the quantification of turbophoresis and centrifugal effects shown in figure~$\ref{mechanism}$, the outer-wall-directed transport mode is expected to dominate under strong-curvature conditions, as evidenced by the comparison between figures~$\ref{mode_flux_1}$ and $\ref{mode_flux_2}$, which shows that $|J^+_1(r,0)|$ is larger than $|J^+_2(r,0)|$ in cases 3$\sim$6.
This is also supported by the substantially larger energy contribution of the first-order mode shown in figure~$\ref{energy}$.
In contrast, for case 2, $|J^+_2(r,0)|$ is relatively larger near the outer wall, suggesting that the wallward transport mode becomes more important when the curvature is mild.
This is likewise reflected in the pronounced energy contribution of the second-order mode for case 2 shown in figure~$\ref{energy}$.
The amplitudes of both fluxes near the inner wall also decrease slightly with decreasing Stokes number, as can be seen from the comparison among cases 3$\sim$6.
This behaviour suggests slower radial particle transport in the near-inner-wall region, in agreement with the trend shown in figure~$\ref{Cw_inner}$.
The identical signs of $J_1^+(r,0)$ and $J_2^+(r,0)$ near the outer wall imply that the two leading modes contribute synergistically in this region, while their opposite signs near the inner wall indicate competition between them.

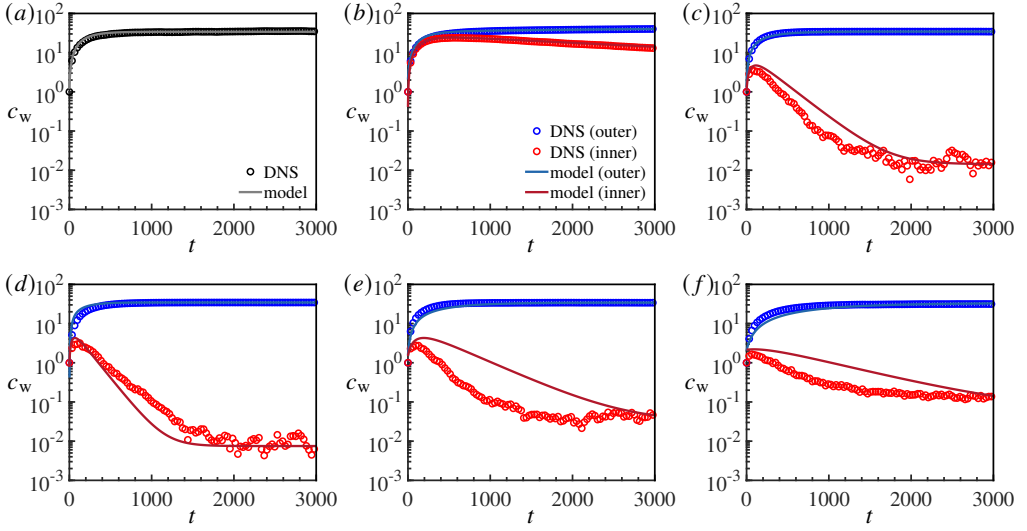
\begin{figure}\small
	\centering \subfigure{
		\begin{minipage}[c]{0.32\textwidth}{}
			\begin{overpic}[width=\textwidth]{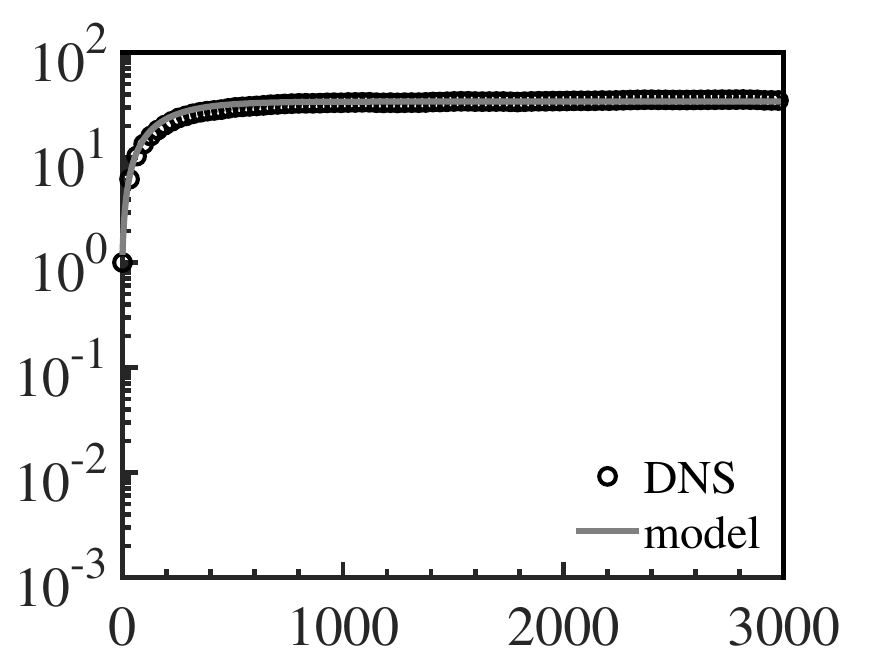}
				\put(-6,67){$(a)$}
				\put(-5,37){$c_\mathrm{w}$}
				\put(50.5,-4){$t$}
			\end{overpic}
		\end{minipage}\label{near_wall_c_eta1}}
	\centering \subfigure{
		\begin{minipage}[c]{0.32\textwidth}{}
			\begin{overpic}[width=\textwidth]{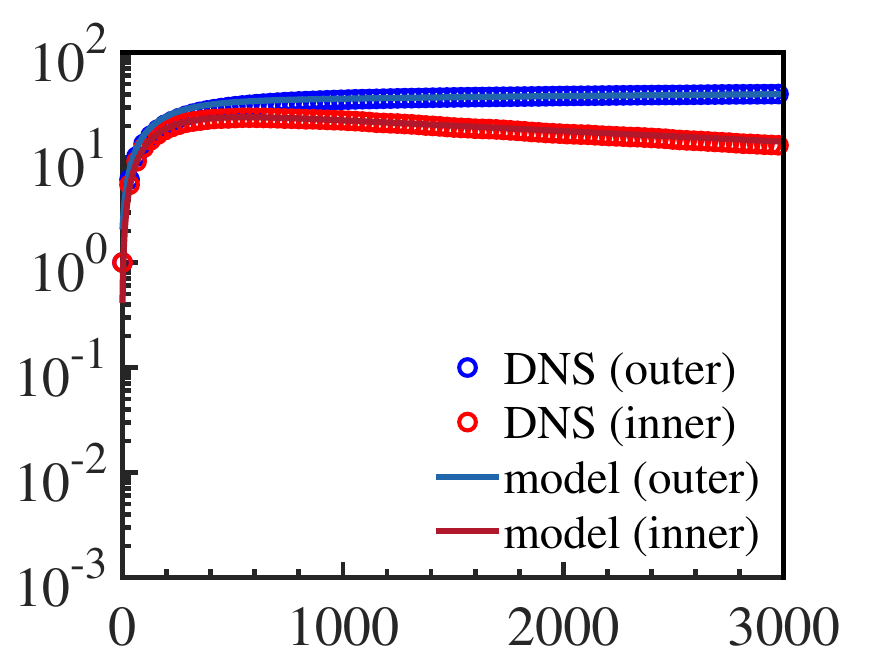}
				\put(-6,67){$(b)$}
				\put(-5,37){$c_\mathrm{w}$}
				\put(50.5,-4){$t$}
			\end{overpic}
		\end{minipage}\label{near_wall_c_eta05}}
	\centering \subfigure{
		\begin{minipage}[c]{0.32\textwidth}{}
			\begin{overpic}[width=\textwidth]{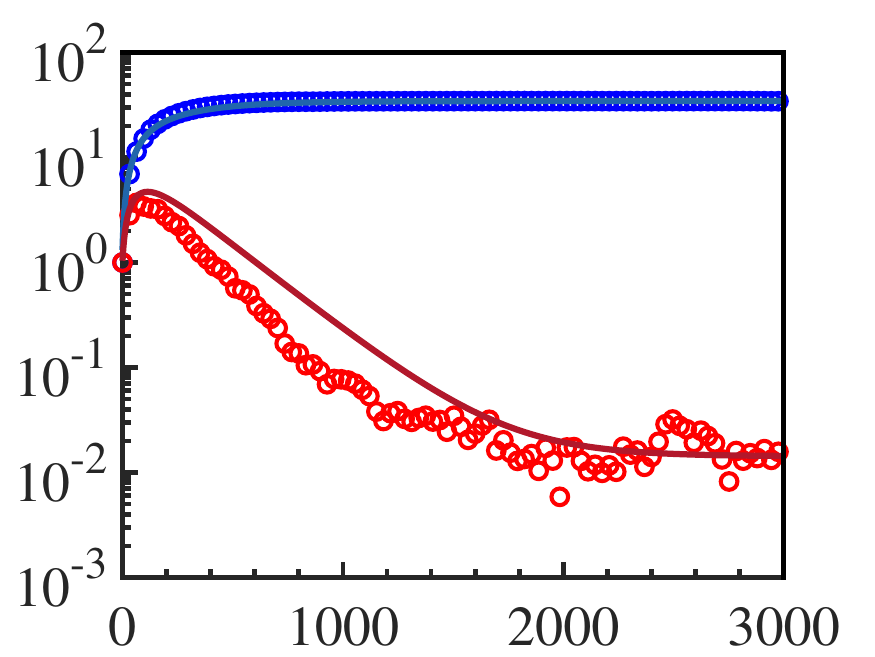}
				\put(-6,67){$(c)$}
				\put(-5,37){$c_\mathrm{w}$}
				\put(50.5,-4){$t$}
			\end{overpic}
		\end{minipage}\label{near_wall_c_eta01_rho1200}} 
	\centering \subfigure{
		\begin{minipage}[c]{0.32\textwidth}{}
			\begin{overpic}[width=\textwidth]{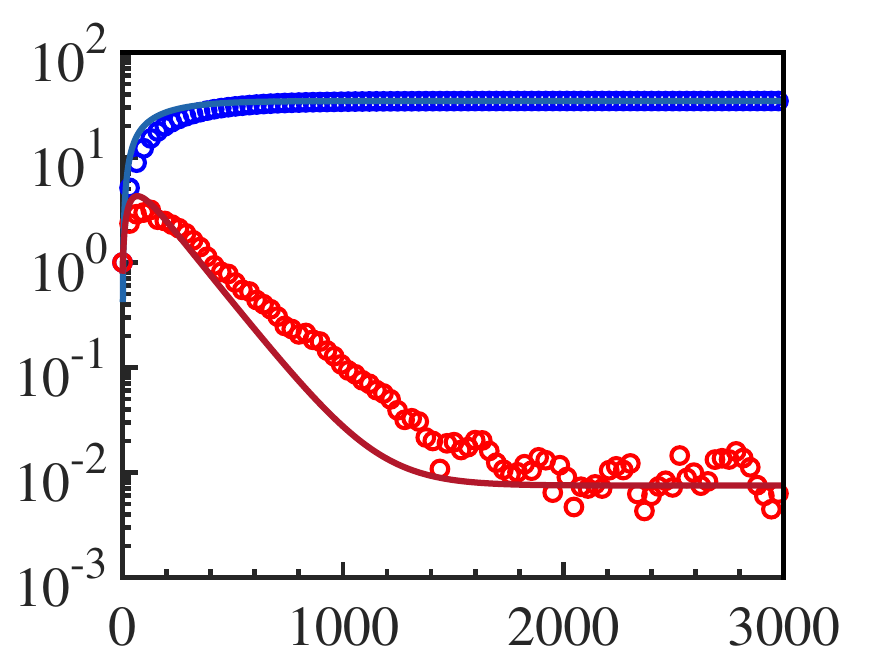}
				\put(-6,67){$(d)$}
				\put(-5,37){$c_\mathrm{w}$}
				\put(50.5,-4){$t$}
			\end{overpic}
		\end{minipage}\label{near_wall_c_eta01_rho2400}} 
	\centering \subfigure{
		\begin{minipage}[c]{0.32\textwidth}{}
			\begin{overpic}[width=\textwidth]{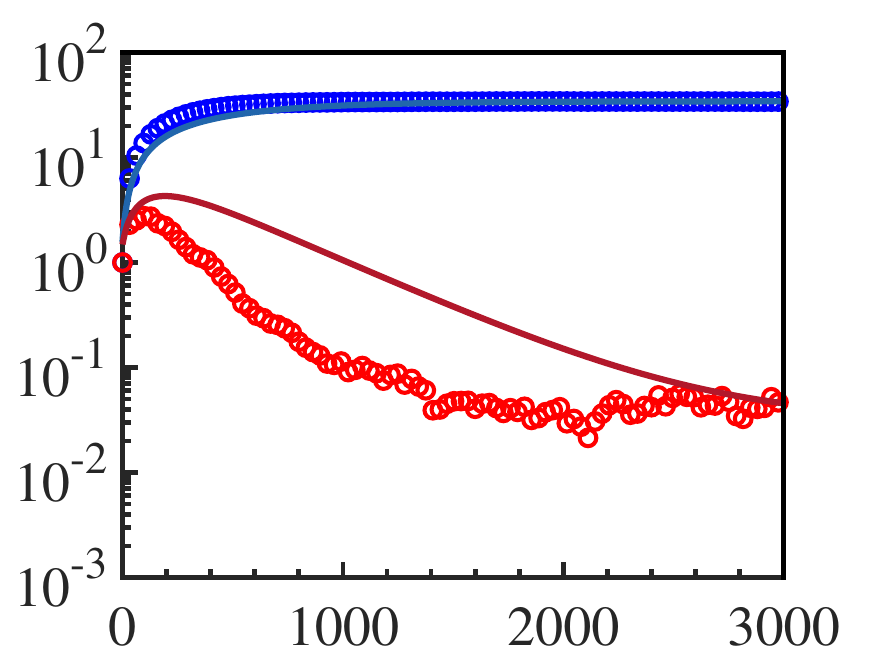}
				\put(-6,67){$(e)$}
				\put(-5,37){$c_\mathrm{w}$}
				\put(50.5,-4){$t$}
			\end{overpic}
		\end{minipage}\label{near_wall_c_eta01_rho0600}} 
	\centering \subfigure{
		\begin{minipage}[c]{0.32\textwidth}{}
			\begin{overpic}[width=\textwidth]{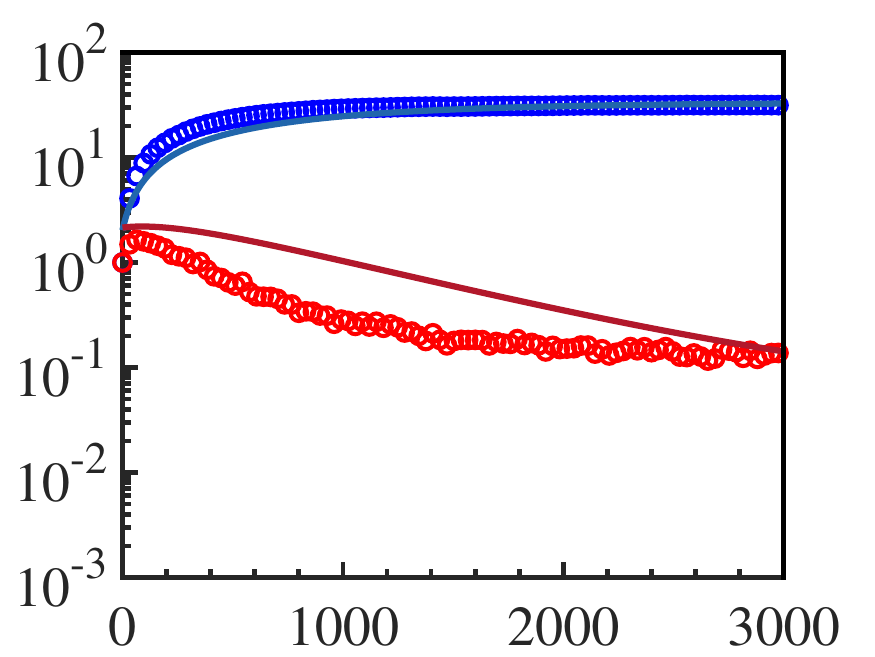}
				\put(-6,67){$(f)$}
				\put(-5,37){$c_\mathrm{w}$}
				\put(50.5,-4){$t$}
			\end{overpic}
		\end{minipage}\label{near_wall_c_eta01_rho0300}}    
	\caption{Reconstructed near-wall concentration $c_\mathrm{w}$ for $(a)$ case 1, $(b)$ case 2, $(c)$ case 3, $(d)$ case 4, $(e)$ case 5 and $(f)$ case 6.}
	\label{near_wall_c}
\end{figure}

To demonstrate the dominant role of these two modes in the transient evolution of particle concentration, the concentration field is reconstructed as follows:
\begin{equation}
c(r,t)\sim c_\infty(r)[
\underbrace{\mathcal{R}_0(r)a_0}_{\substack{\text{constant}\\ \text{mode}}}
+
\underbrace{\mathcal{R}_1(r)a_1\mathrm{e}^{s_1^{+}t}}_{\substack{\text{outer-wall-directed}\\ \text{transport mode}}}
+
\underbrace{\mathcal{R}_2(r)a_2\mathrm{e}^{s_2^{+}t}}_{\substack{\text{wallward}\\ \text{transport mode}}}
].
\label{eq:reconstruction}
\end{equation}
In this reconstruction, only the slowly decaying components of the first three modal orders are retained.
As discussed in figure~$\ref{modes}$, these three leading slowly decaying modes make the dominant contributions to the evolution of the concentration.
The resulting near-wall concentration $c_\mathrm{w}$ is shown in figure~\ref{near_wall_c}.
Remarkably, for cases 1$\sim$3, the first three slowly decaying modes are sufficient to capture the evolution of $c_\mathrm{w}$, particularly its non-monotonic behaviour near the inner wall. 
This indicates that the concentration overshoot is driven by the phase lag between the outer-wall-directed transport mode and the wallward transport mode, which evolve on distinct time scales.
During the initial transient, the wallward transport mode dominates and promotes particle accumulation at the inner wall.
At later times, the more slowly decaying outer-wall-directed transport mode prevails, leading to subsequent particle depletion.
When the concentration dynamics are predominantly governed by the leading slowly decaying modes, the term $\tau_\mathrm{p}\partial J/\partial t\sim \tau_\mathrm{p}s_n^{+}\exp(s_n^{+}t)$ is of smaller order and therefore can be neglected relative to the remaining terms $J$, $D_1c$ and $D_2\partial c/\partial r\sim\exp(s_n^{+}t)$ in equation~\eqref{eq:J-evo}.
This further justifies the proposed modification term \eqref{eq:concentration_modification}.

Considering cases 3$\sim$6, the reduced-order model accurately describes the continuous particle accumulation near the outer wall over a range of Stokes numbers.
The trend of the concentration evolution near the inner wall is also captured reasonably well, despite the less satisfactory quantitative agreement with the DNS data at small $\Stokes^+$. 
Importantly, the model successfully captures the tendency for radial transport to weaken as $\Stokes^+$ decreases and provides a good estimate of the eventual concentration. 
This thus suggests that the reduced-order model captures the dominant physical mechanisms governing the evolution of particle concentration.

The validity of the stationary closure for $D_1$ and $D_2$, which constitutes the only modelling assumption introduced to obtain equation~\eqref{eq:zeta-evolution}, is also re-examined in light of the remaining discrepancies at small $\tau_\mathrm{p}$.
As shown in figure~\ref{steady_closure}, the particle-sampled radial fluid velocity, associated with biased sampling, converges more slowly than the radial and azimuthal particle velocity fluctuations associated with turbophoretic and centrifugal effects.
This suggests that the stationary treatment may be less appropriate for the biased sampling term, which might cause noticeable deviations for smaller $\tau_\mathrm{p}$, as the biased sampling contribution becomes more important as presented in figure~\ref{mechanism}.
Accordingly, the reduced-order model is expected to perform better for cases with larger Stokes numbers, as shown in figure~$\ref{near_wall_c}$.

\section{Conclusions and outlook}\label{sec:Conclusions}

This study reports the first DNS of particle-laden turbulence in concentric annular ducts, revealing how transverse curvature fundamentally alters inertial-particle transport.
The results show that curvature introduces a pronounced asymmetry in the radial transport.
In particular, in the statistical steady state, the curvature-induced competition between turbophoresis and centrifugal effects leads to severe particle depletion at the inner wall under strong curvature, contrasting with the particle accumulation near the outer wall. 
The near-inner-wall concentration also exhibits a distinct non-monotonic evolution over a range of Stokes numbers, characterized by an initial overshoot followed by a subsequent decrease toward its statistical steady state.

To interpret these behaviours, a transport equation for the particle concentration was derived within a statistical framework.
Through a spectral decomposition of the equation, we demonstrate that the concentration dynamics are governed by a limited number of asymmetric transport modes with distinct decay rates.
In particular, the observed overshoot near the inner wall arises from the phase lag between the outer-wall-directed transport mode analogous to centrifugal drift and the wallward transport mode analogous to turbophoresis.
A reduced-order model retaining only the dominant slowly decaying modes is able to capture the main dynamics of particle transport, successfully characterizing the trend of near-wall concentration across different Stokes numbers.
This modal framework not only elucidates the physics of curvature-induced transport, but also offers a basis for modelling non-equilibrium particle dynamics in complex geometries.

In the present study, cases with different Stokes numbers are included to discuss the effects of varying particle response time in the strong curvature configuration.
The range of Stokes number considered, however, is limited. 
Whether the trend observed with decreasing Stokes number, i.e. 
a more uniform particle distribution at steady state, a less pronounced overshoot, and a slower evolution of concentration, persists at even smaller Stokes number, e.g. $\Stokes^+<1$, remains open for future investigation.
Analyses of cases at higher Reynolds numbers are also of great interest, as the turbophoretic effect is expected to strengthen.

\section*{Appendix A. Effects of compressibility}

\begin{figure}\small
	\centering \subfigure{
		\begin{minipage}[c]{0.49\textwidth}{}
			\begin{overpic}[width=\textwidth]{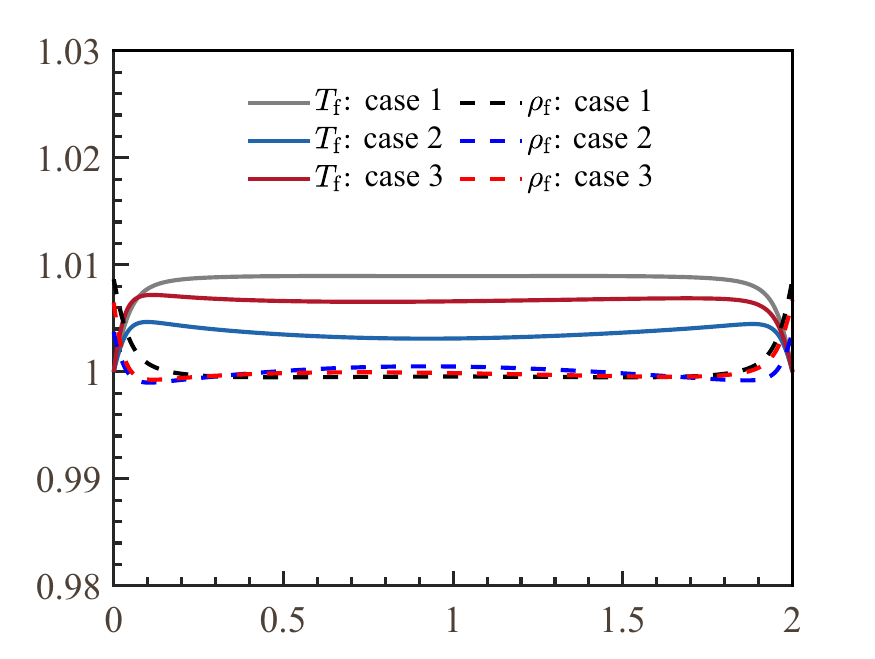}
				\put(-2,68){$(a)$}
				\put(0,34){\rotatebox{90}{$T_\mathrm{f}$, $\rho_\mathrm{f}$}}
				\put(47,-2){$r-R_i$}
			\end{overpic}
		\end{minipage}\label{compressible_ave}}
	\centering \subfigure{
		\begin{minipage}[c]{0.49\textwidth}{}
			\begin{overpic}[width=\textwidth]{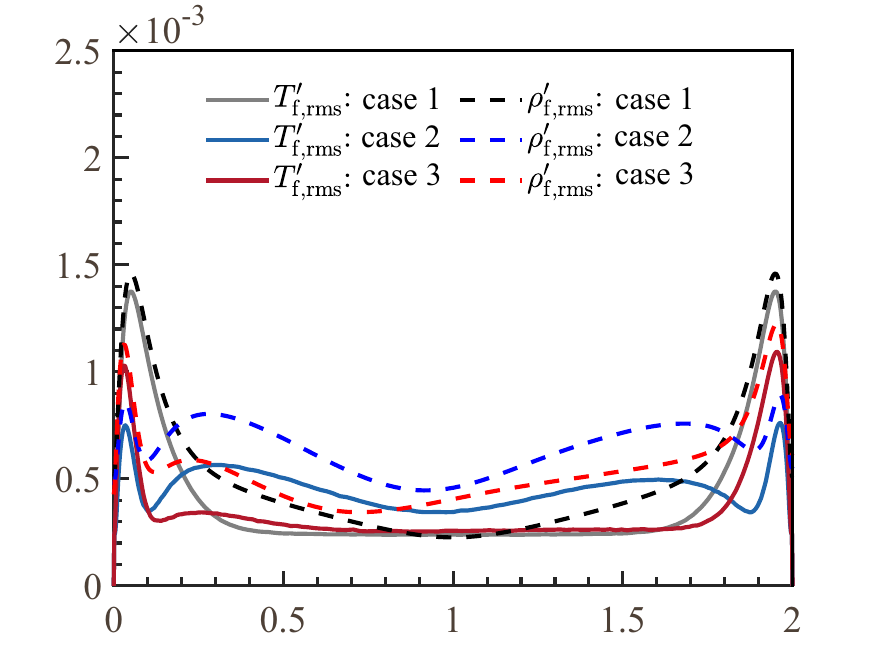}
				\put(-2,68){$(b)$}
				\put(0,27.5){\rotatebox{90}{$T^\prime_\mathrm{f,rms}$, $\rho^\prime_\mathrm{f,rms}$}}
				\put(47,-2){$r-R_i$}
			\end{overpic}
		\end{minipage}\label{compressible_rms}}
	\caption{Profiles of the $(a)$ mean temperature $T_\mathrm{f}$ and density $\rho_\mathrm{f}$ and $(b)$ rms of the corresponding fluctuations $T^\prime_\mathrm{f,rms}$ and $\rho^\prime_\mathrm{f,rms}$.}
	\label{compressiblity}
\end{figure}
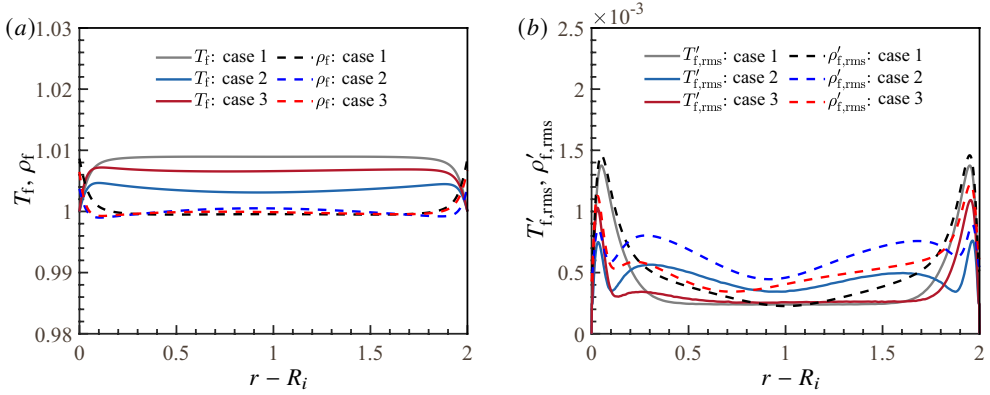

To assess the effects of compressibility, the mean temperature, mean density, and their fluctuations are examined, with detailed profiles presented in figures~$\ref{compressible_ave}$ and $\ref{compressible_rms}$, respectively. 
Both the mean temperature and density remain nearly constant, with deviations confined to within $1\%$ in the near-wall region. 
The fluctuation levels of temperature and density stay below 0.15\%.
Note that the present simulations only consider the drag force \citep{Marchioli2008Statistics} for the particle dynamics, whereas the direct temperature-gradient-induced effect, namely the thermophoretic force \citep{Guha2008Transport,Talbot1980Thermophoresis}, is excluded.
Therefore, the influence of compressibility on particle dynamics is expected to arise predominantly through the slight variation in viscosity, which is considered negligible in the present study.
This is further supported by the close agreement between the present flow and particle statistics and published results, as discussed in figures~\ref{validation} and \ref{particle_statistics}.

\section*{Appendix B. Effects of Reynolds number}

\begin{table}
	\centering
	\begin{tabular}{p{30pt}p{45pt}p{10pt}p{55pt}p{50pt}p{48pt}p{48pt}p{48pt}}
		\multicolumn{1}{c}{\multirow{2}{*}{\centering case}} &
		\multicolumn{1}{c}{$\Rey_\tau$} &
		\multicolumn{1}{c}{\multirow{2}{*}{\centering $\eta$}} &
		\multicolumn{1}{c}{\multirow{2}{*}{\centering $L_r,L_\theta,L_z$}} &
		\multicolumn{1}{c}{\multirow{2}{*}{\centering $N_r,N_\theta,N_z$}} &
		\multicolumn{1}{c}{$\Delta r^+$} &
		\multicolumn{1}{c}{$\Delta\theta^+$} &
		\multicolumn{1}{c}{$\Delta z^+$} \\
		\multicolumn{1}{c}{\multirow{2}{*}{\centering }} &
		\centering inner, outer &
		\multicolumn{1}{c}{\multirow{2}{*}{\centering }} &
		\multicolumn{1}{c}{\multirow{2}{*}{\centering }} &
		\multicolumn{1}{c}{\multirow{2}{*}{\centering }} &
		\centering inner, outer &
		\centering inner, outer &
		\multicolumn{1}{c}{\centering inner, outer} \\
		\hline
  %       \centering $3$ &
		% \centering $194,153$ & 
		% \centering $0.1$ &
		% \centering $2\delta,\frac{22}{9}\pi\delta,12\pi\delta$ &
		% \centering $193,256,566$ &
		% \centering $0.305,0.240$ & 
		% \centering $1.066,8.404$ &  
		% \multicolumn{1}{c}{\centering $13.023,10.263$}\\
		\centering $7$ &
		\centering $233,185$ & 
		\centering $0.1$ &
		\centering $2\delta,\frac{22}{9}\pi\delta,16\delta$ &
		\centering $241,320,300$ &
		\centering $0.292,0.232$ & 
		\centering $1.025,8.140$ &  
		\multicolumn{1}{c}{\centering $12.525,9.950$}
	\end{tabular}
	\caption{Parameters for case 7. Here, $\Rey_\tau$ is the friction Reynolds number, $N$ and $\Delta^+$ denote the number of grid nodes and the grid resolution in wall units in each direction, respectively. `Inner' and `outer' denote quantities normalized by the viscous scales at the inner and outer walls, respectively.}
	\label{Parameters_continuous_Re3062}
\end{table}
\begin{table}
	\centering
	\begin{tabular}{p{15pt}p{22pt}p{22pt}p{40pt}p{45pt}p{30pt}p{48pt}p{48pt}p{48pt}}
		\multicolumn{1}{c}{\multirow{2}{*}{\centering case}} &
        \multicolumn{1}{c}{\multirow{2}{*}{\centering $d_\mathrm{p}/\delta$}} &
		\multicolumn{1}{c}{\multirow{2}{*}{\centering $\rho_\mathrm{p}/\rho_\mathrm{f}$}} &
		\multicolumn{1}{c}{\multirow{2}{*}{\centering $\tau_\mathrm{p}/(\delta/U_b)$}}&
		\multicolumn{1}{c}{\multirow{2}{*}{\centering $\varPhi_\mathrm{p}$}} &
		\multicolumn{1}{c}{\multirow{2}{*}{\centering $N_\mathrm{p}$}} &
		\multicolumn{1}{c}{$d_\mathrm{p}/\eta_\mathrm{w}$} &
		\multicolumn{1}{c}{$d^+_\mathrm{p}$} &
		\multicolumn{1}{c}{$\Stokes^+$} \\
		\multicolumn{1}{c}{\multirow{2}{*}{\centering }} &
		\multicolumn{1}{c}{\multirow{2}{*}{\centering }} &
		\multicolumn{1}{c}{\multirow{2}{*}{\centering }} &
        \multicolumn{1}{c}{\multirow{2}{*}{\centering }} &
		\multicolumn{1}{c}{\multirow{2}{*}{\centering }} &
		\multicolumn{1}{c}{\multirow{2}{*}{\centering }} &
		\centering inner, outer &
		\centering inner, outer &
		\multicolumn{1}{c}{\centering inner, outer} \\
		\hline
  %       \centering 3 &
  %       \centering 0.004 &
		% \centering 1200 &
		% \centering 2.6133 &
		% \centering $2.122\times10^{-5}$ &
		% \centering 366667 &
		% \centering $0.457,0.400$ &  
		% \centering $0.782,0.616$ & 
		% \multicolumn{1}{c}{\centering $40.535,25.179$}\\
		\centering 7 &
		\centering 0.004 &
		\centering 1200 &
		\centering 3.2661 &
		\centering $2.122\times10^{-5}$ &
		\centering 155618 &
		\centering $0.551,0.486$ & 
		\centering $0.939,0.746$ &  
		\multicolumn{1}{c}{\centering $58.461,36.909$}
	\end{tabular}
	\caption{Parameters for case 7. Here, $N_\mathrm{p}$ denotes the total particle number, $\varPhi_\mathrm{p}$ denotes the particle volume fraction, and $\eta_\mathrm{w}$ denotes the smallest eddy scale near the wall.}
	\label{Parameters_dispersed_Re3062}
\end{table}

To assess the robustness of the reported phenomena at a higher Reynolds number, an additional simulation (case 7) is conducted at $\Rey_\mathrm{f}=3062$ for the strong-curvature configuration $\eta = 0.1$. 
To reduce the computational cost, the axial domain length is set to $L_z=16\delta$, consistent with the setup adopted by \citet{Orlandi2025Effects}.
The grid resolution satisfies the requirements reported in previous DNS studies of annular pipe flow \citep{Chung2002Direct, Bagheri2020Effects}, as detailed in table~\ref{Parameters_continuous_Re3062}.
The increase in bulk Reynolds number yields an outer-wall friction Reynolds number of $\Rey_{\tau,o}\approx185$, which is approximately 20\% higher than the value of $\Rey_{\tau,o}\approx153$ in case 3.
Accordingly, case 7 represents a more turbulent flow and is therefore expected to exhibit stronger turbophoretic effects \citep{Bernardini2014Reynolds}.
The particle diameter and density are chosen to be identical to those in case 3 for a direct comparison, and the corresponding parameters are summarized in table~\ref{Parameters_dispersed_Re3062}.

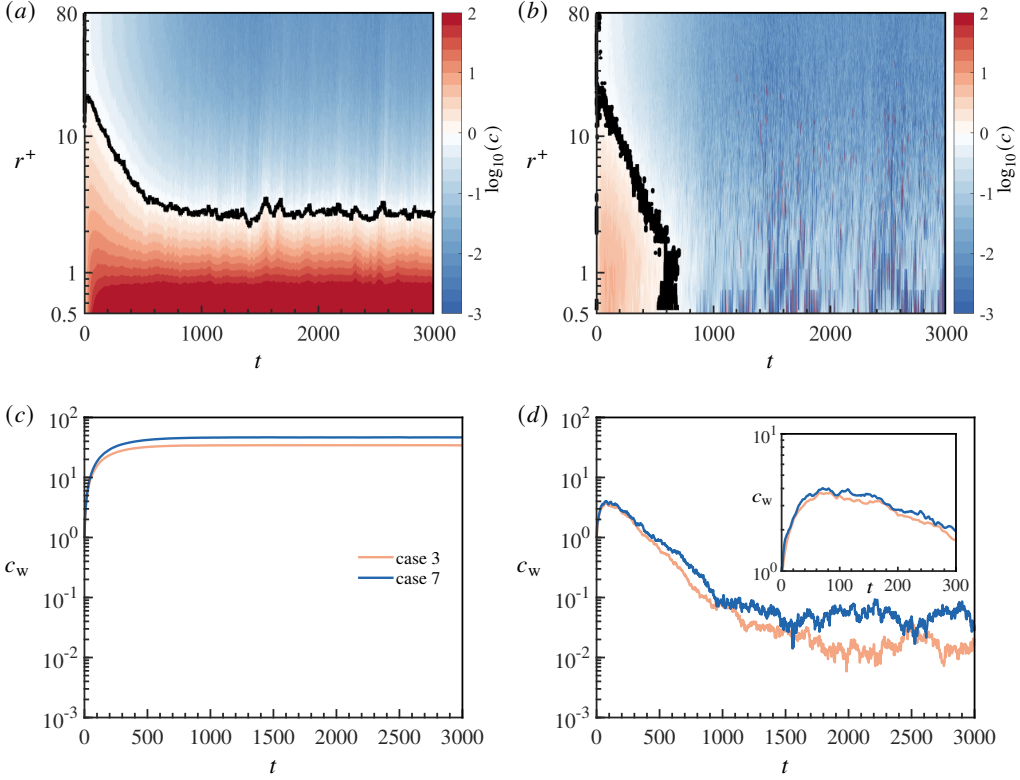
\begin{figure}\small  
	\centering \subfigure{
		\begin{minipage}[c]{0.49\textwidth}{}
			\begin{overpic}[width=\textwidth]{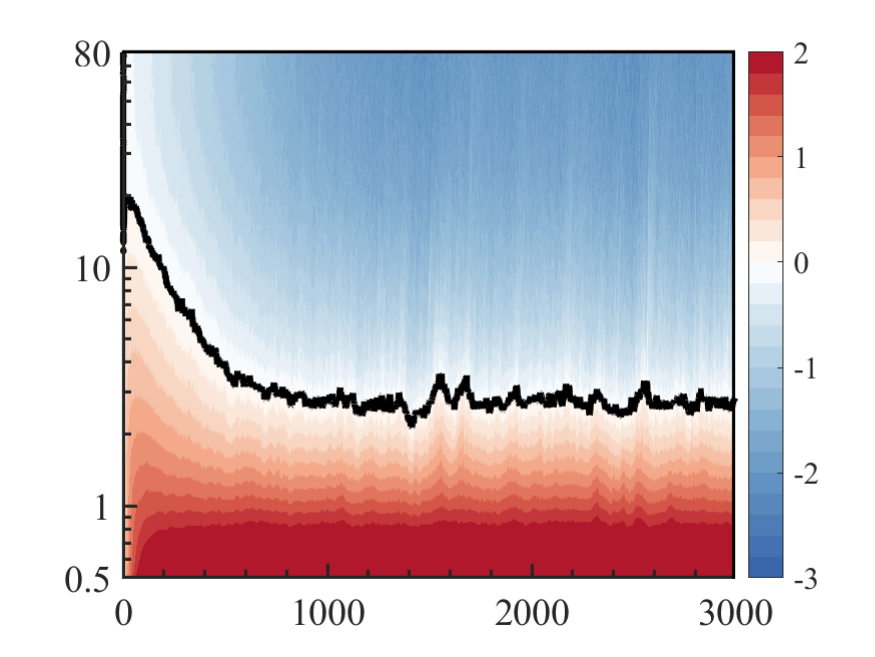}
				\put(-2,68){$(a)$}
				\put(0,38){$r^+$}
				\put(48,-2){$t$}
				
				\put(94,33){\rotatebox{90}{\scriptsize$\log_{10}(c)$}}
			\end{overpic}
		\end{minipage}\label{C_eta01_Re3062_outer}}
	\centering \subfigure{
		\begin{minipage}[c]{0.49\textwidth}{}
			\begin{overpic}[width=\textwidth]{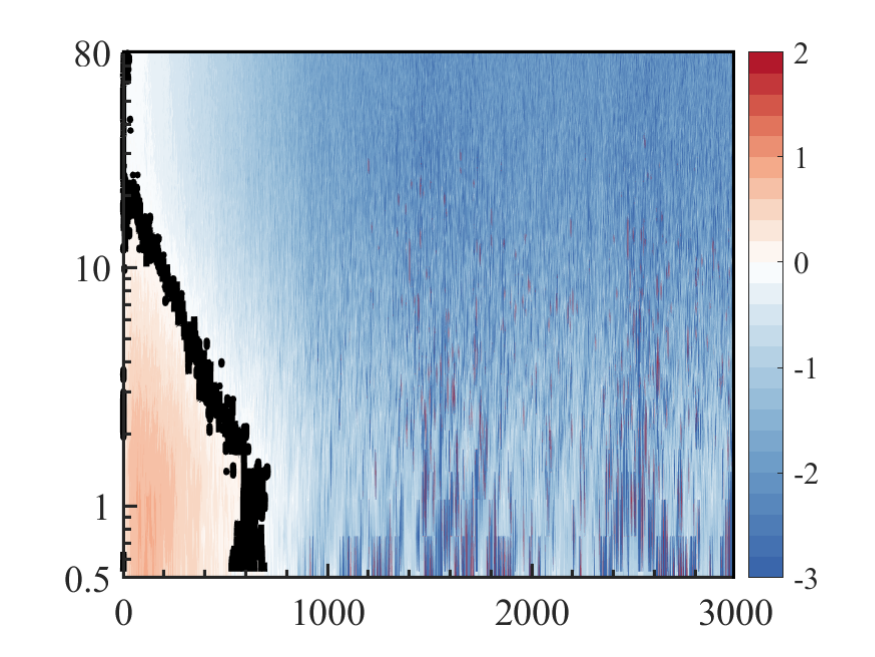}
				\put(-2,68){$(b)$}
				\put(0,38){$r^+$}
				\put(48,-2){$t$}
				
				\put(94,33){\rotatebox{90}{\scriptsize$\log_{10}(c)$}}
			\end{overpic}
		\end{minipage}\label{C_eta01_Re3062_inner}}
	\centering \subfigure{
		\begin{minipage}[c]{0.49\textwidth}{}
			\begin{overpic}[width=\textwidth]{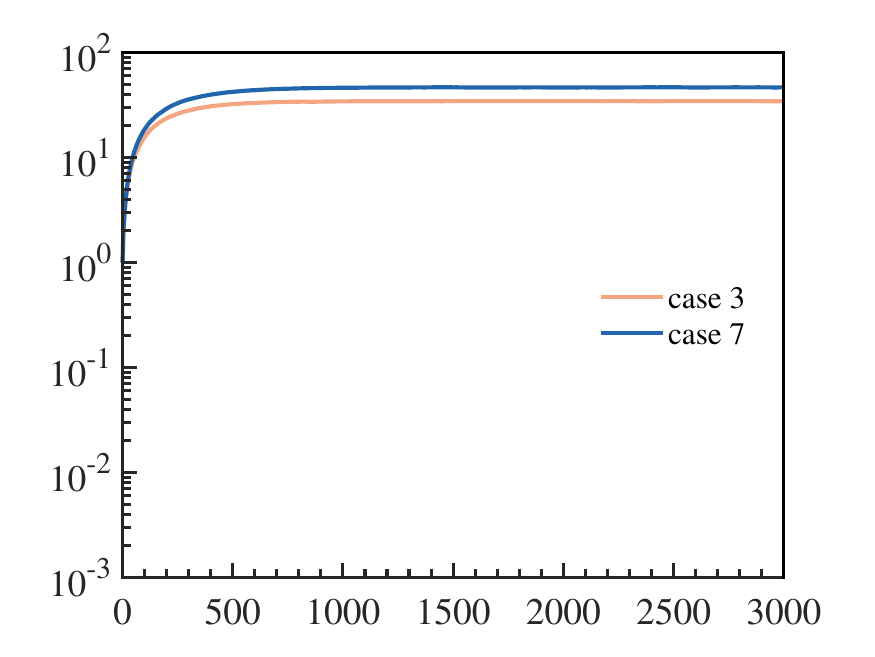}
				\put(-2,68){$(c)$}
				\put(-2,38){$c_\mathrm{w}$}
				\put(51,-2){$t$}
			\end{overpic}
		\end{minipage}\label{Cw_outer_Re3062}}
	\centering \subfigure{
		\begin{minipage}[c]{0.49\textwidth}{}
			\begin{overpic}[width=\textwidth]{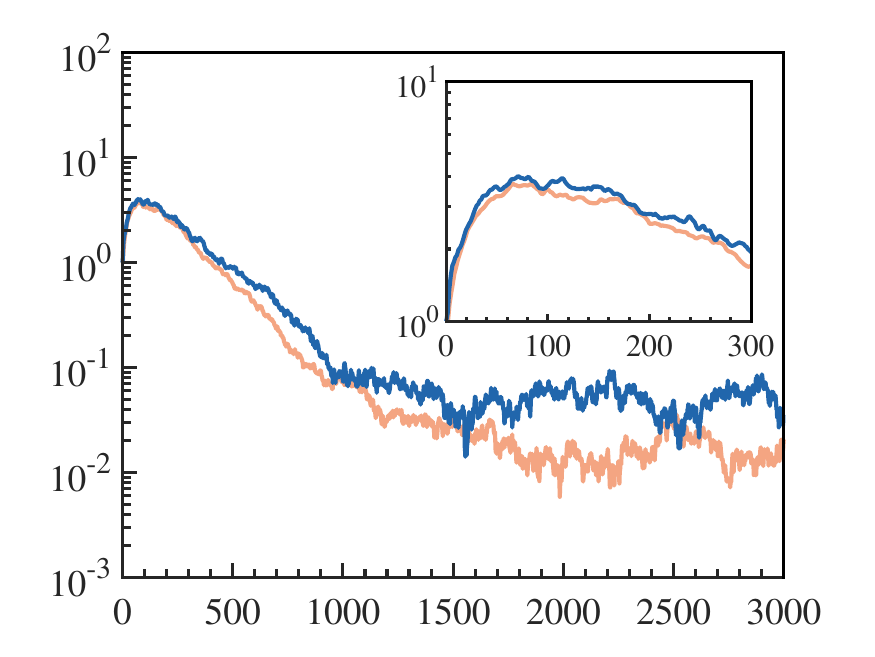}
				\put(-2,68){$(d)$}
				\put(-2,38){$c_\mathrm{w}$}
				\put(51,-2){$t$}

                \put(45,52){\scriptsize$c_\mathrm{w}$}
                \put(68,34){\scriptsize$t$}
			\end{overpic}
		\end{minipage}\label{Cw_inner_Re3062}}  
	\caption{Spatial and temporal evolution of the dimensionless particle concentration $c(r,t)$ near the $(a)$ outer and $(b)$ inner walls for case 7, and $(c,d)$ the corresponding dimensionless mean particle concentration within the viscous sublayer ($r^+<5$). Here, $r^+$ denotes the viscous-scaled radial distance, and contours represent the particle concentration $c(r,t)$, with black iso-lines denoting $c(r,t)=1$. A zoomed-in view of inner-wall concentration $c_\mathrm{w}$ is added in figure~$\ref{Cw_inner_Re3062}$.}
	\label{C_evolution_Re3062}
\end{figure}

Figure~\ref{C_evolution_Re3062} presents the spatial and temporal evolution of the particle concentration near both walls for case 7. 
Noticeable differences are observed between the concentration distributions near the two walls in case 7, further confirming the asymmetric nature of the radial transport. 
Particles also continue to accumulate near the outer wall, as indicated by the monotonic increase of $c_\mathrm{w}$ in figure~$\ref{Cw_outer_Re3062}$, whereas the inner-wall concentration exhibits the characteristic non-monotonic overshoot identified in \S~\ref{subsec:transport}. 
Consistent with the stronger turbophoretic effect at the higher Reynolds number, the eventual near-wall concentration levels at both walls are higher than those in case 3, and the overshoot amplitude at the inner wall is also slightly increased, as shown in the inset of figure~$\ref{Cw_inner_Re3062}$.

\begin{figure}\small
	\centering \subfigure{
		\begin{minipage}[c]{0.49\textwidth}{}
			\begin{overpic}[width=\textwidth]{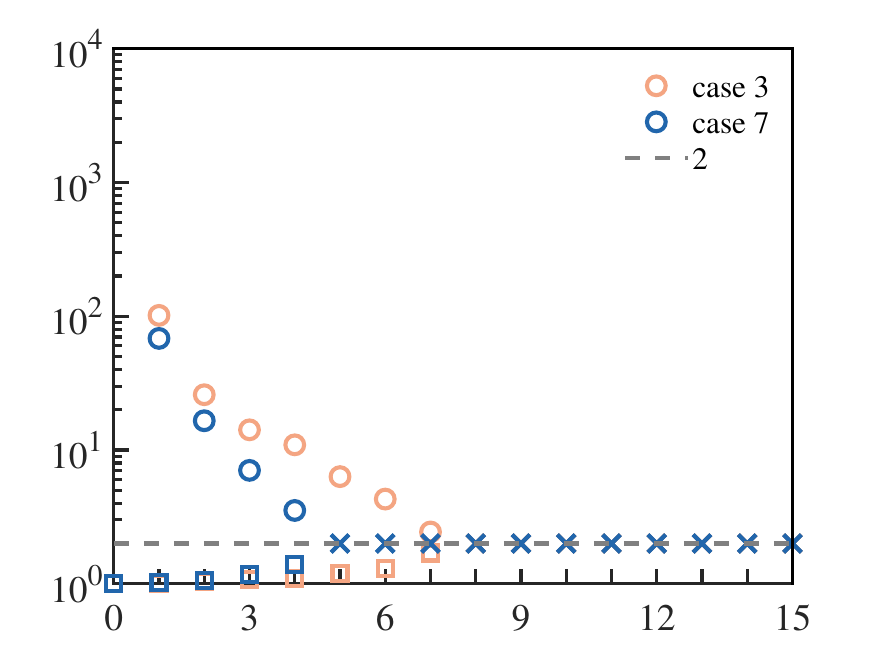}
				\put(-2,68){$(a)$}
				\put(-1,25){\rotatebox{90}{${|\mathrm{Re}(s_n^\pm)|}^{-1}/\tau_\mathrm{p}$}}
				\put(50.5,-2){$n$}
			\end{overpic}
		\end{minipage}\label{time_scales_Re3062}}
    \centering \subfigure{
		\begin{minipage}[c]{0.49\textwidth}{}
			\begin{overpic}[width=\textwidth]{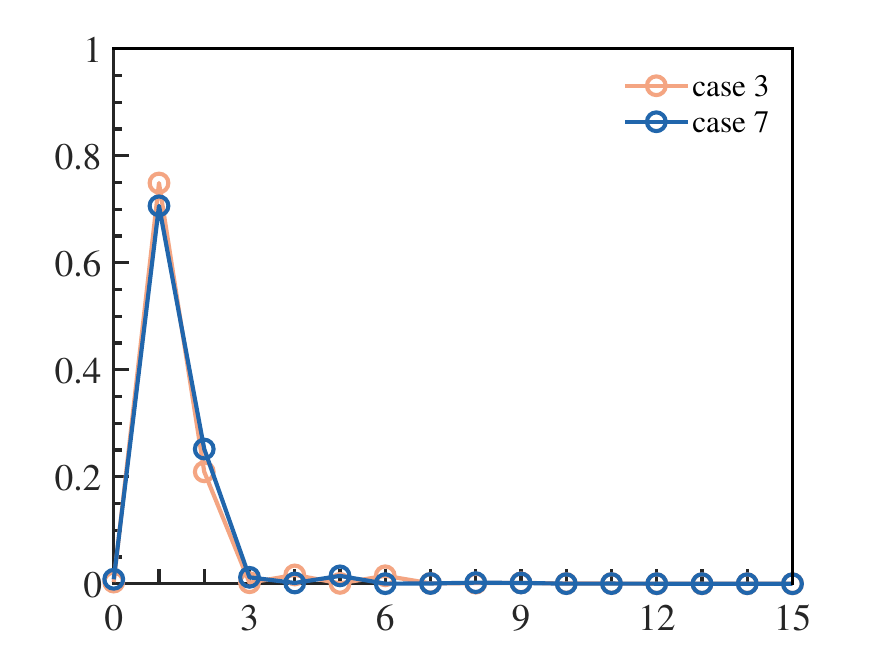}
				\put(-2,68){$(b)$}
                \put(-1,30){\rotatebox{90}{$E_{k,n}/E_k$}}
				\put(50.5,-2){$n$}
			\end{overpic}
		\end{minipage}\label{energy_Re3062}}    
    \centering \subfigure{
		\begin{minipage}[c]{0.49\textwidth}{}
			\begin{overpic}[width=\textwidth]{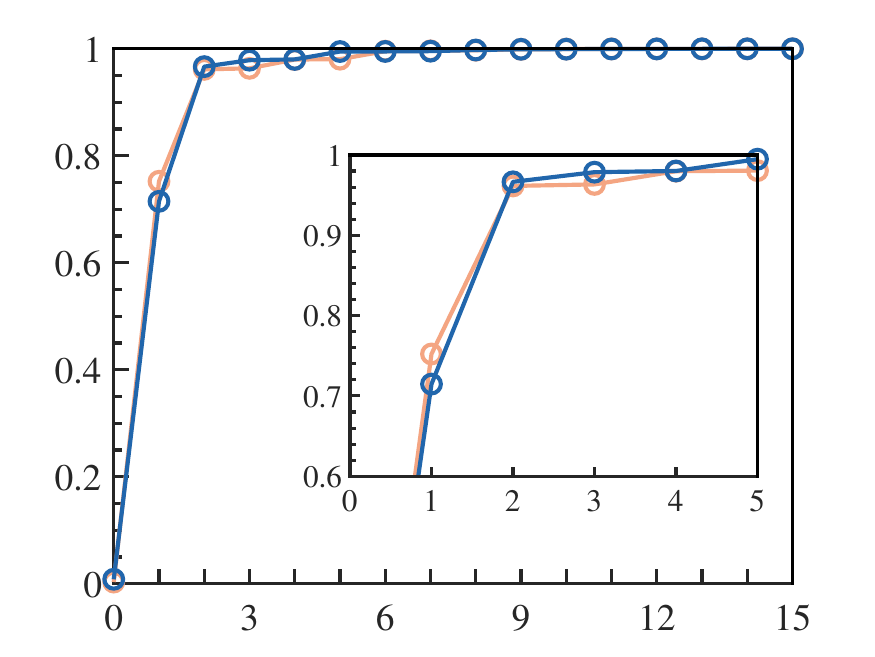}
				\put(-2,68){$(c)$}
                \put(-1,26){\rotatebox{90}{$\sum_{i=0}^{n}E_{k,i}/E_k$}}
				\put(50.5,-2){$n$}

                \put(62,15){\scriptsize$n$}
			    \put(29,28){\scriptsize\rotatebox{90}{$\sum_{i=0}^{n}E_{k,i}/E_k$}}
			\end{overpic}
		\end{minipage}\label{truncation_Re3062}}    
    \centering \subfigure{
		\begin{minipage}[c]{0.49\textwidth}{}
			\begin{overpic}[width=\textwidth]{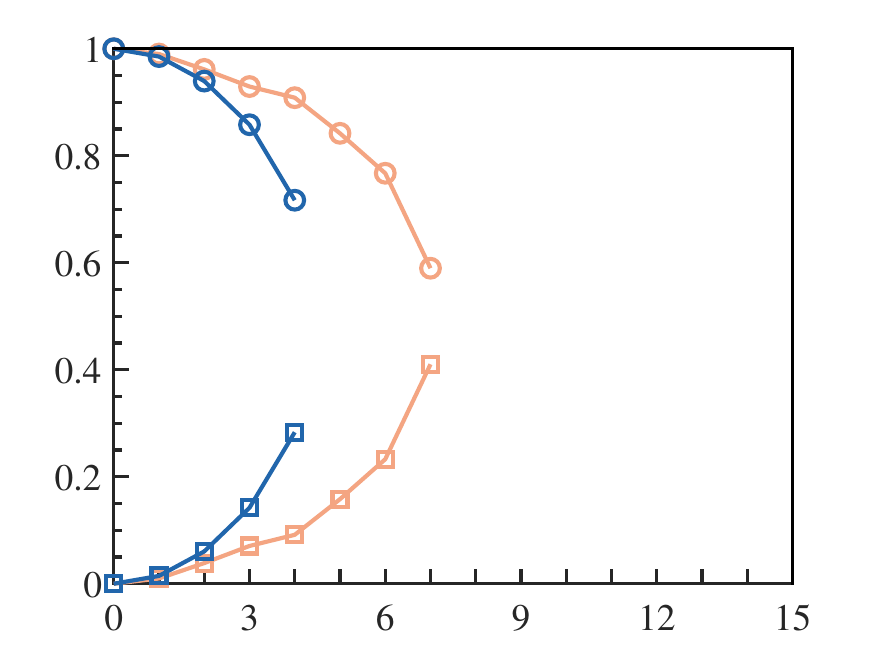}
				\put(-2,68){$(d)$}
                \put(-1,32.5){\rotatebox{90}{$\alpha_n$, $\beta_n$}}
				\put(50.5,-2){$n$}
			\end{overpic}
		\end{minipage}\label{amplititudes_relative_Re3062}}
	\caption{Characteristics of the modes: $(a)$ the time scales normalized by $\tau_\mathrm{p}$ of the slowly decaying modes (circle), rapidly decaying modes (square), and oscillatory decaying modes (cross); $(b)$ the initial modal energy contribution $E_{k,n}/E_{k}$; $(c)$ the cumulative modal energy contribution $\sum_{i=0}^{n}E_{k,i}/E_k$; $(d)$ the relative amplitudes of slowly decaying modes $\alpha_n=|a_n|/(|a_n|+|b_n|)$ (circle) and rapidly decaying modes $\beta_n=|b_n|/(|a_n|+|b_n|)$ (square).}
	\label{modes_Re3062}
\end{figure}
\begin{figure}\small
	\centering \subfigure{
		\begin{minipage}[c]{0.49\textwidth}{}
			\begin{overpic}[width=\textwidth]{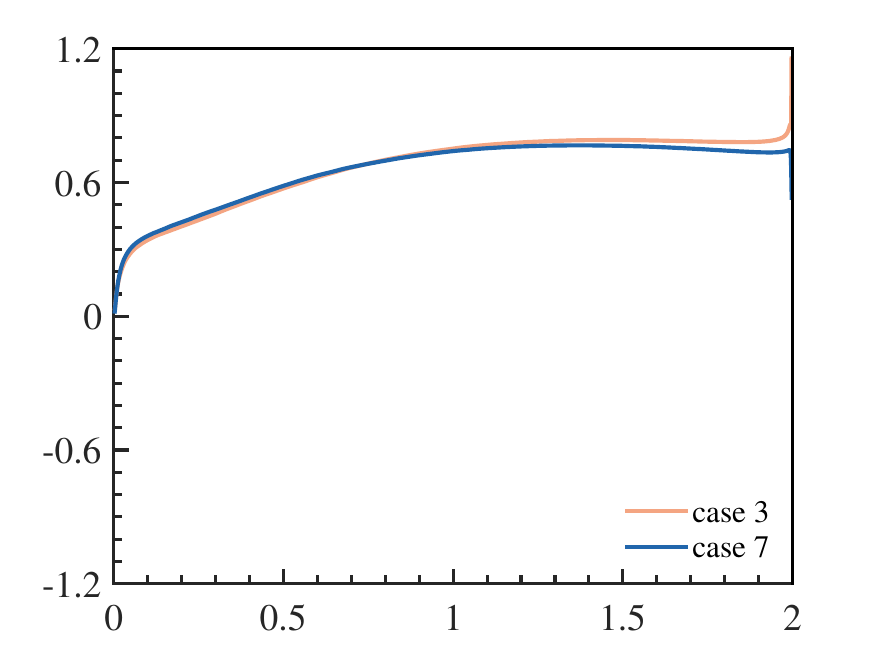}
				\put(-2,68){$(a)$}
				\put(0,31){\rotatebox{90}{$J^+_1(r,0)$}}
				\put(47,-2){$r-R_i$}
			\end{overpic}
		\end{minipage}\label{mode_flux_1_Re3062}}
	\centering \subfigure{
		\begin{minipage}[c]{0.49\textwidth}{}
			\begin{overpic}[width=\textwidth]{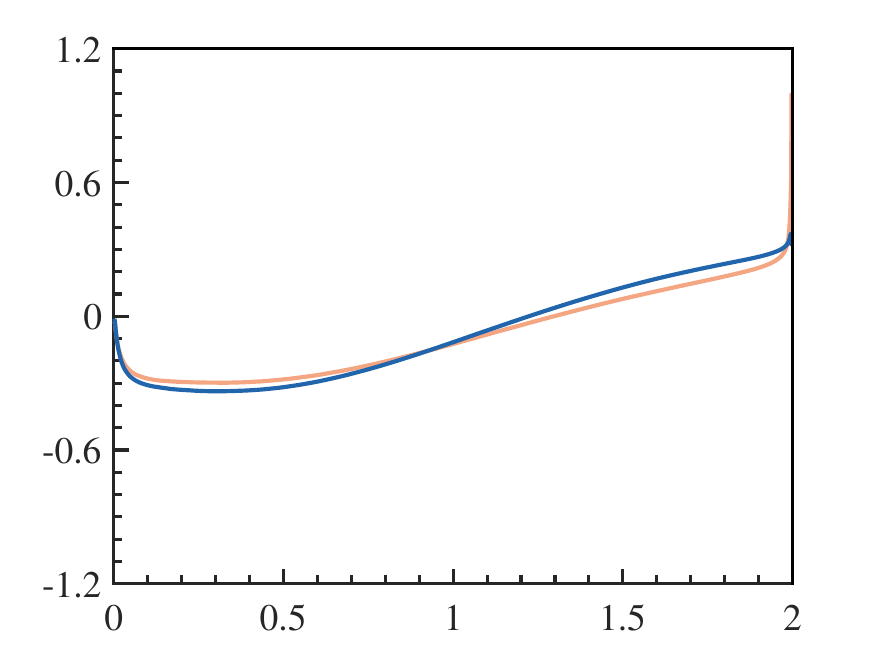}
				\put(-2,68){$(b)$}
				\put(0,31){\rotatebox{90}{$J^+_2(r,0)$}}
				\put(47,-2){$r-R_i$}
			\end{overpic}
		\end{minipage}\label{mode_flux_2_Re3062}}
	\caption{Radial profile of initial flux $J^+_n(r,0)$ for $(a)$ the outer-wall transport mode and $(b)$ the wallward transport mode.}
	\label{mode_flux_Re3062}
\end{figure}

The characteristics of the Sturm-Liouville modes are revisited in figure~\ref{modes_Re3062}.
As shown in figure~$\ref{time_scales_Re3062}$, the increase in Reynolds number reduces the time scales of the slowly decaying modes, indicating enhanced radial particle transport.
Figure~$\ref{energy_Re3062}$ shows that the leading modes, especially the first- and second-order modes, retain larger initial energy than the higher-order modes.
Notably, a higher $\Rey_\tau$ is associated with a decrease in $E_{k,1}$ but an increase in $E_{k,2}$, suggesting that the contribution of the curvature-induced centrifugal mechanism weakens while that of turbophoresis strengthens, consistent with the overall physical picture.
The modes of the first three orders account for more than 90\% of the total initial energy, as shown in figure~$\ref{truncation_Re3062}$.
Most of this contribution is associated with the slowly decaying modes, as indicated by figure~$\ref{amplititudes_relative_Re3062}$, confirming their dominant role in the transient evolution.
Consistent with the conclusion drawn from figure~$\ref{energy_Re3062}$, a moderate increase in $\Rey_\tau$ slightly weakens the outer-wall-directed transport mode near the outer wall, while strengthening the wallward transport mode, as reflected by the initial flux profiles $J^+_n(r,0)$ presented in figure~$\ref{mode_flux_Re3062}$.

\begin{figure}\small
	\centering \subfigure{
		\begin{minipage}[c]{0.49\textwidth}{}
			\begin{overpic}[width=\textwidth]{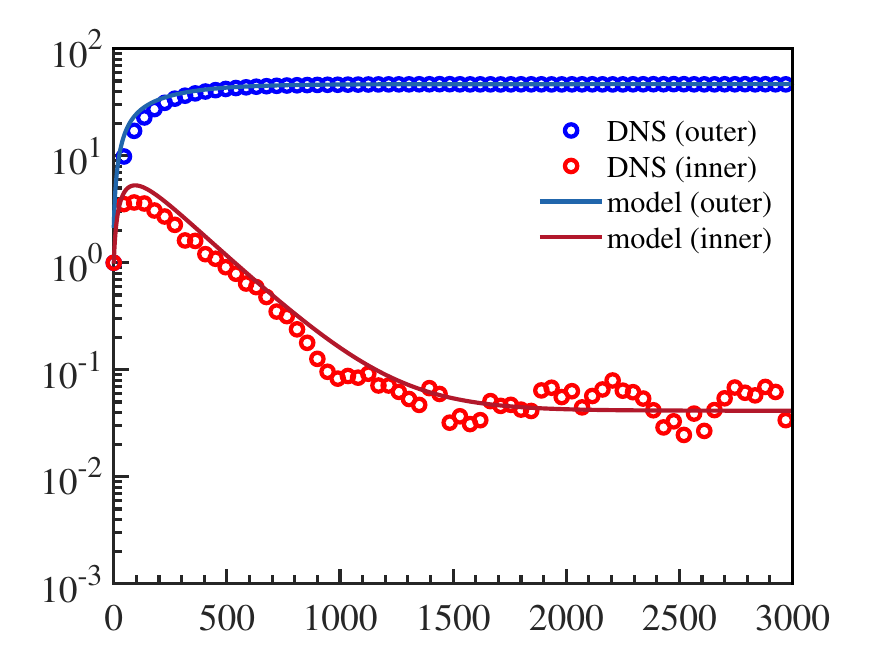}
				\put(-2,38){$c_\mathrm{w}$}
				\put(51,-2){$t$}
			\end{overpic}
		\end{minipage}}
	\caption{Reconstructed near-wall concentration $c_\mathrm{w}$ for case 7.}
	\label{near_wall_c_Re3062}
\end{figure}

The proposed three-mode reconstruction in equation~\eqref{eq:reconstruction} is then applied to case 7, and the corresponding results are presented in figure~\ref{near_wall_c_Re3062}.
The reduced-order model remains applicable at the higher Reynolds number. 
The model accurately captures the inner-wall overshoot, agreeing well with the DNS data.
This is expected, as the turbophoretic effect is of great importance in this high Reynolds number case (see discussion in \S~\ref{subsec:modes}).

\backsection[Acknowledgements]{We wish to thank Prof. Sandberg for the license of using HiPSTAR. This work has been supported by National Natural Science Foundation of China (Grant Nos. 12588201, 12572247 and 12432010).}

\backsection[Declaration of interests]{The authors report no conflict of interest.}

\bibliographystyle{jfm}
\bibliography{tyw}

\end{document}